\documentclass[acmtog]{acmart}

\AtBeginDocument{%
  \providecommand\BibTeX{{%
    \normalfont B\kern-0.5em{\scshape i\kern-0.25em b}\kern-0.8em\TeX}}}

\setcopyright{acmcopyright}
\copyrightyear{2021}
\acmYear{2021}
\acmDOI{10.1145/1122445.1122456}

\acmJournal{TOG}

\usepackage{microtype}
\usepackage{textcomp}
\usepackage{dblfloatfix}
\usepackage{color}
\usepackage{todonotes}
\usepackage[normalem]{ulem}
\usepackage[utf8]{inputenc}
\usepackage{soul}

\newlength{\picturewidth}
\newlength{\pictureheight}
\newlength{\imageheight}
\newlength{\imagewidth}
\graphicspath{{images/}{images/new_images/}{images/new_images0/}{images/new_images1/}{images/mould/}{images/original/}{images/crops/}{images/pdfs/}}

\newcommand{\eg}{e.\,g.}
\newcommand{\ie}{i.\,e.}

\usepackage{algorithmicx,algorithm}
\usepackage{float}
\newfloat{algorithm}{t}{lop}
\usepackage{algpseudocode}
\usepackage{amsmath}
\usepackage{mathtools}
\usepackage{graphicx}
\usepackage{enumitem}
\usepackage{ccicons}
\usepackage{nicefrac}

\usepackage[hang,small,sf,SF]{subfigure}
\usepackage{alphalph}
\begin{document}

\title{Interpolation of Point Distributions for Digital Stippling}

\author{Germ{\'a}n Arroyo}
\email{arroyo@ugr.es}
\orcid{0000-0001-7229-5029}
\affiliation{%
  \institution{University of Granada}
  \streetaddress{}
  \city{}
  \state{}
  \postcode{}
  \country{Spain}
}

\author{Domingo Mart{\'i}n}
\email{dmartin@ugr.es}
\orcid{0000-0002-4088-0554}
\affiliation{%
  \institution{University of Granada}
  \streetaddress{}
  \city{}
  \state{}
  \postcode{}
  \country{Spain}
}

\author{Tobias Isenberg}
\email{tobias.isenberg@inria.fr}
\orcid{0000-0001-7953-8644}
\affiliation{%
  \institution{Universit{\'e} Paris-Saclay, CNRS, Inria, LISN}
  \streetaddress{}
  \city{}
  \country{France}
}


\begin{abstract}
We present a new way to merge any two point distribution approaches using distance fields. Our new process allows us to produce digital stippling that fills areas with stipple dots without visual artifacts as well as includes clear linear features without fussiness. Our merging thus benefits from past work that can optimize for either goal individually, yet typically by sacrificing the other. The new possibility of combining any two distributions using different distance field functions and their parameters also allows us to produce a vast range of stippling styles, which we demonstrate as well.
\end{abstract}

%

\ccsdesc[500]{Computing methodologies~Non-photorealistic rendering}

%

\keywords{point distributions, digital stippling.}


\maketitle
\makeatletter
\global\@topnum\z@
\global\@botnum\z@
\makeatother

\section{Introduction}
\label{sec:intro}

Digital stippling \cite{Deussen:2013:HAS,Martin:2017:SDS} is a non-photorealistic rendering technique \cite{Gooch:2001:NPR,Strothotte:2002:NPC,Rosin:2013:IVA} that emulates hand-made stippling by means of distributing stipple dots based on target images. The traditional art form has been and continues to be used for illustrations in several scientific domains such as biology, entomology, or archeology. For artists, while it seemingly only requires to place dots on paper using an ink pen, the traditional technique is an arduous and extremely repetitive task. It requires of time to produce relatively small drawings (\eg, A4) because the artist must place thousands of dots and has to correctly reproduce tone, shape, and texture. Ultimately, this illustration technique is thus increasingly being replaced by others due to its high cost.

Research in NPR, however, has led to numerous approaches for producing digital stippling without such long production times \cite{Martin:2017:SDS}, taking into account not only the actual dot distribution but also the image resolutions \cite{Martin:2010:EBS,Martin:2011:SDE}, the shape of the dots \cite{Martin:2015:DCR,Martin:2019:ADC}, and properties such as the introduced abstraction \cite{Spicker:2017:QVA,Spicker:2019:QVA}. One of the main driving forces of digital stippling has been the stipple dot placement without visual artifacts, as early approaches based on Centroidal Voronoi Diagrams \cite{Lloyd:1982:LSQ,McCool:1992:HPD} lead to unintended chains of dots which stipple artists are trained to avoid \cite{Hodges:2003:TGH}. While several approaches have since addressed this artifact issue (see the overview in \citeauthor{Martin:2017:SDS}'s [\citeyear{Martin:2017:SDS}] survey), sometimes it is still important that the dots are, in fact, aligned to intentionally represent dedicated features. Consequently, some approaches (\eg, \cite{Mould:2007:SPD,Li:2011:SPS,Li:2017:PSS}) focus specifically on placing dots in a structure-aware fashion.

The fundamental problem is now that any given point distribution technique usually either produces nice point distributions for filling areas of an image, or it produces adequate feature stippling. More generally, to produce a digital stippling one may want to combine any two dot distribution techniques that one chooses for a given visual goal. In this article we thus describe a new approach to realize such a smooth interpolation between two point distributions such that their respective properties are maintained. For this purpose we consider stippling algorithms as discrete probabilistic functions, so we can create a new algorithm based on the interpolated function of their distributions. Our approach to combine two distributions is general enough such that it not only allows us to solve our particular problem. Instead, it also allows us to produce a great variety of results by only selecting different distributions and/or adjusting the distance field, which we see as a novel way of controlling the stylization for digital stippling.

\begin{figure*}[t!]
	\begin{center}
		\includegraphics[width=\linewidth]{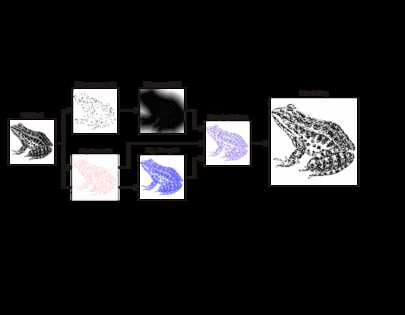}\vspace{-2ex}
		\caption{Overview of our method (while we use color here to distinguish the distributions, the result only uses grayscale or black and white).}\vspace{-1ex}
		\label{figure:scheme}
	\end{center}
\end{figure*}

\section{Previous Work}
\label{sec:review}

A complete survey of digital stippling is beyond the scope of this section and has recently been presented elsewhere \cite{Martin:2017:SDS}, so here we only briefly mention some main related work and those techniques that specifically apply to our approach. The reproduction of stippling using computers began in the late 1990s. Researchers initially tried to place dots appropriately to represent a picture. These initial solutions were based on the concept of Centroidal Voronoi Diagrams (also called Lloyd's method \cite{Lloyd:1982:LSQ,McCool:1992:HPD}); the main idea is to equalize the energy between Voronoi regions by moving the Voronoi sites to the cell's centroids. The first implementation was a painting systems with brushes to modify the dot distribution \cite{Deussen:1999:CGS,Deussen:2000:FPM}. \citet{Secord:2002:WVS} later extended this approach by using the tone of the original image as a weight to control the distance metric.



Original CVDs and their extensions have the mentioned problem of producing patterns due to the regularity of the distribution, which must be avoided for realistic reproduction \cite{Hodges:2003:TGH}. Several solutions have been investigated to reduce or solve this problem. \citet{Schlechtweg:2005:RMA}, for instance, used autonomous agents, the Renderbots, to place stipples with some random processing. Another idea is to use distributions with blue noise properties. \citet{Kopf:2006:RWT}, for example, presented a method based on Wang tiles and Poisson disk sampling, which has blue noise characteristics and thus avoids patterns and, moreover, allows them to smoothly and indefinitely zoom into stipple images. \citet{Balzer:2009:CPD} presented a capacity-constrained way to create point distributions based on Lloyd's method that also possess blue noise characteristics. A generalization of the original CVD but using an optimization of the energy was described by \citet{Deussen:2009:APP}. Another blue noise approach that is also fast was presented by \citet{Ascencio-Lopez:2010:AIS}. Finally, \citet{Arroyo:2010:SGD} used a Monte Carlo technique for sampling an adaptive probability density function. 

Another way to avoid visible patterns is to use example-based techniques and hand-made stippling as input. For example, \citet{Barla:2006:IHS,Barla:2006:SPA} synthesized different hatching and stippling styles using techniques from texture synthesis. Later, \citet{Kim:2009:SBE} synthesized dot distributions using gray-level co-occurrence matrices (GLCM) for a statistical analysis of hand-made samples. Finally, \citet{Martin:2010:EBS,Martin:2011:SDE} showed that the use of resolution-dependent halftoning for dot distributions combined with scanned dots can also achieve satisfactory results.

In contrast to the approaches discussed thus far that aim to avoid visual details, others attempt to use stipples that line up to form linear structures on purpose. One form is called hedcut stippling \cite{Kim:2008:FGI,Kim:2010:AHI,Son:2011:SGD} and is somewhat related to hatching. More related to our own work, however, are structure-aware techniques that aim to highlight particular sparse features in an image using aligned stipples, while reproducing the remainder of the image without such artifacts. For example, \citet{Mould:2007:SPD} used a graph whose edge weights recorded the magnitude of the local image gradient. He then used a version of Disjkstra's algorithm to determine stipple dot positions, emphasizing particularly edges in an image. Later, \citet{Li:2011:SPS,Li:2017:PSS} based another structure-aware stippling technique on an approach for contrast-aware halftoning \cite{Chang:2009:SAE,Li:2010:CAH}.

None of the techniques are ideal, however. For example, while \citet{Li:2011:SPS,Li:2017:PSS} are able to nicely emphasize linear structures, the regularity of their dot placement is sometimes noticeable, in particular, in dark regions. In contrast, example-based approach such as those by \citet{Kim:2009:SBE} and \citet{Martin:2011:SDE} can produce convincing distributions for relatively dark or relatively light regions, but fail to properly reproduce linear structures.

\section{Overview}
\label{sec:overview}

Our visual goal is that of hand-made stippling. Here, artists avoid any visible artifact in dot placement for most of the stippled regions \cite{Hodges:2003:TGH}. To represent features, edges, or borders, however, they align the dots or even use proper lines instead of chained dots (\eg, see the examples in \citeauthor{Hodges:2003:TGH}'s [\citeyear{Hodges:2003:TGH}] book or in Figure~1 in \citeauthor{Martin:2017:SDS}'s [\citeyear{Martin:2017:SDS}] survey). To be able to achieve a similar effect with digital stippling we need to address the mentioned problem of recent related work which either focuses on artifact avoidance or on feature representation. In this paper we thus demonstrate how we can interpolate between any given two distributions to be able to use them for those parts where they shine, and use others where they would not be ideal. We base our approach on a dedicated distance field that we use to determine which of the two distributions to use. Usually we derive this field using some edge detector, but other fields are possible. With this approach we can smoothly transition between any two given point distributions, without introducing new patterns to the stippling.

We show a graphical overview of our approach in
\autoref{figure:scheme}. In general terms, our method takes two stippling
algorithms and a distance field such as one that is computed from an edge detector applied to the source image.
We then construct a stochastic function from this information to get the Probability Density Functions (PDFs) of both algorithms. We then interpolate the two PDFs according to the distance field, and then rendering the result to obtain the final illustration.
Next, we first explain the general method to combine any two point distributions, before we demonstrate how this approach allows us to solve the artifact issue for stippling.

\section{Distributions and Interpolation Functions}
\label{sec:distribution}

Any stippling algorithm can be considered as a function that
takes a source image (as well as certain parameters) as input and then
determines dot positions on a virtual paper (ignoring the dot aspects
\cite{Martin:2015:DCR,Martin:2019:ADC} for now). We can consider the
process of placing the dots as a random function that places a dot at a
particular location. In the case of those stippling algorithms that use
random numbers (\eg, \cite{Schlechtweg:2005:RMA,Arroyo:2010:SGD,Li:2011:SPS,Li:2017:PSS}), the probability of placing the dots is given by the techniques' respective Probability Density Functions (PDFs) directly. For the rest of deterministic stippling and point distribution algorithms, we can simulate a corresponding stochastic version simply by sampling the stippling result according to the density of stippling dots per area and deriving a respective PDF.


A PDF is a function that makes use of a random variable $X$ and has a density function associated
$f_{X}$, where $f_{X}$ is a non-negative Lebesgue-integrable function. It can thus be defined as
\begin{equation*}
  \Pr[a\leq X\leq b]=\int _{a}^{b}f_{X}(x)\,dx\:.
\end{equation*}

The 2D nature of the virtual stippling canvas makes it convenient to redefine the
PDF in terms of area instead of some interval as
\begin{equation*}
  \Pr[X \in A]=\int _{A}f_{X}(x)\,dx\:.
\end{equation*}

In our case, $X$ represents some random stippling dot so, intuitively, one can think of $f_{X}(x)\,dx$ as being the probability of some random stippling dot $X$ falling within some infinitesimal area $A$.

The linear interpolation of two different PDFs ($f$ and $g$) for the same area $A$ can be written \cite{bursal1996interpolating} as
\begin{equation}\label{eq:PDF-interpolation}
I_{X}(\alpha) = (1-\alpha)f_{X} + \alpha g_{X} \text{, with } X \in A
\end{equation}
for some scalar value $\alpha \in [0,1]$. We thus have
%
%
\begin{equation*}
  Pr[X \in A] = (1-\alpha) \int_A f_{X}(x)\; dx + \alpha \int_A  g_{X}(x)\;dx \text{, with } X \in A\:.
\end{equation*}

For those stippling algorithms that directly use continuous, integrable functions it is enough to compute the interpolation of the respective two PDFs to merge both algorithms. But even those algorithms that use PDFs for the dot placement typically redefine the PDF in terms of time: the PDF is constructed while the dots are placed. Therefore, such algorithms are terribly slow due their need of re-adjusting the PDF for each iteration, changing the probabilities every time that a new dot is placed on the canvas (normally by reducing the probability in the area where the dot was placed).

For this reason, and for these particular cases, the interpolation
between the PDFs cannot be performed until all dots have been placed and
the PDF is finalized. An additional downside of this approach
is that it increases the algorithm's computational complexity. 
 Moreover, it is impossible to combine such techniques with approaches
 that lack a well-defined PDF such as Voronoi-based algorithms. To
 address this problem, we propose to use a discrete probability function
 (DPF) instead of a regular PDF, we detail next. 

\subsection{Discrete Probability Functions}
\label{sec:dpf}

One possibility of converting the PDF into a DPF is by the discretization of the canvas, that would give us a random variable $X$ that specifies cells of the resulting grid instead of the continuous positions in the 2D space of the stippling canvas.
We can go a step further, however, and directly consider the grid cells to contain the respective probability value of the cell receiving the next stipple dot or not.
%
%
This representation has the advantage that we can model any purely deterministic stippling algorithm, guiding the probabilities for each cell of the grid as if it was a DPF. Moreover, with a sufficiently fine resolution of the grid, for deterministic algorithms we produce virtually the same result as before the discretization.

Let us assume such a sufficiently fine-grained grid placed over the stippled image, with $N$ stipple points to be placed. We then mark any cell that has no stipple point location in it as a white cell ($W$), and any cell that has one or more stipple point locations in it as a black cell ($B$). White cells have no probability to be stippled, so we assign them a probability of 0\%. Black cells (their total number is $M$ with $M \le N$) have some probability greater than $0\%$ of being stippled, so they get a non-zero probability. With the probability of a cell $Pr[B_i]$ we refer to the chance that the next stipple dot that is placed is assigned to the given cell. We can thus state that $\sum_{i=1}^M Pr[B_i] = 1$ because the next stipple has a 100\% probability to be placed into one of the remaining black cells. We can derive the real initial probability of each black cell by sampling the PDF for non-deterministic algorithms, by using some heuristic (\eg, considering the amount of dots per cell), or simply by assigning the same probability to all black cells ($Pr[B_i] = 1/M$).

This way, a simple stochastic algorithm gives us a dot distribution virtually identical to that the original algorithm as follows:
\begin{enumerate}
\item Cast a single stipple dot to the grid. Randomly determine one black cell into which it is placed, considering the probabilities of each black cell.
\item For the chosen cell, reduce the probability of this black cell getting future stipples by subtracting the probability of the current stipple $1/N'$, $N'$ being the number of remaining stipples to be placed including the current one, from the cell, and distribute this probability equally to the rest of the remaining black cells. This means that these remaining cells now have a higher probability to get the next stipple point than before.
\item If a black cell reaches $0\%$ probability it turns into a white cell. If it would become negative, only subtract the remaining positive amount in the previous step and distribute it.
\item Repeat from step (1) until there is only one black cell (with probability equal to $1$), the last stipple dot is placed into this cell and the algorithm ends.
\end{enumerate}

\begin{figure*}[t!]
  \centering
	\setlength{\subfigcapskip}{0ex}
  \subfigure[Distribution results of the DPF-based interpolation from a 2D uniform random distribution in a grid of cells (left) to a 2D normal distribution (right).]{\label{fig:iterpolation:a}\includegraphics[width=\textwidth]{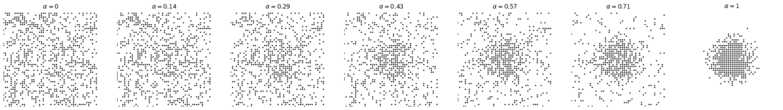}}\\[2.5ex]
	\setlength{\subfigcapskip}{-3ex}
  \subfigure[Distribution results of the DPF-based interpolation from a 2D normal random distribution in a grid of cells (left) to a sampled annulus in a grid (right).]{\label{fig:iterpolation:b}\includegraphics[width=\textwidth]{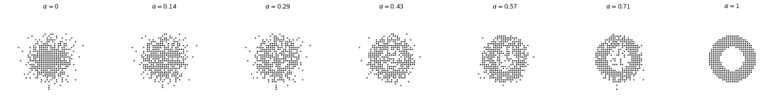}}\\[2.5ex]
	\setlength{\subfigcapskip}{0ex}
  \subfigure[Same DPF-based interpolation as in \subref{fig:iterpolation:a}, only using a high, practically relevant resolution.]{\label{fig:iterpolation:c}\includegraphics[width=\textwidth]{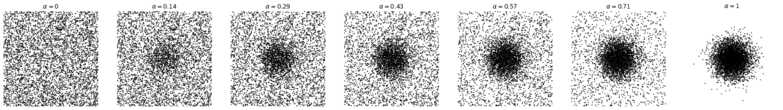}}\\[2.5ex]
	\setlength{\subfigcapskip}{-3ex}
  \subfigure[Same DPF-based interpolation as in \subref{fig:iterpolation:b}, using a high, practically relevant resolution.]{\label{fig:iterpolation:d}\includegraphics[width=\textwidth]{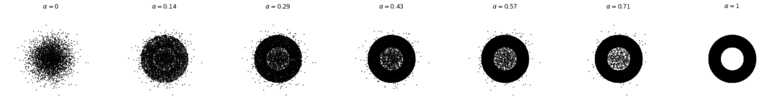}}\\[2.5ex]
	\setlength{\subfigcapskip}{0ex}
  \subfigure[Distribution results of the PDF-based interpolation from a 2D uniform random distribution (left) to a 2D normal distribution (right).]{\label{fig:iterpolation:e}\includegraphics[width=\textwidth]{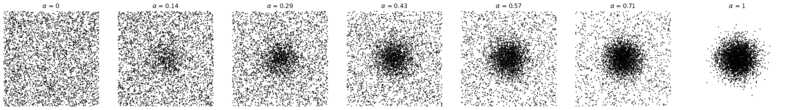}}\\[2.5ex]
	\setlength{\subfigcapskip}{-3ex}
  \subfigure[Distribution results of the PDF-based interpolation from a 2D normal random distribution (left) to an annulus-shaped stochastical PDF sampling (right).]{\label{fig:iterpolation:f}\includegraphics[width=\textwidth]{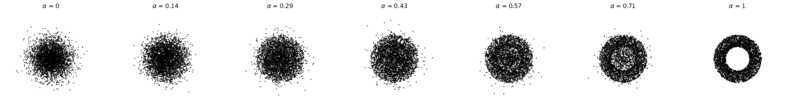}}
   \caption{Interpolation examples of two couples of DPFs, first in \subref{fig:iterpolation:a}--\subref{fig:iterpolation:b} at low and later in \subref{fig:iterpolation:c}--\subref{fig:iterpolation:d} at high resolution; \subref{fig:iterpolation:e}--\subref{fig:iterpolation:f} interpolation of their PDF counterparts.}\vspace{-1ex}\label{fig:iterpolation}
 \end{figure*}

\setlength{\imagewidth} {0.135\textwidth}
\setlength{\imageheight}{0.00785\textwidth}
\begin{figure*}[t!]
	\setlength{\subfigcapskip}{-1.5ex}
  \subfigure[On the left, a sampled annulus in a grid; on the right, a sampled logo in a grid; in between, the distributions result of the interpolation of their DPFs.]{\label{fig:iterpolationDPF:a}\includegraphics[width=\textwidth]{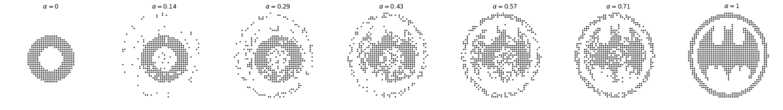}}\\[2.5ex]
	\setlength{\subfigcapskip}{0ex}
  \subfigure[A more realistic example of the interpolation of two DPFs with a high-resolution grid, from $\alpha=0$ on the left to $\alpha=1$ on the right, with an increment of $1/7$ units per step. The left is a halftoned detail section of the frog in \autoref{figure:scheme}, the right is the same picture after applying weighted Voronoi stippling.]{\label{fig:iterpolationDPF:b}%
	\includegraphics[width=\imagewidth]{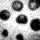}\hspace{\imageheight}%
	\includegraphics[width=\imagewidth]{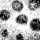}\hspace{\imageheight}%
	\includegraphics[width=\imagewidth]{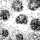}\hspace{\imageheight}%
	\includegraphics[width=\imagewidth]{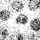}\hspace{\imageheight}%
	\includegraphics[width=\imagewidth]{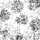}\hspace{\imageheight}%
	\includegraphics[width=\imagewidth]{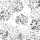}\hspace{\imageheight}%
	\includegraphics[width=\imagewidth]{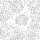}\hspace{\imageheight}%
	}
  \caption{Examples of DPF interpolation for which PDF-based interpolation is unfeasible.}\vspace{-0.5ex}
	\label{fig:iterpolationDPF}
\end{figure*}

This algorithm produces a stipple result equivalent and virtually identical to that of a deterministic technique---up to the chosen grid resolution and with the possibility that the number of dots in a given grid cell may differ occasionally due to the probabilistic approach. Moreover, if we had some way of computing the actual probability (as opposed to deriving it from an input stipple distribution), we would be able to maintain the exact same distribution as described by the PDF. Our goal, however, is not to run the algorithm. Instead, we want to derive a DPF for any given stippling technique, which is simply the grid with black and white cells and their associated discrete probabilities. We now use this DPF to be able to smoothly interpolate between two arbitrary dot distributions.

\subsection{Linear Interpolation}
\label{sec:dpf-interpolation}


Given two DPFs $f$ and $g$ corresponding to two different dot distributions, we can express the interpolation of both DPF as
\begin{equation}\label{eq:interpAlpha}
  interpolated\_DPF(\alpha) = f \cdot (1 - \alpha) + g \cdot \alpha\:;\:\: \alpha \in [0,1]\:.
\end{equation}

We just derived the input DPFs as a set of cells with probabilities (denoted $\Omega$ as the set of all cells in the grid) and we assume that both DPFs use the same resolution and alignment. To be able to compute \autoref{eq:interpAlpha} we thus need to do a pair-wise interpolation of the respective probabilities in the cells in both DPFs.

We do this interpolation also in a probabilistic way, such that the chance that the cell value from $f$ has influence is $1 - \alpha$ (let us call this event F) and the chance that the value from $g$ takes effect is $\alpha$ (which we call event G). We model the event of a cell's probability taking effect using a uniform/rectangular distribution $U_{cell}$ in the range $[0, 1]$, which we can compare to $1 - \alpha$ or to $\alpha$ and which we need to define for each cell in the grid. We are interested in dots being placed, so we want to know the union of events F and G, so we can specify the interpolation of a given cell as
\begin{equation}\label{eq:interpAlphaCell}
\begin{split}
  Pr[cell &\in \Omega\text{, }\alpha \in [0,1]] =  \\
  &\Pr\Big[\Big(f_{cell} \cap (U_{cell} \geq \alpha)\Big) \cup \Big(g_{cell} \cap (U_{cell} < \alpha)\Big)\Big]\:.
\end{split}
\end{equation}

In probability theory, given two different probability events $A$ and $B$, and if $A$ and $B$ are independent, the probability of having the two events happen at the same time is $\Pr(A \cap B) = \Pr(A) \cdot \Pr(B)$. The probability that events $A$ or $B$ occur is the probability of the union of $A$ and $B$ and can be written as $\Pr(A \cup B) = \Pr(A) + \Pr(B) + \Pr(A \cap B)$, with the last term $\Pr(A \cap B)$ being 0 in our case.
In an example where $f_{cell}$ has a 50\% probability of placing a dot and $g_{cell}$ has 10\% and for an $\alpha$ of 0.1, we would thus get an interpolated probability $Pr[cell\text{, }\alpha =0.1] = 0.5 \cdot (1 - 0.1) + 0.1 \cdot 0.1 = 0.46$.

We can thus apply the interpolation of the grid of probabilities by means of \autoref{eq:interpAlphaCell}, leading to results as demonstrated in \autoref{fig:iterpolation}. \autoref{fig:iterpolation:a}--\subref{fig:iterpolation:d} show the direct application for the interpolation of two DPFs each, with a very coarse grid size to demonstrate the effect. \autoref{fig:iterpolation:a} shows the interpolation between DPFs derived from a random distribution and a normal distribution, while \autoref{fig:iterpolation:b} interpolates from the DPF of the normal distribution to the DPF of a procedural figure. Notice that, due to the coarse grid size of the example, the single stipple of each cell is placed in the resulting regular grid pattern. Next, \autoref{fig:iterpolation:c}--\subref{fig:iterpolation:d} show the DPF-based interpolation for equivalent distributions at a high resolution (sampled from the original PDFs). For comparison, \autoref{fig:iterpolation:e}--\subref{fig:iterpolation:f} then demonstrate the PDF-based interpolation for the same distribution (\ie, before their discretization to DPF) by following \autoref{eq:PDF-interpolation} and then sampling it directly (\ie, using a Monte Carlo algorithm \cite{Arroyo:2010:SGD}). As one can see, the resulting DPF-based interpolation and the PDF-based interpolation produce visually similar results, with the exception of the density of the coverage of the ring image that we explain below.

\begin{figure*}[h!]
  \centering
    \includegraphics[width=\textwidth]{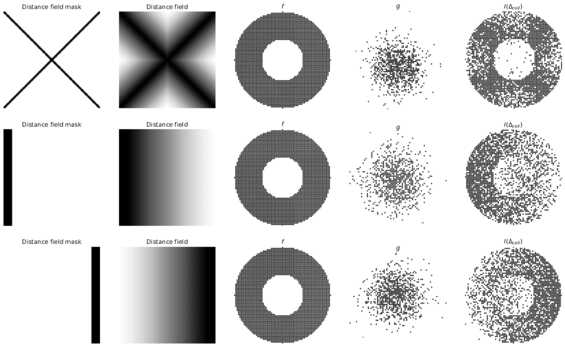}\vspace{-1.5ex}
    \caption{Different masks, their respective distance field, two distributions ($f$ and $g$), and their interpolation based on the distance fields.}\vspace{-1ex}
		\label{fig:distance-field}
\end{figure*}

This last example shows that two PDFs can be interpolated. However, as argued at the beginning of this section we do not actually have a PDF for all input distributions that we could use, in particular when we want to represent an arbitrary input image such as in stippling. Here the best approach is to represent the input resolutions as DPFs and to interpolate between them, as shown for a toy example in \autoref{fig:iterpolationDPF:a}: while the distribution on the right is derived from a PDF, the distribution on the left is derived from an input image and \autoref{fig:iterpolationDPF:a} shows the DPF-based interpolation between them. The argument holds even more for the interpolation between two image-based distributions as shown in the realistic example in \autoref{fig:iterpolationDPF:b}, in which we interpolate using DPFs between two different stochastic algorithms (a halftoning algorithm on the left and a weighted Voronoi algorithm \cite{Secord:2002:WVS} on the right).

One important property of the DPF-based interpolation, in contrast to the use of PDFs, is that it is computed in terms of cells not in term of samples. This means that, while in the case of the sampling of a PDF the number of samples (or stipple points) can remain the same during the interpolation, in DPF-based interpolation we cannot guarantee that the number of samples remains constant during the interpolation process. We even cannot guarantee a constant sample count if the two algorithms to be interpolated each have the same amount of samples to start. Such situation is rare when dealing with real algorithms, however, so this constraint has little practical implications such as when applying the DPF-based interpolation to digital stippling. Moreover, PDF sampling cannot guarantee that we get completely black areas are covered by the samples as shown in the last image on the right of \autoref{fig:iterpolation:f}: we instead get noisy images. The grid of the DPF, in contrast, forces us to always sample all the cells of the grid, which in turn allows us to get close to the original as shown in the last image on the right of \autoref{fig:iterpolation:d}.

\begin{figure*}[h!]
  \centering
    \includegraphics[width=\textwidth]{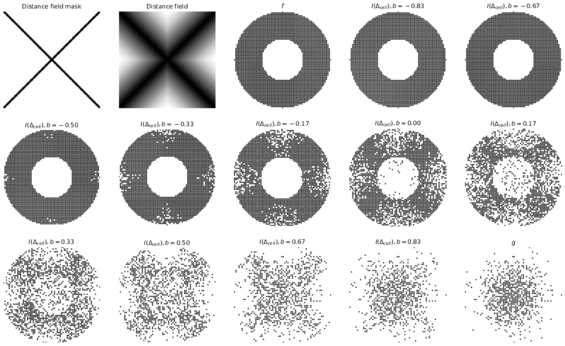}\vspace{-1.5ex}
    \caption{Different masks, their respective distance field, two distributions ($f$ and $g$), and their interpolation based on the distance fields.}\vspace{-1ex}
		\label{fig:distance-field-bias}
\end{figure*}

\subsection{DPF Interpolation based on a Distance Field}

The problem with the linear interpolation of DPFs (or, similarly, PDFs) as discussed thus far is that the interpolation is applied globally for the entire canvas. We cannot yet control the interpolation depending on the original picture's composition to combine different distributions as we motivated in \autoref{sec:intro}. We thus now define a new function $\Delta$ that controls the interpolation based on the 2D distance between a specific cell and a given subset of cells as
\begin{equation}\label{eq:distance-field}
  \Delta(x,\partial \Omega ):=\frac{\underset{{\scriptscriptstyle y\in \partial \Omega }}{\inf} d(x,y)}{\max(\Omega)} \ ;\  x \in \Omega \:.
\end{equation}
Here, $\Omega$ represents the complete space, \ie, the set of all the cells in the grid. This set can be considered to be a partially ordered set because we can order the cells based on some arbitrary function such as the distance to a given, selected cell $x$. We can thus use the infimum to get the smallest value of possible distances $d(x,y)$ between $x$ and any cell in a subset of cells $\partial\Omega$.
The distance $d(x,y)$ could use any distance metric, we use a simple Euclidean distance.
While $\partial\Omega$ could be equal to $\Omega$ such that it contains all its cells, in practice it will be just a part of the cells. The reason is that our goal is to construct a distance field from some mask that could guide the interpolation of the two distributions. We thus construct a mask by marking some of the cells of $\Omega$ as black cells, constructing the subset $\partial\Omega$. With the Euclidean distance $d(x,y)$, $\inf _{y\in \partial \Omega }d(x,y)$ gives us the distance from $x$ to the closest black cell. We then normalize the result based on the maximum distance in $\Omega$, namely $\max(\Omega)$. Overall, $\Delta_{cell}$ thus calculates the normalized distance field from each cell $x$ to the closest black cell in the grid ($\Omega$).

\autoref{fig:distance-field} shows the effect of three different masks (\ie, different subsets $\partial\Omega \subset \Omega$) and their resulting distance fields on the interpolation between a uniform random distribution ($g$) and a deterministic algorithm drawing a 2D torus ($f$). In these examples the masks are independent from the used point distributions. Ultimately, however, we want to control the interpolation based on features in the original image, which also led to the two point distributions (in the case of digital stippling). We thus make the interpolated distribution $I$ depend on $\Delta$ instead of $\alpha$ and rewrite \autoref{eq:interpAlpha} as
\begin{equation*}
  I(\Delta_{cell}) = f_{cell} \cdot (1 - \Delta_{cell}) + g_{cell} \cdot \Delta_{cell} \:.
\end{equation*}

In addition to this distance-based interpolation we also want to provide users with control for when $f$ or $g$ become dominant. We thus add a bias $b \in [-1,1]$ to $\Delta_{cell}$, so \autoref{eq:interpAlpha} finally becomes
\begin{equation}\label{eq:interp-delta}
  I(\Delta_{cell}, b) = f_{cell} \cdot (1 - \Delta_{cell}-b) + g_{cell} \cdot (\Delta_{cell} + b) \:.
\end{equation}
\autoref{fig:distance-field-bias} shows the effect of this bias $b$ on the interpolation between a uniform random distribution ($g$) and a deterministic algorithm drawing a 2D torus ($f$). When $b = 0$ we get function $f$, whereas when $b = 1$ the returned values correspond to $g$.

So three functions are involved in the process of the interpolation. On one hand we have the two DPFs that can be implemented as simple masks (for deterministic algorithms these mask are binary). On the other hand we have the distance field mask. This is another binary mask that represents how the interpolation is performed. As noted before, one advantage of using DPFs instead of PDFs is that we can directly use the result of any stippling algorithm. We can thus take advantage of the parallelism of modern GPUs due to the cells in the masks being independent from each other. Another advantage is the simplicity of the computation of the distance field mask. The problem with the use of DPFs is the regularity of the resulting cells, but for that we can use some post-processing in the render stage, as we explain later.

\section{Artifact-controlled digital stippling}
\label{sec:results1}

As we had discussed, many existing techniques are successful in shading areas in images. While some early techniques based CVDs created unwanted linear features, most recent techniques produce artifact-free distributions. With our new distribution interpolation technique we could thus select any of these techniques and use it to shade larger regions, and combine it with another technique that can represent borders and features well. Next, however, we discuss specific choices for the algorithms and show how they efficiently facilitate artifact-controlled stippling.

\subsection{Border Stippling}
\label{subsec:border}

We begin with the stippling of borders, for which we could now use a dedicated structure-aware stippling technique (\eg, by \citet{Mould:2007:SPD} or \citet{Li:2011:SPS,Li:2017:PSS}). With our new ability to combine any two dot distributions, however, we can use the easier early stippling techniques, which are simpler to implement. \citeauthor{Secord:2002:WVS}'s WCVD [\citeyear{Secord:2002:WVS}], for instance, produces results with chained stipples even in shaded areas, but as long as we only use the feature lines it produces for which we do want stipple chaining this does not pose a problem.

The key insight for the stippling approach we describe next is that, using our previously described interpolation technique for dot distributions, we would first express both as DPFs, assuming that each resulting black pixel represents only one dot of the stippling result. For some of them, including \citeauthor{Secord:2002:WVS}'s WCVD [\citeyear{Secord:2002:WVS}], this process involves taking an image as an input, doing some image processing such as contrast adjustment or edge detection, and then running the actual dot distribution process, before extracting the DPF from the generated dot distribution. Second, we would interpolate the two generated DPFs as described in \autoref{sec:distribution}, and then place stipple dots for the interpolated DPF as described in \autoref{sec:rendering}. It turns out, however, that we can optimize the conversion from input image to filtered image to analytic dot positions to DPF (which is also essentially an image) by using the filtered image, after some additional (image) processing, directly as the input to our interpolation. Next we thus explain this process for generating the edge-representing DPF for our hybrid stippling process, before we discuss the shading of areas and the mixing of both approaches using a distance field.

One of our main goals is that we want those stipple dots that represent single lines in an image or that are borders of a stippled region to be aligned. Our first step is thus to compute an adjusted input image that emphasizes these edges. For this purpose we can employ established edge detection filters (\eg, by \citet{Canny:1986:CAE}, by \citet{Sobel:2014:HDS}, by \citet{Prewitt:1970:OEE}, the Laplacian of Gaussians---LoG \cite{Marr:1982:VIS}, or the Difference of Gaussians---DoG \cite{Marr:1980:TED}), and past NPR work \cite{Hertzmann:1999:INP,Isenberg:2003:DGS} has discussed that some of these filters need to be combined to better cover those edges typically used in hand-drawn line images. Nonetheless, these filters are not always able to cover all lines a human artist would draw \cite{Cole:2008:WDP,Cole:2009:HWD}, and they also often require difficult fine-tuning or parameters to the chosen input image. In recent years, researchers have tried to address these problems by using deep neural networks (DNNs) (\eg, \cite{Liang:2015:SIS,Shen:2015:DDC,Kokkinos:2015:SHB,Xie:2017:HED}). Although these solutions can detect edges successfully, they need many training images and expensive human input to annotate the valid edges of the images. In addition, they require powerful hardware provide interactive response times. In this work we thus restrict ourselves to using traditional image filters as they are quite successful and because we combine them with stippled areas; yet they could be easily replaced by another technique in the future.
\enlargethispage{.5pt}

\begin{figure}[t!]
	\begin{center}
		\includegraphics[width=0.25\linewidth]{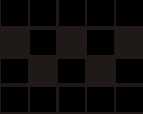}
		\caption{Possible zig-zag pixel distribution due to a 2-pixel-wide edge.}\vspace{-2ex}
		\label{figure:2pixel_width}
	\end{center}
\end{figure}

One essential prerequisite is to obtain edges of one-pixel thickness. We need this constraint to align stipple points later-on along the edge. For example, lines with a width of two pixels may end up being stippled in a zig-zag fashion as shown in \autoref{figure:2pixel_width}, for an average stipple distance of $\sqrt{2}$ pixels.
For these reasons we use \citeauthor{Canny:1986:CAE}'s [\citeyear{Canny:1986:CAE}] filter for the examples in the remainder of our work to guide the placement of stipples along one-dimensional feature edge paths.

We thus now have a filtered input image for the first of the stippling techniques (\citeauthor{Secord:2002:WVS}'s [\citeyear{Secord:2002:WVS}] WCVD stippling) that we use in our hybrid approach. As indicated above, similar to the binary image we ultimately need in our DPF-based interpolation of stippling distributions, this input image also essentially is a binary image, containing roughly the same information: black pixels that indicate locations for placing (much larger) stipple dots after the interpolation. The use of WCVD stippling just to then extract the DPF from the resulting distribution thus seems like an unnecessary detour, even with a fast GPU-based WCVD implementation.


\setlength{\picturewidth}{.22\textwidth}
\begin{figure*}[t]
	\centering
	\setlength{\subfigcapskip}{-3.0ex}%
	\mbox{~}\mbox{~}\mbox{~}\mbox{~}\mbox{~}\includegraphics[width=\picturewidth]{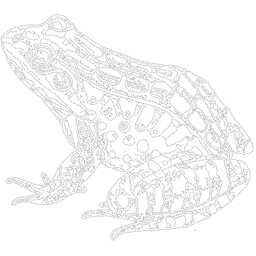}\hfill	
	\mbox{~}\mbox{~}\mbox{~}\mbox{~}\mbox{~}\includegraphics[width=\picturewidth]{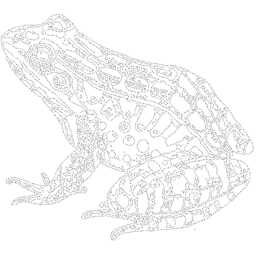}\hfill		
	\mbox{~}\mbox{~}\mbox{~}\mbox{~}\mbox{~}\includegraphics[width=\picturewidth]{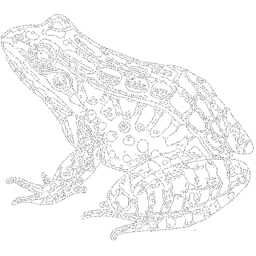}\hfill		
	\mbox{~}\mbox{~}\mbox{~}\mbox{~}\mbox{~}\includegraphics[width=\picturewidth]{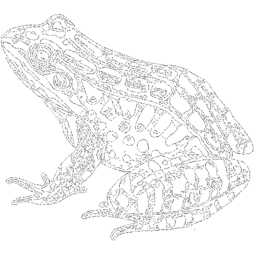}\\
	\subfigure[\hspace{\linewidth}]{\mbox{~}\mbox{~}\mbox{~}\mbox{~}\mbox{~}\label{fig:new_method_noise:a}\includegraphics[width=\picturewidth,trim=200 200 150 150,clip]{canny_frog_new_method_d3}}\hfill		
	\subfigure[\hspace{\linewidth}]{\mbox{~}\mbox{~}\mbox{~}\mbox{~}\mbox{~}\label{fig:new_method_noise:b}\includegraphics[width=\picturewidth,trim=200 200 125 125,clip]{canny_frog_new_method_d3_n33}}\hfill	
	\subfigure[\hspace{\linewidth}]{\mbox{~}\mbox{~}\mbox{~}\mbox{~}\mbox{~}\label{fig:new_method_noise:c}\includegraphics[width=\picturewidth,trim=200 200 125 125,clip]{canny_frog_new_method_d3_n66}}\hfill		
	\subfigure[\hspace{\linewidth}]{\mbox{~}\mbox{~}\mbox{~}\mbox{~}\mbox{~}\label{fig:new_method_noise:d}\includegraphics[width=\picturewidth,trim=200 200 125 125,clip]{canny_frog_new_method_d3_n100}}\vspace{-1.5ex}
	\caption{Adding noise (full results and details), using $d_0 = 3.5$ pixels: \subref{fig:new_method_noise:a} 0\%, \subref{fig:new_method_noise:b} +/-16.5\%, \subref{fig:new_method_noise:c} +/-33\%, \subref{fig:new_method_noise:d} +/-50\%. Note that we clamp $d$ to be at least 1 pixel.}
	\label{fig:new_method_noise}
\end{figure*}

In our approach for producing the edge DPF we thus do not actually run \citeauthor{Secord:2002:WVS}'s WCVD but use image processing techniques to generate a similar result, one that also emphasizes edges and ensures an even distribution of stipple dots along them. Nonetheless, the result of first stippling and then extracting the DPF is not identical to the edge image we used as the input: the DPF image would not contain a consecutive series of pixels for every edge, in contrast to the filtered edge image. We thus need to add a controllable amount of space in-between the aligned stipples, a gap that would normally be ensured by the WCVD stippling, which limits the overall number of stipple dots in the image and thus also the number of black pixels that represent an edge in the DPF. Yet we can use a similar ink-constraining process and apply it directly to the edge image---in a way a simplified 1D case of WCVD stippling to distribute the remaining pixels evenly along edges.


For this purpose we determine all edge pixels and treat them as paths. Starting with a given black pixel, we find all its black neighbors in the 8-neighborhood. Given the direction that we came from the previous pixel, we then select that black pixel among the current pixel's neighbors that best continues the path (or a random neighbor if we started a new path). We continue this stepping along pixel paths until a path cannot be continued anymore (\ie, there are no more black neighbors). 
Once this is completed, we start a new path tract by selecting a new random pixel that was not visited yet---until all pixels have been visited. To best emphasize long, outline strokes, we begin by searching the image from top-to-bottom and left-to-right until we find the first black pixel---our initial starting pixel. We also use the heuristic of using the left-most pixel (w.r.t.\ our incoming direction) in the 8-neighborhood as we select the next pixel
to ensure walking around shapes first. We also tested the addition of a backward search after the forward search to follow paths in both directions, but the results were visually equivalent to only doing a forward search, so we kept this simpler approach.


The process as we have described it so far simply yields all black pixels, and the result is identical to the edge buffer. To simulate the effect of WCVD stippling, we introduce sections of white pixels between the black ones. We thus only output a new pixel once we have covered a user-specified distance $d$ along the path (counting $\Delta_d=1$ for horizontal or vertical steps and $\Delta_d=\sqrt{2}$ for diagonal steps). To avoid regular patterns, we add noise to the distance. While we could have used a normally distributed noise offset around an average distance $d_0$, our test have shown that we can produce visually satisfying results by adding noise to $d_0$ in form of a random offset from $[-d_n, d_n]$. We demonstrate the effect of this noise being added to $d_0$ in \autoref{fig:new_method_noise}, for different amounts of noise.

\begin{figure}[t!]
	\begin{center}
		\includegraphics[width=\linewidth]{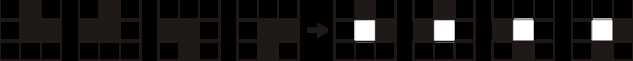}
		\caption{Corner problem produced by Canny's filter.}\vspace{-2ex}
		\label{fig:corners}
	\end{center}
\end{figure}

\setlength{\wd0}{\picturewidth}
\setlength{\ht0}{\picturewidth}
\setlength{\picturewidth}{.220\textwidth}
\begin{figure*}[t]
	\centering
	\setlength{\subfigcapskip}{-3.0ex}%
	\subfigure[\hspace{\linewidth}]{\label{fig:canny:a}\includegraphics[width=\picturewidth]{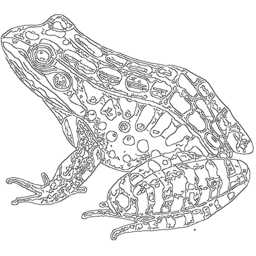}}\hfill
	\subfigure[\hspace{\linewidth}]{\mbox{~}\mbox{~}\mbox{~}\mbox{~}\mbox{~}\label{fig:canny:b}\includegraphics[width=\picturewidth,clip,trim=188 188 235 235]{canny_frog1}}\hfill
	\subfigure[\hspace{\linewidth}]{\mbox{~}\mbox{~}\mbox{~}\mbox{~}\mbox{~}\label{fig:canny:c}\includegraphics[width=\picturewidth,clip,trim=188 188 235 235]{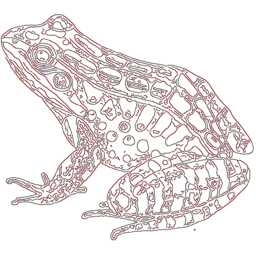}}\hfill
	\subfigure[\hspace{\linewidth}]{\mbox{~}\mbox{~}\mbox{~}\mbox{~}\mbox{~}\label{fig:canny:d}\includegraphics[width=\picturewidth,clip,trim=188 188 235 235]{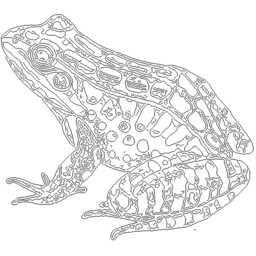}}\vspace{-.8ex}
	\caption{Problems of the Canny filter and how we address them: \subref{fig:canny:a} result of the Canny filter, \subref{fig:canny:b} detail section showing the problems, \subref{fig:canny:c} detail section with the problematic pixels in red, and \subref{fig:canny:d} detail showing the corrected version.}
	\label{fig:canny}
\end{figure*}

\setlength{\picturewidth}{.235\textwidth}
\begin{figure*}[t]
	\centering
	\setlength{\subfigcapskip}{-3.0ex}%
	\settowidth{\pictureheight}{\footnotesize{}(b)}%
	\addtolength{\pictureheight}{2pt}%
	\subfigure[\hspace{\linewidth}]{\label{fig:wcvd_vs_new_method:a}\includegraphics[width=\picturewidth]{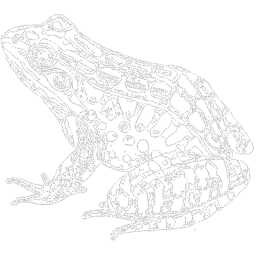}}\hfill%
	\subfigure[\hspace{\linewidth}]{\label{fig:wcvd_vs_new_method:b}\includegraphics[width=\picturewidth,clip,trim=110 222 262 150]{canny_frog_wcvd}}\hfill%
	\subfigure[\hspace{\linewidth}]{\label{fig:wcvd_vs_new_method:c}\includegraphics[width=\picturewidth]{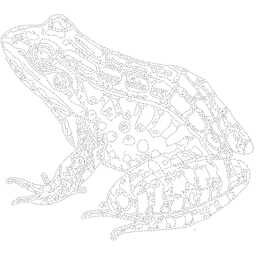}}\hfill%
	\subfigure[\hspace{\linewidth}]{\label{fig:wcvd_vs_new_method:d}\includegraphics[width=\picturewidth,clip,trim=110 222 262 150]{canny_frog_new_method}}\vspace{-.8ex}
	\caption{Comparison of WCVD stippling with our pixel-based heuristic: \subref{fig:wcvd_vs_new_method:a} WCVD, \subref{fig:wcvd_vs_new_method:b} detail from \subref{fig:wcvd_vs_new_method:a}, \subref{fig:wcvd_vs_new_method:c} our heuristic, \subref{fig:wcvd_vs_new_method:d} detail from \subref{fig:wcvd_vs_new_method:c}.}
	\label{fig:wcvd_vs_new_method}
\end{figure*}

One issue with this process is that the Canny filter, which we use, produces combinations of three black pixels (see \autoref{fig:corners}), which resembles a zig-zag line in the final image instead of a straight one or resembles a block. We thus remove these cases prior to stepping along the pixel paths. We show a more realistic example in \autoref{fig:canny}. 


For an example edge input as the one in \autoref{fig:canny:a}, this process then yields results as we show in \autoref{fig:wcvd_vs_new_method:c}. Compared to the WCVD stippling result for the same input shown in \autoref{fig:wcvd_vs_new_method:a}, we argue that the result is visually equivalent. Figures~\ref{fig:wcvd_vs_new_method:d} and~\ref{fig:wcvd_vs_new_method:b} show detail sections for this comparison, respectively.

\subsection{Stippling shaded regions}
\label{subsec:shaded}

In shaded areas, it is well established that CVD and WCVD are counterproductive due to the regularity of these distributions. As we discussed in \autoref{sec:review}, other approaches produce similar blue noise distributions to avoid visual patterns and artifacts. To achieve our goal of realism in digital stippling, however, we base our solution on \citeauthor{Martin:2011:SDE}'s [\citeyear{Martin:2010:EBS,Martin:2011:SDE}] use of scanned grayscale dots and halftoning-based placement because their process considers the target image's physical size. The packing factor and the noise added to dot locations in this process produce good results for shaded areas (\ie, areas with a low gradient). Moreover, it is fast to compute because its dot distribution is based on a halftoned image \cite{Ostromoukhov:2001:SEE}, which also has blue noise properties, and its results resemble hand-made distributions according to the statistical analysis by \citet{Maciejewski:2008:MSA}.

The main problem of this approach is that it uses a grid to distribute points and adds random noise to remove the regularity. While this leads to good results in shaded regions and makes stippled edges appear fuzzy and unclear, with our distribution interpolation we can now combine it with the edge stippling discussed before. We thus only need to derive the needed mixing function $\alpha$ to control the interpolation, as we describe next.

\subsection{Mixing functions}
\label{subsec:mixing_function}

Now that we have the two distributions in place, we need to define the specific function that controls the interpolation between them. This function needs to encode our initial goal of using the edge-emphasizing stippling from \autoref{subsec:border} for the edges, and the distribution mentioned in \autoref{subsec:shaded} for representing filled areas. In \autoref{eq:interpAlpha} we initially controlled the interpolation by means of a parameter $\alpha$, but later improved upon this concept in \autoref{eq:interp-delta} to a interpolation based on a distance field $\Delta$ and with a bias $b$.

The distance field $\Delta$ intuitively appears to be well suited to achieve our goal, provided that we use the initial edge Canny-filtered image generated in \autoref{subsec:border} as the distance mask $\partial\Omega$. We may want to, however, achieve a number of different goals as follows:
\begin{itemize}[leftmargin=2\parindent]
\item show both border and the area stipples at the same time, with different densities depending on the distance from the edge,
\item leave some white space around the border to enhance it,
\item produce an inversion effect in which edges are represented by the lack of stipples, etc.
\end{itemize}

Our current linear computation of the distance $\Delta$ is too inflexible to achieve these goals. We thus update \autoref{eq:interp-delta} that we used for the interpolation to also employ a general function $\Gamma$ that maps the distance field, the domain, to the same codomain but with a different shape to fulfill the desired goal:
\begin{equation}\label{eq:final-interp}
  I(\Delta_{cell}, b) = f_{cell} \cdot \Big(1 - \Gamma(\Delta_{cell})-b\Big) + g_{cell} \cdot \Big(\Gamma(\Delta_{cell}) + b\Big) \:.
\end{equation}

%


By default, $\Gamma$ is defined as a linear function ($\Gamma(cell) = \Delta_{cell}$). The possibilities of the $\Gamma$ function can be shown with an example:  If we want to control, for instance, the density of pixels for producing the combination of the borders and the area at the same time and to produce a band surrounding the borders, we could use this $\Gamma$ function, given the parameters $L_1$ and $L_2$ with $L_1,L_2 \in [0,1]$:
\begin{equation}\label{eq:elaborate-gamma}
  \Gamma(\Delta_{cell}) = \left\{ \begin{array}{ll}
0 &: \Delta_{cell} \le L_1\\
\frac{\Delta_{cell}-L_1}{L_2-L_1} &: L_1 \le \Delta_{cell} \le L_2 \\
1 &: \Delta_{cell} > L_2
\end{array}\right.\:.
\end{equation}

\subsection{Stipple Rendering}
\label{sec:rendering}

We can now render the resulting interpolated distribution, which is represented as grid of black and white pixels and with black pixels to receive dots. Following our stochastic algorithm from \autoref{sec:dpf} we can then place dots into black pixel regions, until we treated all black pixels. In contrast to recent approaches which freely re-arrange stipple dots based on an initial positions (\eg, \citep{Balzer:2009:CPD}), we thus have to place dots based on the black pixels and consider the dot sizes and shapes. For the realistic capture of dot distributions such as in traditional stippling, in particular, the grid of the DPF has to be highly detailed, while a placed dot is ultimately much larger than a grid pixel.

As we based the stippling of shaded regions on \citeauthor{Martin:2011:SDE}'s [\citeyear{Martin:2010:EBS,Martin:2011:SDE}] halftoning-based dot placement, we also use it for the overall rendering of stipple dots. This allows us to take care of the target resolution of the output, the physical size of the stipple points, and overlapping of stipple dots. \citeauthor{Martin:2011:SDE} removed the regularity of the halftoning-based dot placement by adding noise, which worked great for areas but made their stippled linear features look fuzzy. In our new method, we resolve this problem by controlling the added noise individually for areas and linear features. In most cases, the added noise to linear features is zero, but it depends on needs of the artist and the configuration of the pixels. 

We use the same approach to control the range of sizes for dots in a discrete or continuous way. The discrete option allows us to simulate the use of real technical pens, which have a limited set of tip sizes and allow us to use scanned dots (\ie, textures). Both discrete and continuous control can be used to produce vector results in SVG format. The selection of the size for each individual pixel can be computed in a uniform random way, or by modulating it for the tone of the corresponding pixel in the original image. This control provides us with a great level of flexibility in the appearance of the results, as we discuss and showcase next.

\section{Results}
\label{sec:results}

To be able to illustrate the spectrum of possibilities with our new interpolated point distributions (henceforth IPD), we first compare them with established digital stippling approaches and, second, discuss examples to showcase the effect of different parametrization.

\subsection{Comparison}

For the comparison we selected three techniques from the literature that placed particular emphasis on either representing features or areas with digital stippling. First, we selected \citeauthor{Secord:2002:WVS}'s [\citeyear{Secord:2002:WVS}] weighted centroidal Voronoi stippling (henceforth WVS) because it is able to capture linear features well, despite creating line artifacts in areas due to being based on centroidal Voronoi diagrams. Second, we compare with \citeauthor{Li:2011:SPS}'s [\citeyear{Li:2011:SPS,Li:2017:PSS}] structure-preserving stippling (henceforth SPS) which placed particular emphasis on linear structures. Third, we compare with \citeauthor{Martin:2010:EBS}'s [\citeyear{Martin:2010:EBS,Martin:2011:SDE}] example-based grayscale stippling (henceforth EBG) because it managed to avoid linear structures in areas and came close to artistic examples, yet at the expense of only being able to produce fuzzy linear structures. These techniques differ from each other as follows:
\begin{description}[leftmargin=\parindent]
\item[Color:] WVS and SPS are B\&W only processes. EBG is a grayscale process that is inspired by the process of ink accumulation.
\item[Dot shape:] WVS and SPS use circles. EBG uses scanned dots, \ie, irregular shapes recorded as textures.
\item[Dot size:] WVS can use dots with a constant size, but can also modulate the dot size depending on the Voronoi region's area and the tone of the input image \cite{Secord:2002:RMP}. This modulation maximizes the perceived effect of each dot. SPS produces dots of constant size. EBG is based on scanning original hand-made stippling artwork and extracting individual dots, thus uses random stipple sizes within a given range. 
\item[Output size:] EBG is a raster method with the goal of producing realistic stipple output for a given physical output size and pixel resolution, using matching scans of dots from stipple artwork. 
WVS and SPS are vector methods, their dot size can be controlled independently and continuously. 
\end{description}

We selected four target images for our tests: \emph{City} and \emph{Headlamp} from \citeauthor{Mould:2017:DAB}'s [\citeyear{Mould:2017:DAB}] benchmark set: these showcase straight and curved linear features. We also use \emph{Lenna} and \citeauthor{Secord:2002:WVS}'s [\citeyear{Secord:2002:WVS}] \emph{Plant} because they were used in the past for studying struc\-ture-pre\-ser\-ving stippling. For a fair comparison between the re\-so\-lu\-tion-de\-pen\-dent EBG with the re\-so\-lu\-tion-in\-de\-pen\-dent WVS, SPS, and IPD techniques we set a physical target equivalent of A5 output size at 300\,ppi resolution. Based on these constraints, we used the original EBG code to produce stipple images, noting the stipple count. We then used \citeauthor{Secord:2002:WVS}'s [\citeyear{Secord:2002:WVS}] original implementation of WVS to produce results with the same target stipple counts (Figures~\ref{fig:results_secord_constant} and~\ref{fig:results_secord_modulation}), 
and with the help of \citeauthor{Li:2011:SPS}'s original implementation also obtained stippled versions for SPS
(\autoref{fig:results_mould_constant}), again with the same target stipple count per image.\footnote{\emph{City}: 171,172, \emph{Headlight}: 137,327, \emph{Lenna}: 1,134,741, and \citeauthor{Secord:2002:WVS}'s \emph{Plant}: 149,327.} For IPD we used the combination of WVD with EBG described in \autoref{sec:results1}.

With all four techniques, we created several versions of the four benchmark images shown in Figures~\ref{fig:results_constant_bw}--\ref{fig:results_random_ebg_scanned_dots}. To save space we do not show the whole A5 images (find them in the additional material in Figures~\ref{fig:results_constant_bw_WVS}--\ref{fig:results_random_ebg_scanned_dots_IPD}), but only use a representative section. Specifically, we show vector images with constant, small, black, and circular dots in \autoref{fig:results_constant_bw} to show the distributions without any dot overlap. In \autoref{fig:results_modulated_ebg_scanned_dots} we still use black circles and vector output, but modulate the dot size by the tone color of the corresponding pixel in the original image such that darker tones produce bigger and brighter tones produce smaller dots. Finally, \autoref{fig:results_random_ebg_scanned_dots} shows a raster output with random dot sizes and scanned dots, with the actual dot size range being derived for all images using the EBG approach.

\setlength{\picturewidth}{.245\textwidth}
\begin{figure*}[t]
	\centering
	\setlength{\subfigcapskip}{-3.5ex}%
	\subfigure[\hspace{\linewidth}]{\label{fig:results_constant_bw:a}\includegraphics[width=\picturewidth]{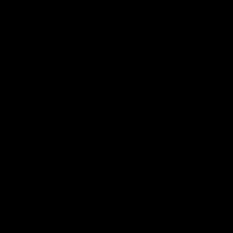}}\hfill%
	\subfigure[\hspace{\linewidth}]{\label{fig:results_constant_bw:b}\includegraphics[width=\picturewidth]{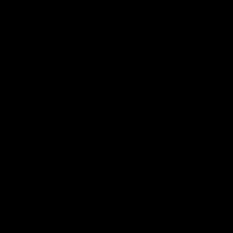}}\hfill%
	\subfigure[\hspace{\linewidth}]{\label{fig:results_constant_bw:c}\includegraphics[width=\picturewidth]{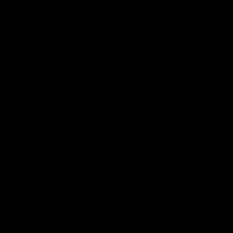}}\hfill%
	\subfigure[\hspace{\linewidth}]{\label{fig:results_constant_bw:d}\includegraphics[width=\picturewidth]{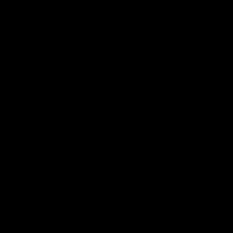}}\\[.5ex]%
	\subfigure[\hspace{\linewidth}]{\label{fig:results_constant_bw:e}\includegraphics[width=\picturewidth]{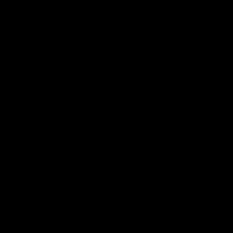}}\hfill%
	\subfigure[\hspace{\linewidth}]{\label{fig:results_constant_bw:f}\includegraphics[width=\picturewidth]{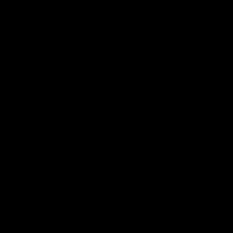}}\hfill%
	\subfigure[\hspace{\linewidth}]{\label{fig:results_constant_bw:g}\includegraphics[width=\picturewidth]{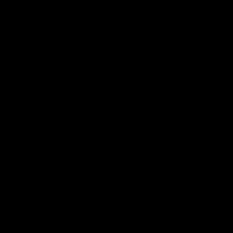}}\hfill%
	\subfigure[\hspace{\linewidth}]{\label{fig:results_constant_bw:h}\includegraphics[width=\picturewidth]{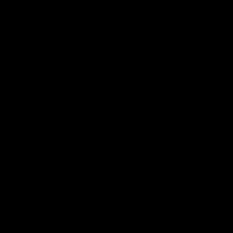}}\\[.5ex]%
	\subfigure[\hspace{\linewidth}]{\label{fig:results_constant_bw:i}\includegraphics[width=\picturewidth]{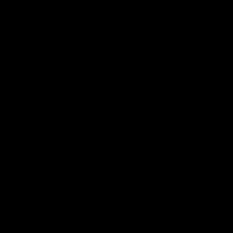}}\hfill%
	\subfigure[\hspace{\linewidth}]{\label{fig:results_constant_bw:j}\includegraphics[width=\picturewidth]{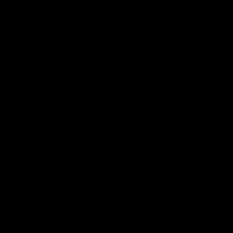}}\hfill%
	\subfigure[\hspace{\linewidth}]{\label{fig:results_constant_bw:k}\includegraphics[width=\picturewidth]{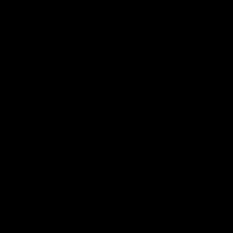}}\hfill%
	\subfigure[\hspace{\linewidth}]{\label{fig:results_constant_bw:l}\includegraphics[width=\picturewidth]{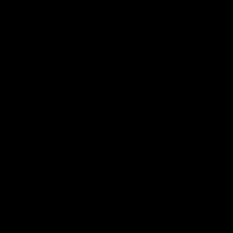}}\\[.5ex]%
	\subfigure[\hspace{\linewidth}]{\label{fig:results_constant_bw:m}\includegraphics[width=\picturewidth]{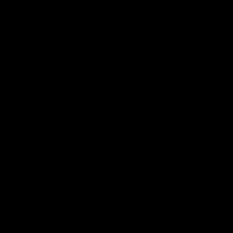}}\hfill%
	\subfigure[\hspace{\linewidth}]{\label{fig:results_constant_bw:n}\includegraphics[width=\picturewidth]{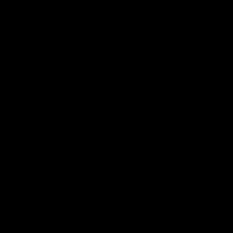}}\hfill%
	\subfigure[\hspace{\linewidth}]{\label{fig:results_constant_bw:o}\includegraphics[width=\picturewidth]{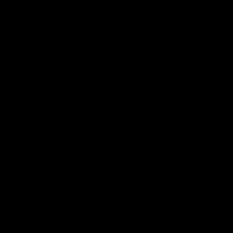}}\hfill%
	\subfigure[\hspace{\linewidth}]{\label{fig:results_constant_bw:p}\includegraphics[width=\picturewidth]{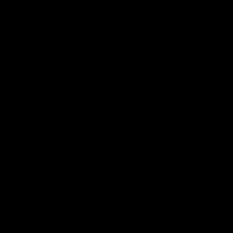}}\vspace{-0.8ex}
	\caption{Constant size dots, radius 0.5, B\&W. First row WVS, second row SPS, third row EBG, and fourth row IPD.}\vspace{-1.5ex}
	\label{fig:results_constant_bw}
\end{figure*}

\setlength{\picturewidth}{.245\textwidth}
\begin{figure*}[t]
	\centering
	\setlength{\subfigcapskip}{-3.5ex}%
	\subfigure[\hspace{\linewidth}]{\label{fig:results_modulated_ebg_scanned_dots:a}\includegraphics[width=\picturewidth]{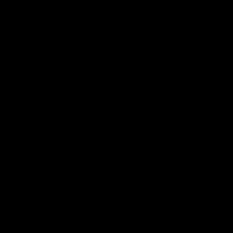}}\hfill%
	\subfigure[\hspace{\linewidth}]{\label{fig:results_modulated_ebg_scanned_dots:b}\includegraphics[width=\picturewidth]{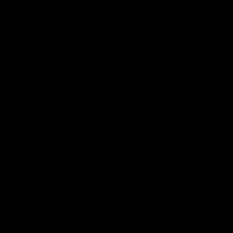}}\hfill%
	\subfigure[\hspace{\linewidth}]{\label{fig:results_modulated_ebg_scanned_dots:c}\includegraphics[width=\picturewidth]{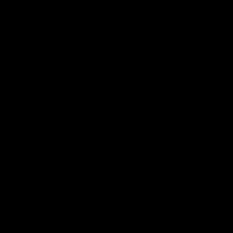}}\hfill%
	\subfigure[\hspace{\linewidth}]{\label{fig:results_modulated_ebg_scanned_dots:d}\includegraphics[width=\picturewidth]{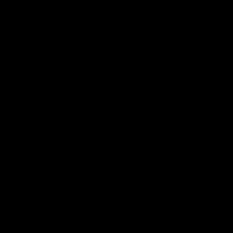}}\\[.5ex]%
	\subfigure[\hspace{\linewidth}]{\label{fig:results_modulated_ebg_scanned_dots:e}\includegraphics[width=\picturewidth]{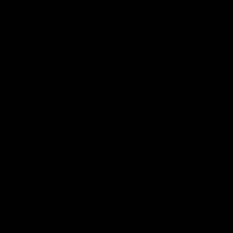}}\hfill%
	\subfigure[\hspace{\linewidth}]{\label{fig:results_modulated_ebg_scanned_dots:f}\includegraphics[width=\picturewidth]{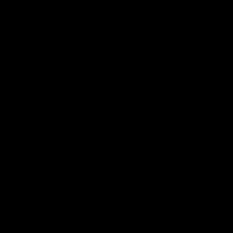}}\hfill%
	\subfigure[\hspace{\linewidth}]{\label{fig:results_modulated_ebg_scanned_dots:g}\includegraphics[width=\picturewidth]{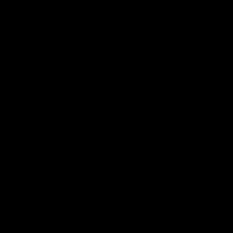}}\hfill%
	\subfigure[\hspace{\linewidth}]{\label{fig:results_modulated_ebg_scanned_dots:h}\includegraphics[width=\picturewidth]{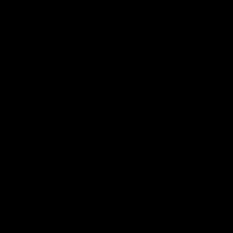}}\\[.5ex]%
	\subfigure[\hspace{\linewidth}]{\label{fig:results_modulated_ebg_scanned_dots:i}\includegraphics[width=\picturewidth]{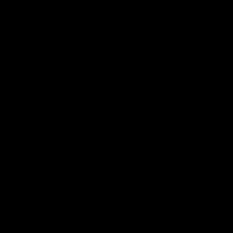}}\hfill%
	\subfigure[\hspace{\linewidth}]{\label{fig:results_modulated_ebg_scanned_dots:j}\includegraphics[width=\picturewidth]{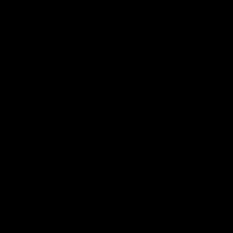}}\hfill%
	\subfigure[\hspace{\linewidth}]{\label{fig:results_modulated_ebg_scanned_dots:k}\includegraphics[width=\picturewidth]{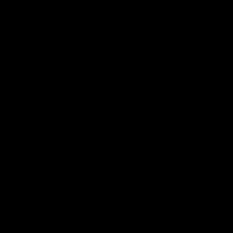}}\hfill%
	\subfigure[\hspace{\linewidth}]{\label{fig:results_modulated_ebg_scanned_dots:l}\includegraphics[width=\picturewidth]{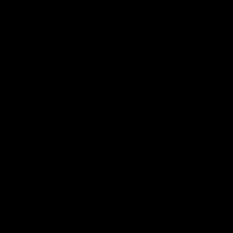}}\\[.5ex]%
	\subfigure[\hspace{\linewidth}]{\label{fig:results_modulated_ebg_scanned_dots:m}\includegraphics[width=\picturewidth]{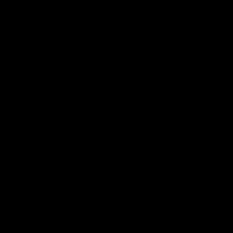}}\hfill%
	\subfigure[\hspace{\linewidth}]{\label{fig:results_modulated_ebg_scanned_dots:n}\includegraphics[width=\picturewidth]{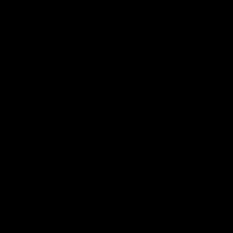}}\hfill%
	\subfigure[\hspace{\linewidth}]{\label{fig:results_modulated_ebg_scanned_dots:o}\includegraphics[width=\picturewidth]{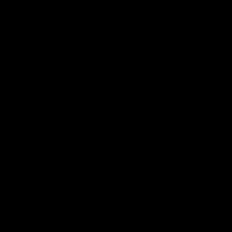}}\hfill%
	\subfigure[\hspace{\linewidth}]{\label{fig:results_modulated_ebg_scanned_dots:p}\includegraphics[width=\picturewidth]{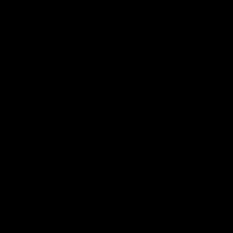}}\vspace{-0.8ex}
	\caption{Range between 2 and 4, continuous, modulated by tone, B\&W. First row WVS, second row SPS, third row EBG, and fourth row IPD.}\vspace{-1.5ex}
	\label{fig:results_modulated_ebg_scanned_dots}
\end{figure*}

\setlength{\picturewidth}{.245\textwidth}
\begin{figure*}[t]
	\centering
	\setlength{\subfigcapskip}{-3.5ex}%
	\subfigure[\hspace{\linewidth}]{\label{fig:results_random_ebg_scanned_dots:a}\includegraphics[width=\picturewidth]{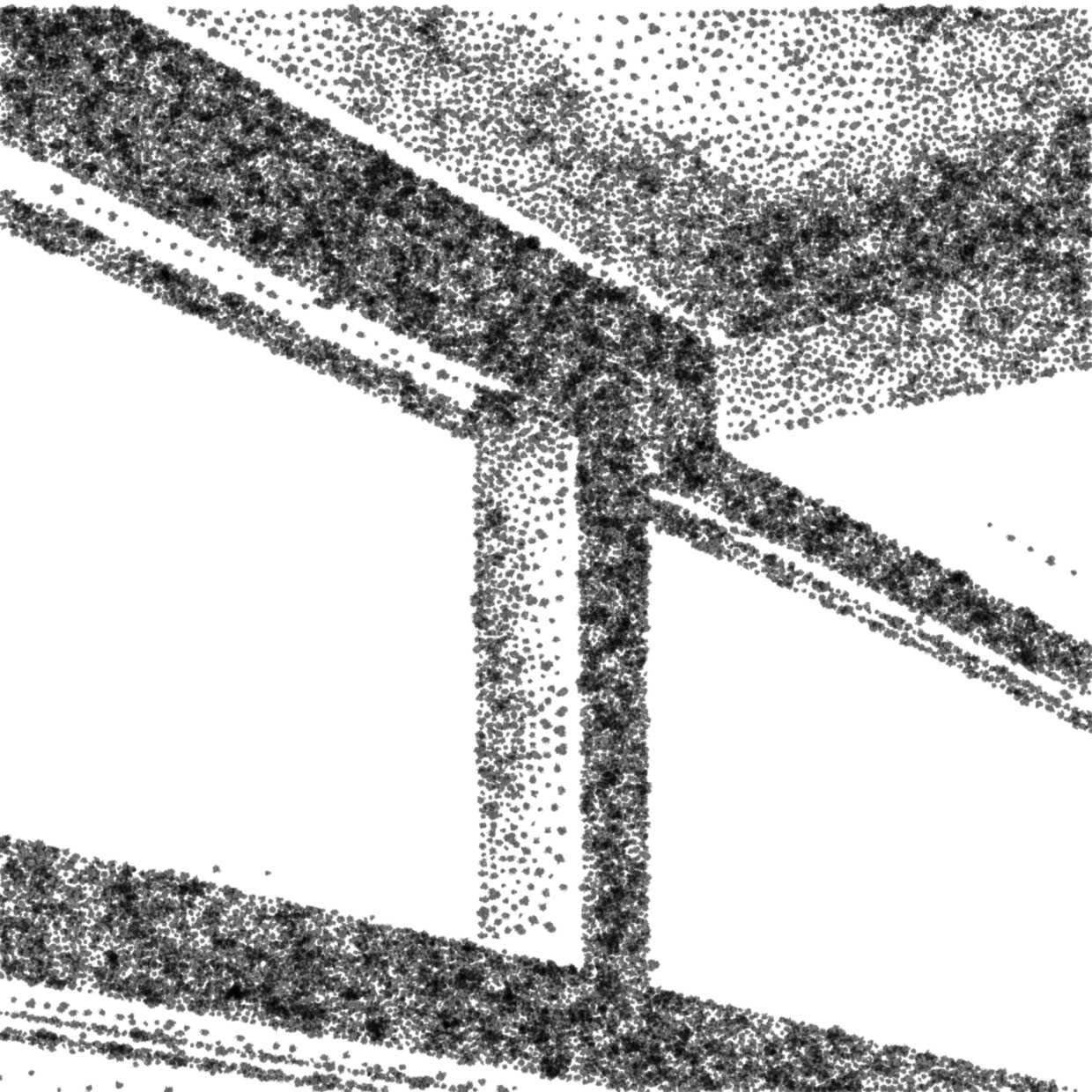}}\hfill%
	\subfigure[\hspace{\linewidth}]{\label{fig:results_random_ebg_scanned_dots:b}\includegraphics[width=\picturewidth]{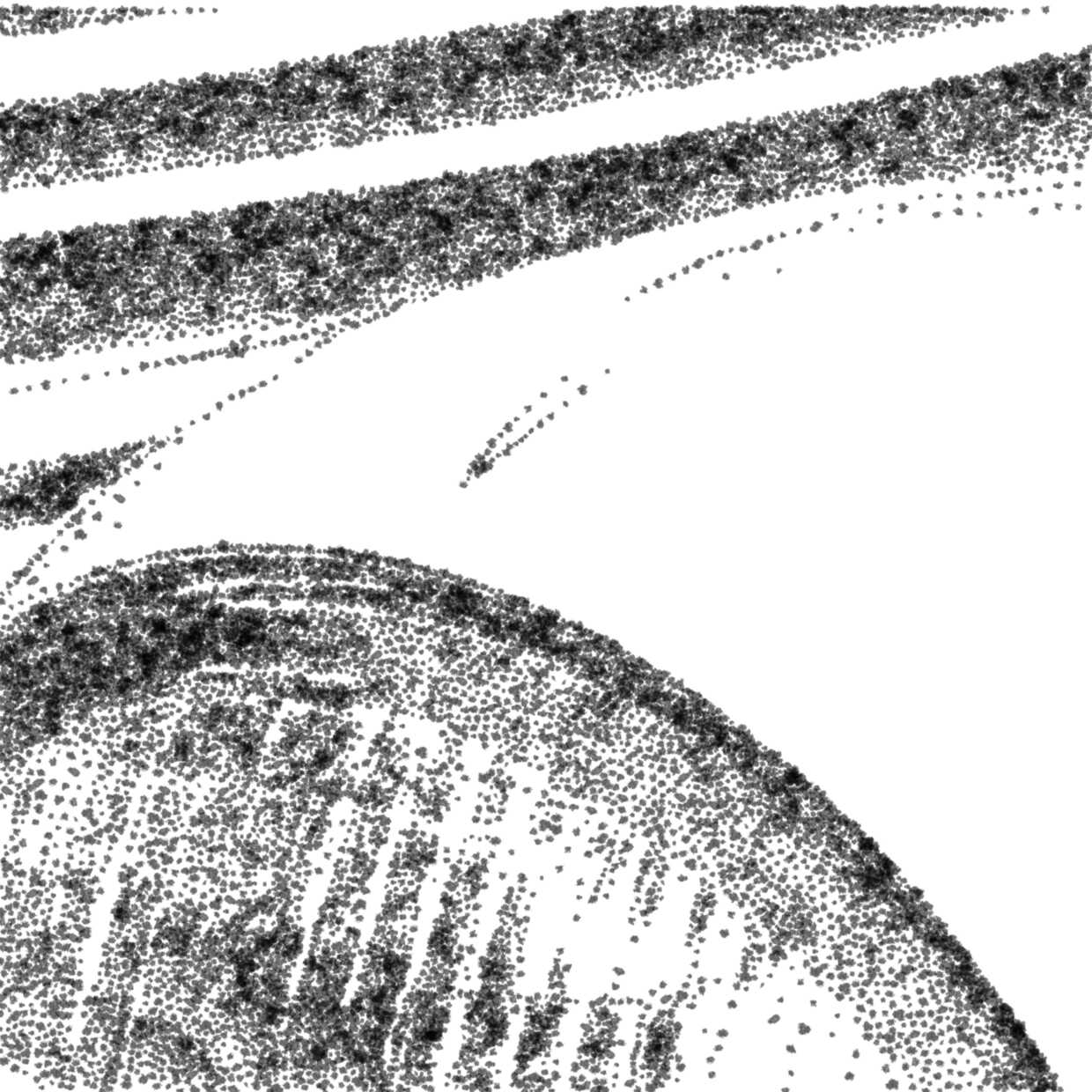}}\hfill%
	\subfigure[\hspace{\linewidth}]{\label{fig:results_random_ebg_scanned_dots:c}\includegraphics[width=\picturewidth]{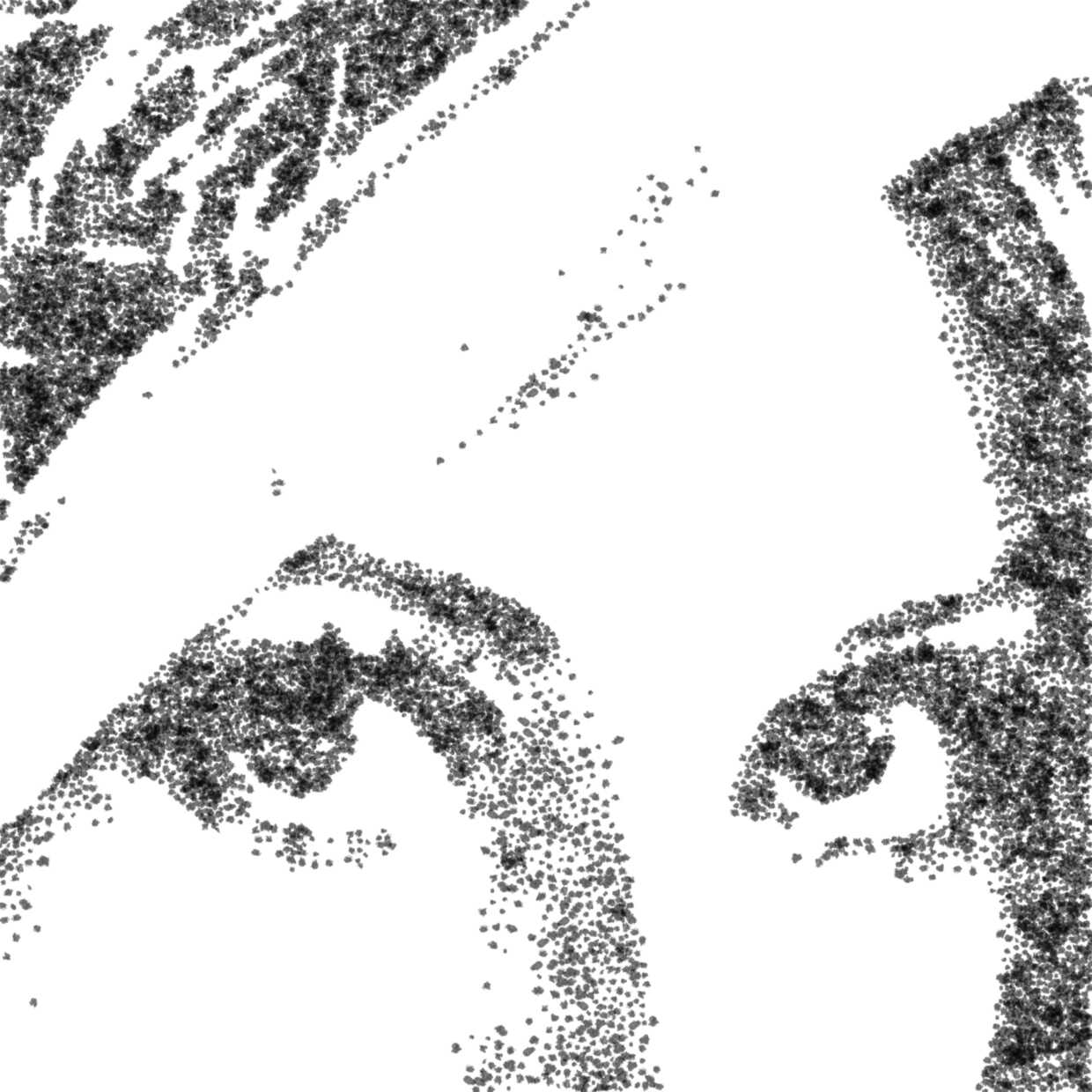}}\hfill%
	\subfigure[\hspace{\linewidth}]{\label{fig:results_random_ebg_scanned_dots:d}\includegraphics[width=\picturewidth]{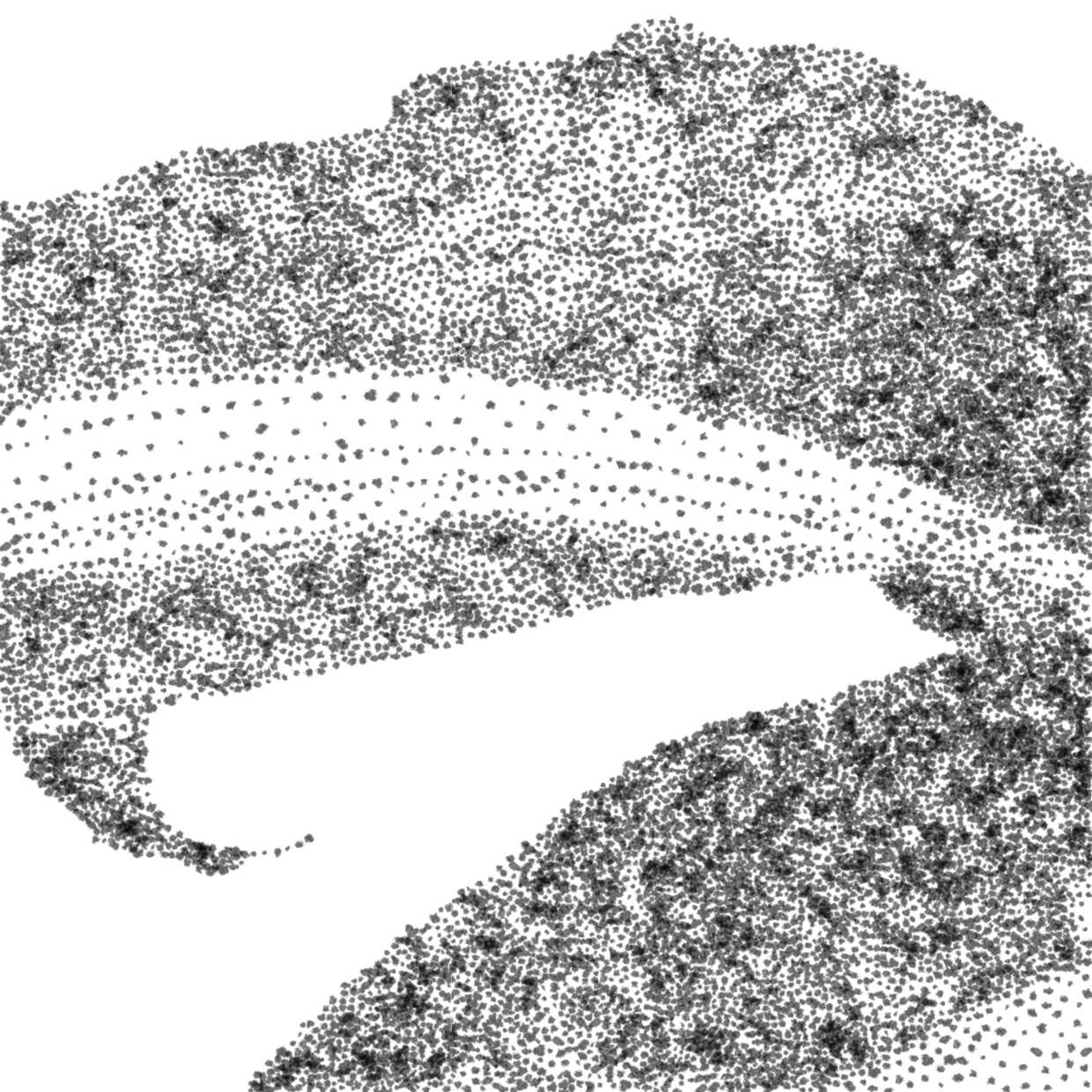}}\\[.5ex]%
	\subfigure[\hspace{\linewidth}]{\label{fig:results_random_ebg_scanned_dots:e}\includegraphics[width=\picturewidth]{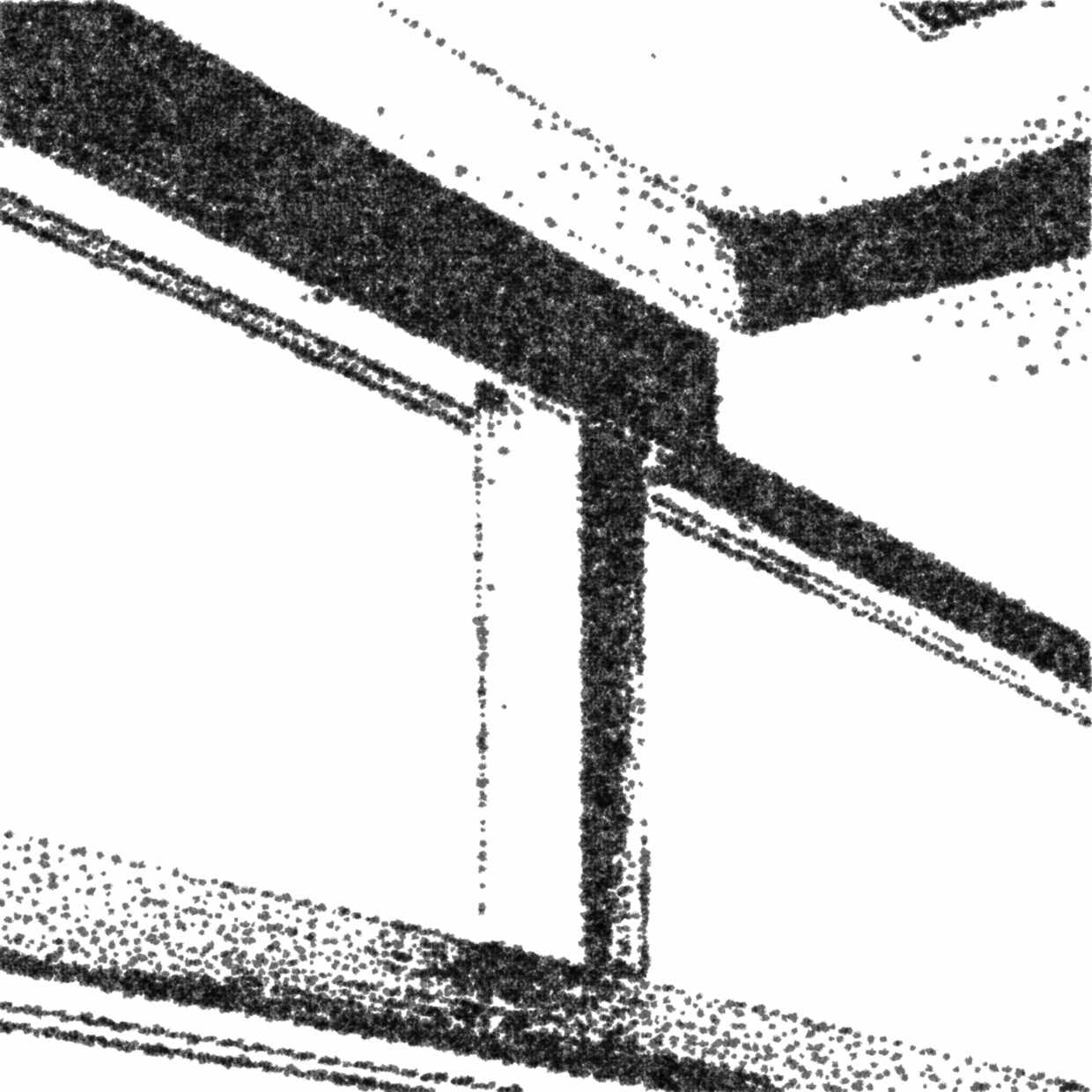}}\hfill%
	\subfigure[\hspace{\linewidth}]{\label{fig:results_random_ebg_scanned_dots:f}\includegraphics[width=\picturewidth]{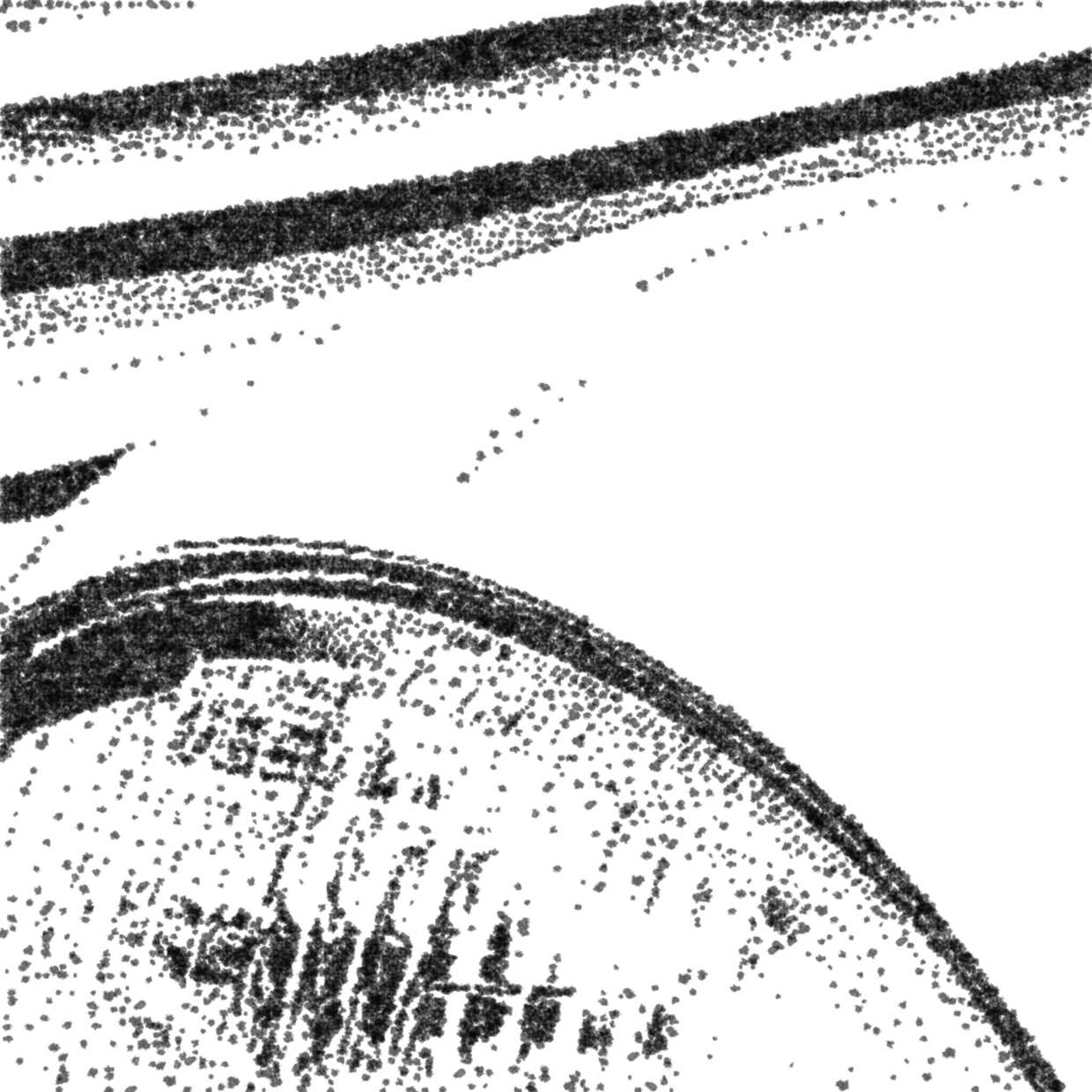}}\hfill%
	\subfigure[\hspace{\linewidth}]{\label{fig:results_random_ebg_scanned_dots:g}\includegraphics[width=\picturewidth]{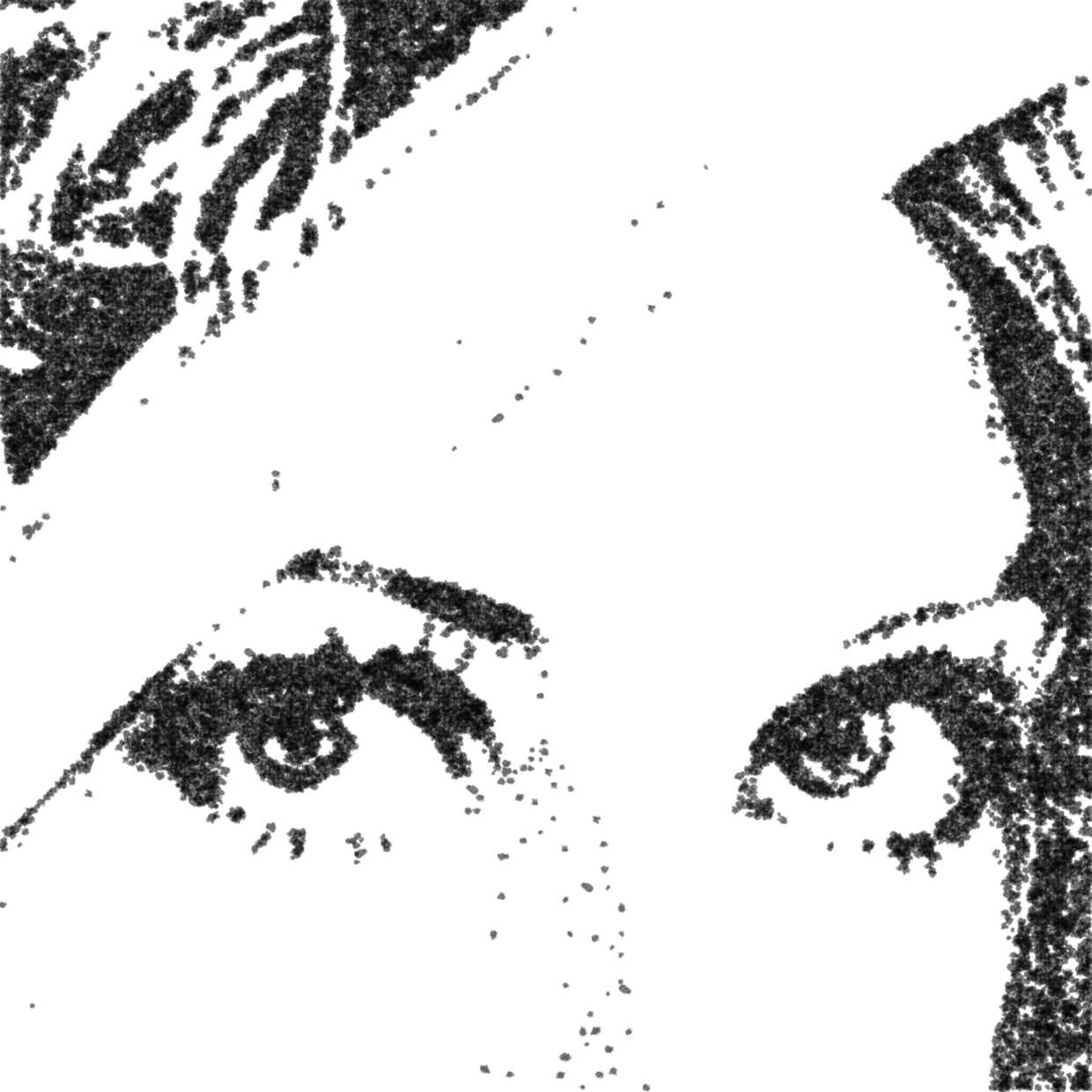}}\hfill%
	\subfigure[\hspace{\linewidth}]{\label{fig:results_random_ebg_scanned_dots:h}\includegraphics[width=\picturewidth]{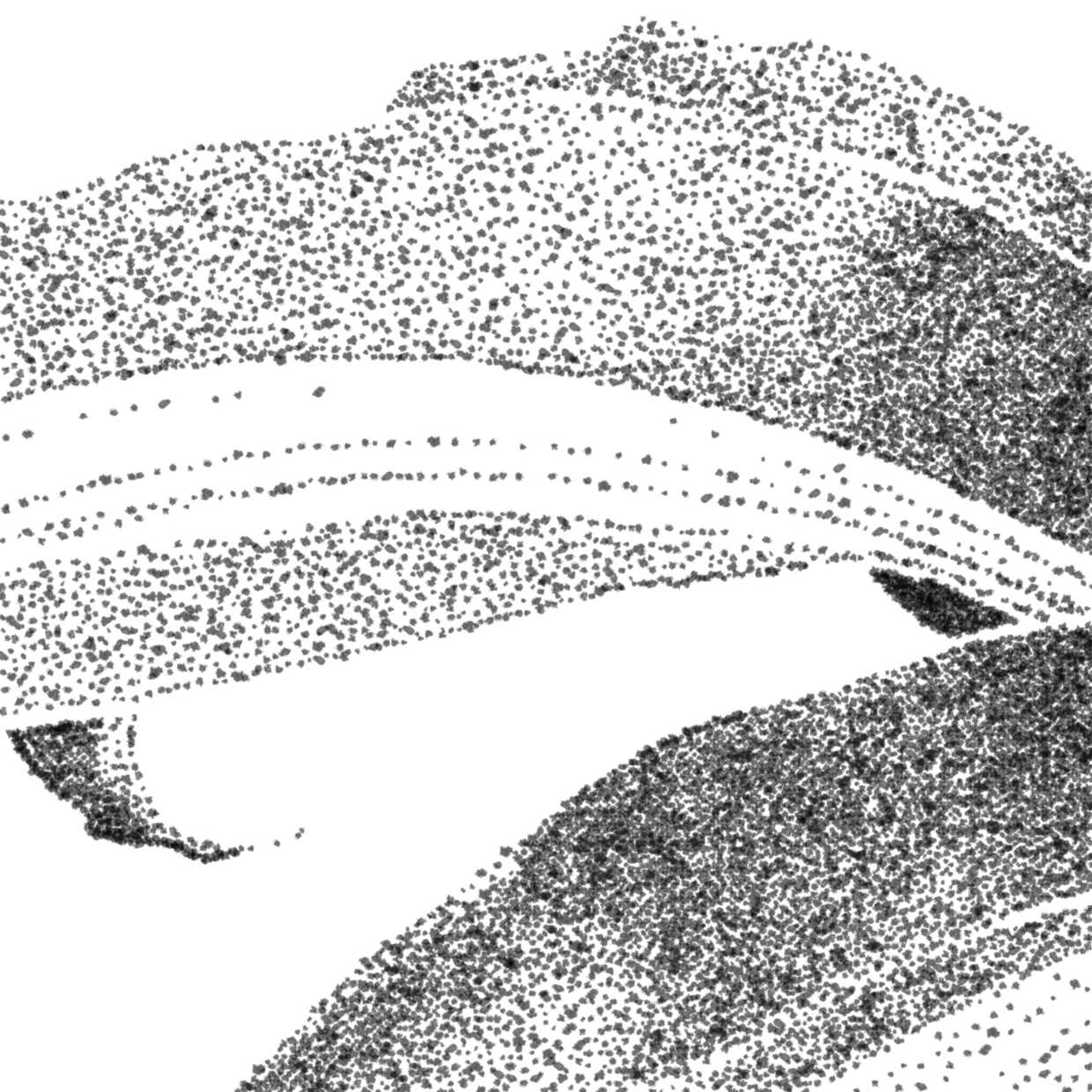}}\\[.5ex]%
	\subfigure[\hspace{\linewidth}]{\label{fig:results_random_ebg_scanned_dots:i}\includegraphics[width=\picturewidth]{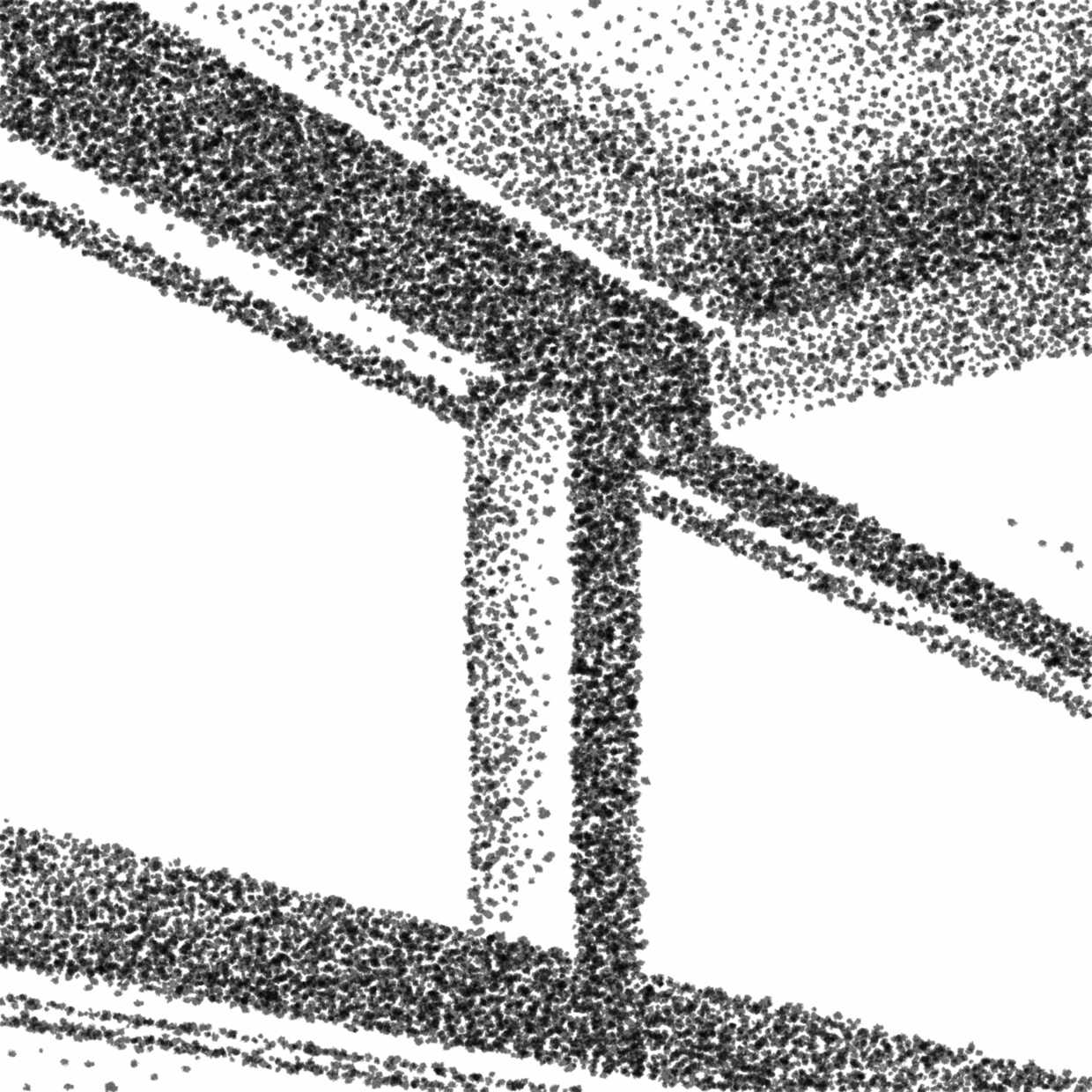}}\hfill%
	\subfigure[\hspace{\linewidth}]{\label{fig:results_random_ebg_scanned_dots:j}\includegraphics[width=\picturewidth]{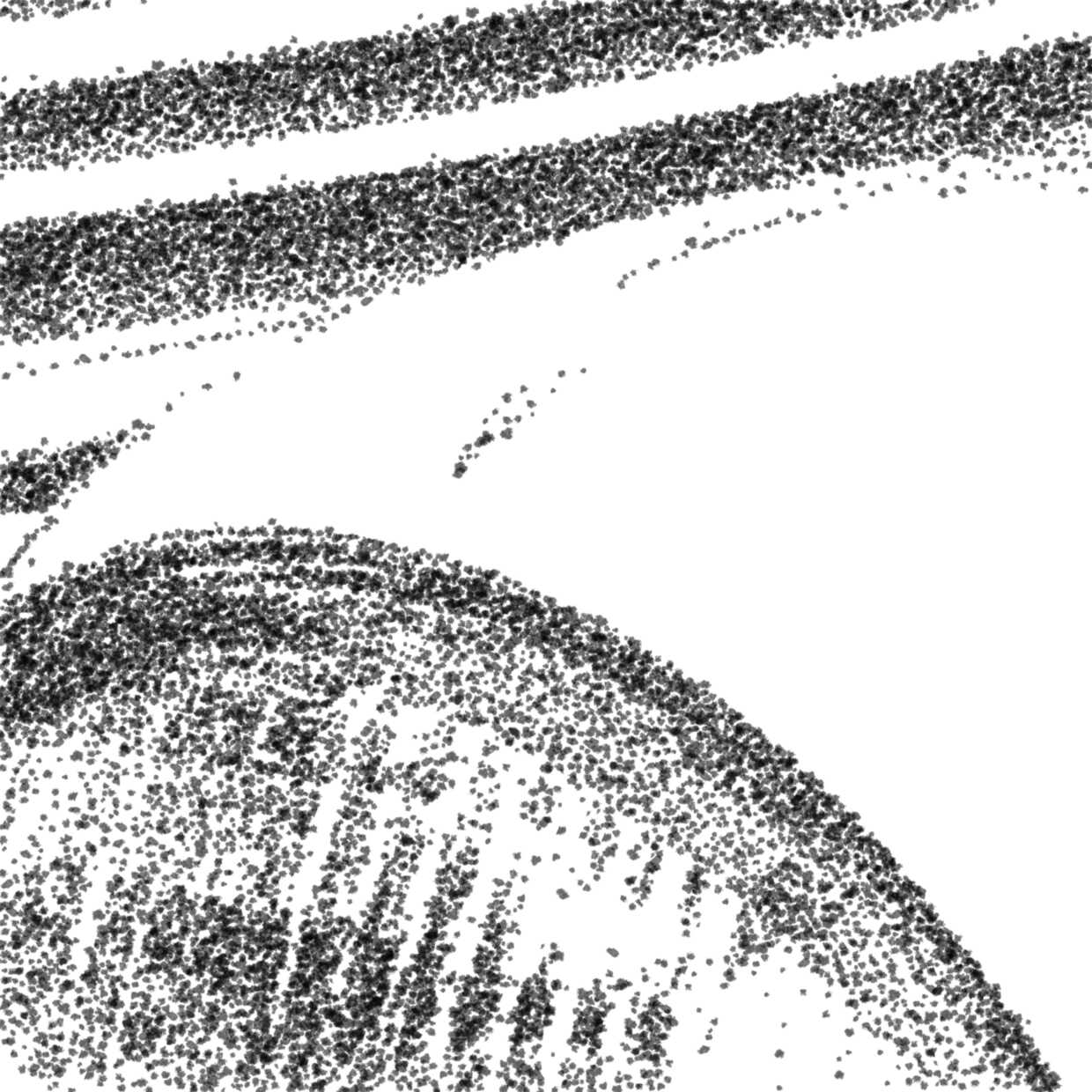}}\hfill%
	\subfigure[\hspace{\linewidth}]{\label{fig:results_random_ebg_scanned_dots:k}\includegraphics[width=\picturewidth]{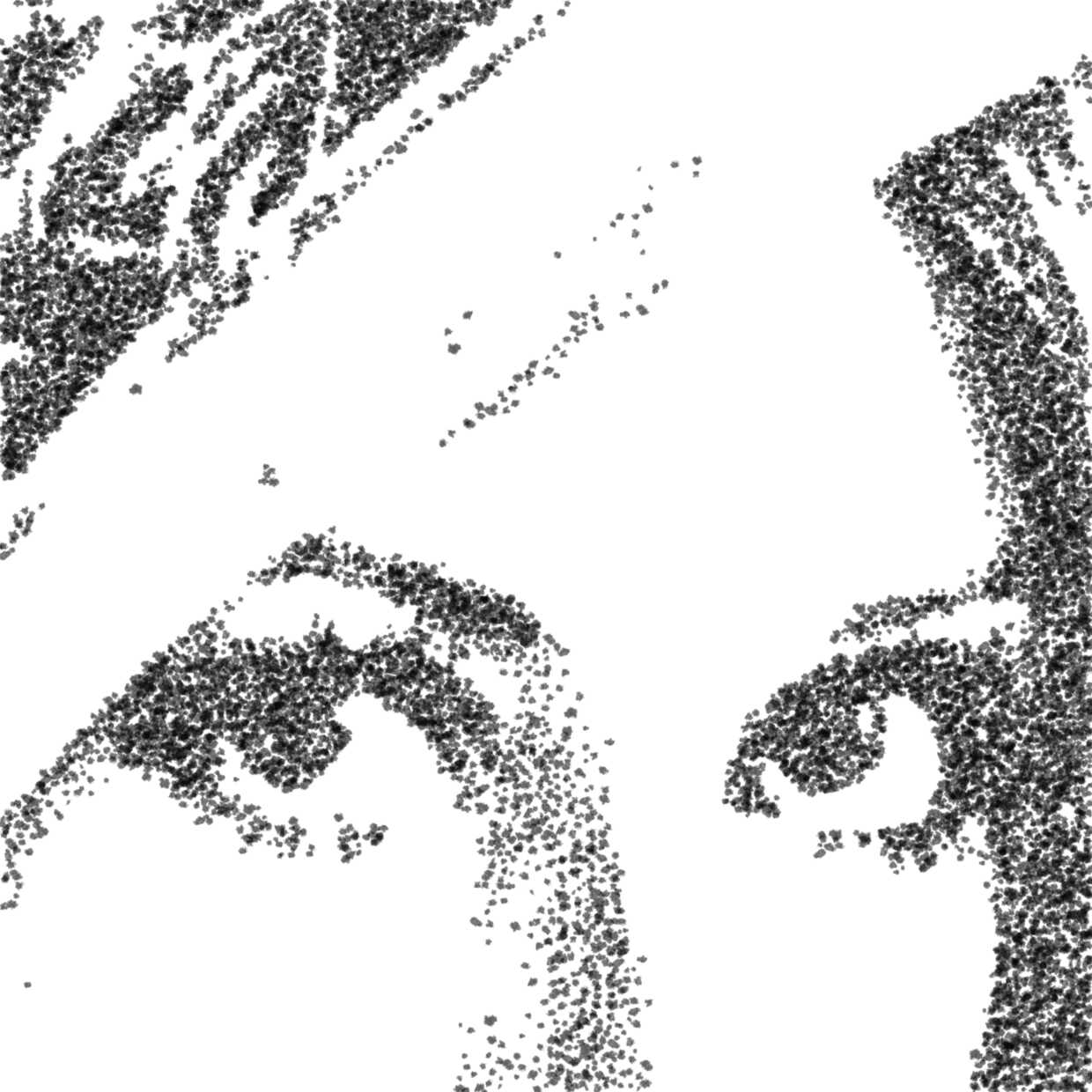}}\hfill%
	\subfigure[\hspace{\linewidth}]{\label{fig:results_random_ebg_scanned_dots:l}\includegraphics[width=\picturewidth]{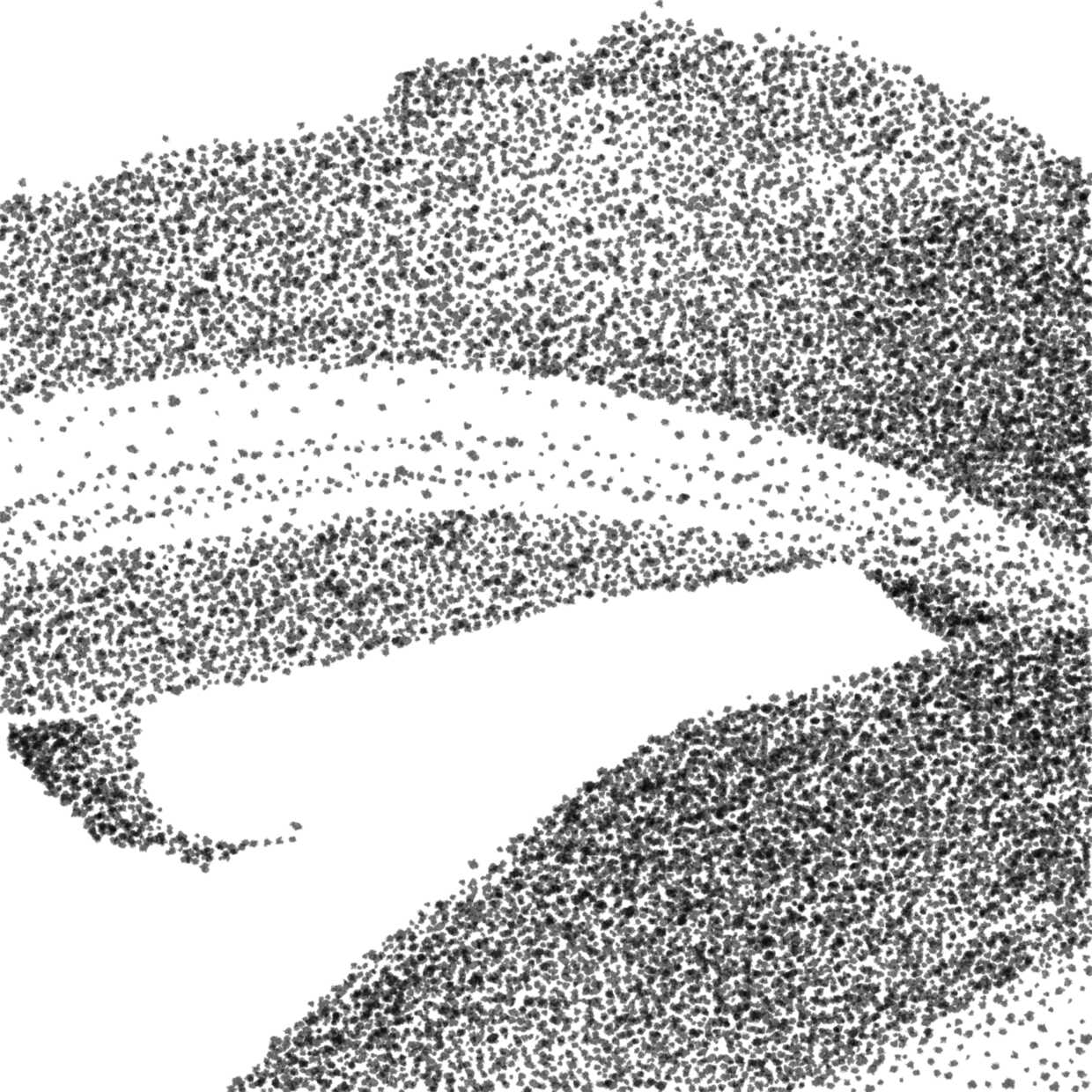}}\\[.5ex]%
	\subfigure[\hspace{\linewidth}]{\label{fig:results_random_ebg_scanned_dots:m}\includegraphics[width=\picturewidth]{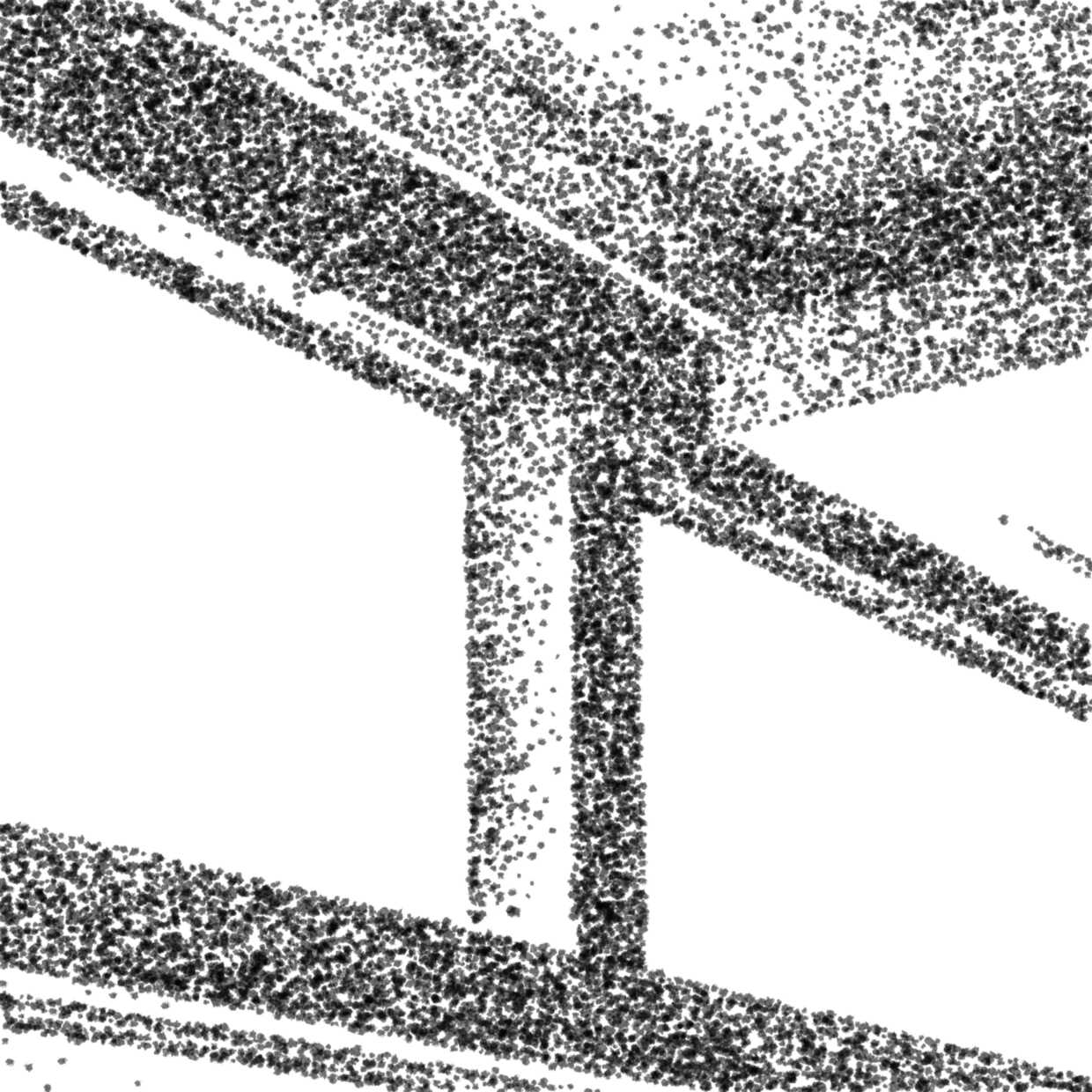}}\hfill%
	\subfigure[\hspace{\linewidth}]{\label{fig:results_random_ebg_scanned_dots:n}\includegraphics[width=\picturewidth]{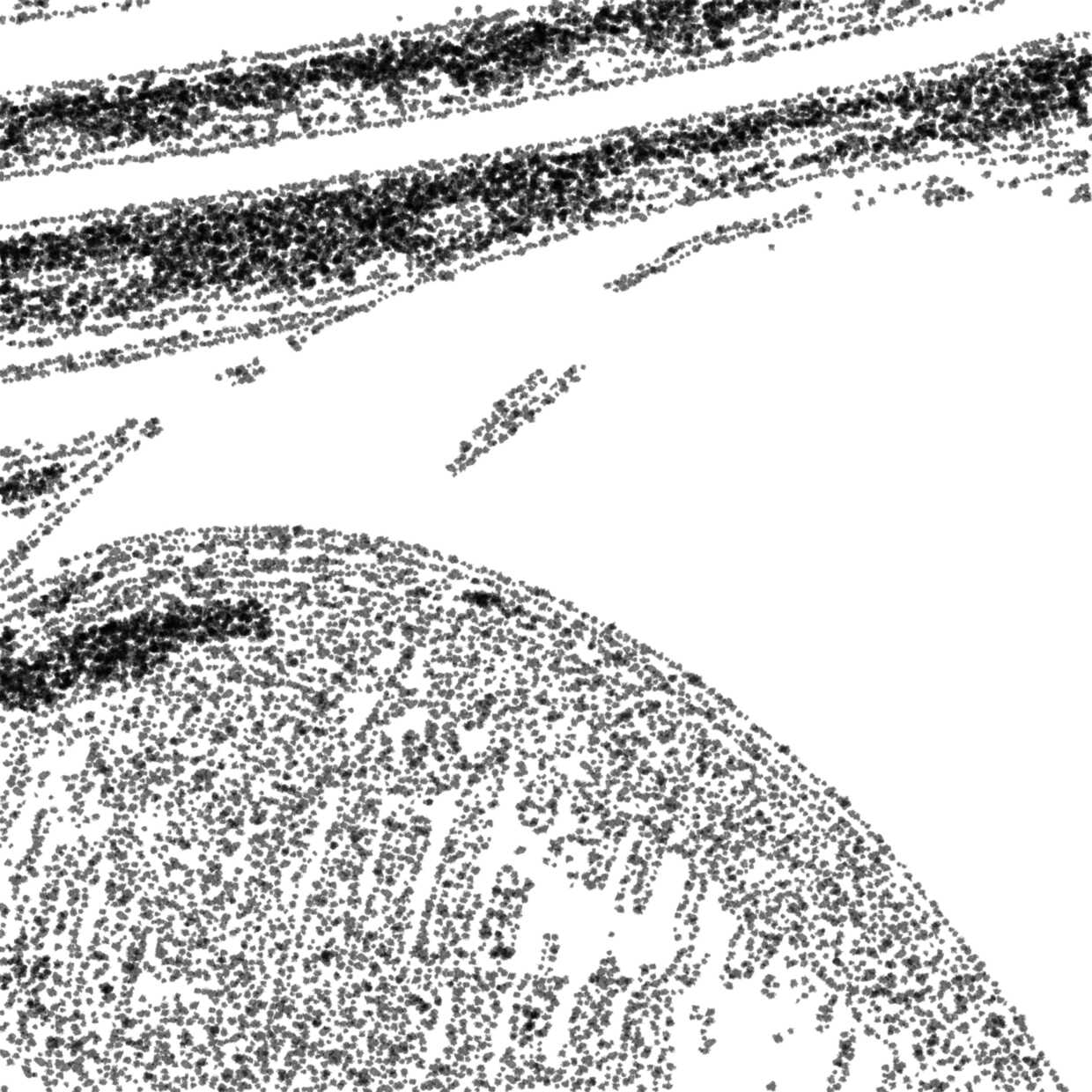}}\hfill%
	\subfigure[\hspace{\linewidth}]{\label{fig:results_random_ebg_scanned_dots:o}\includegraphics[width=\picturewidth]{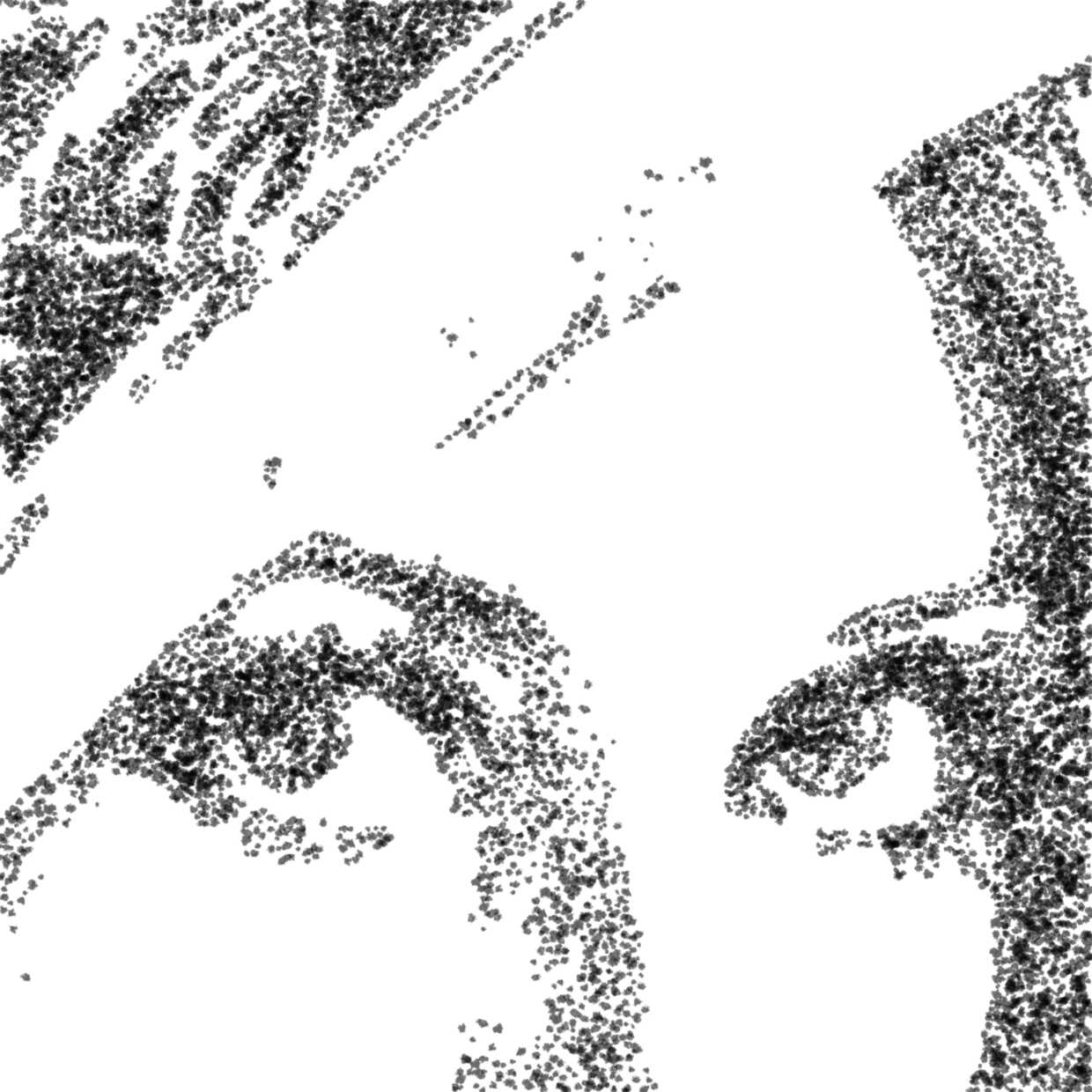}}\hfill%
	\subfigure[\hspace{\linewidth}]{\label{fig:results_random_ebg_scanned_dots:p}\includegraphics[width=\picturewidth]{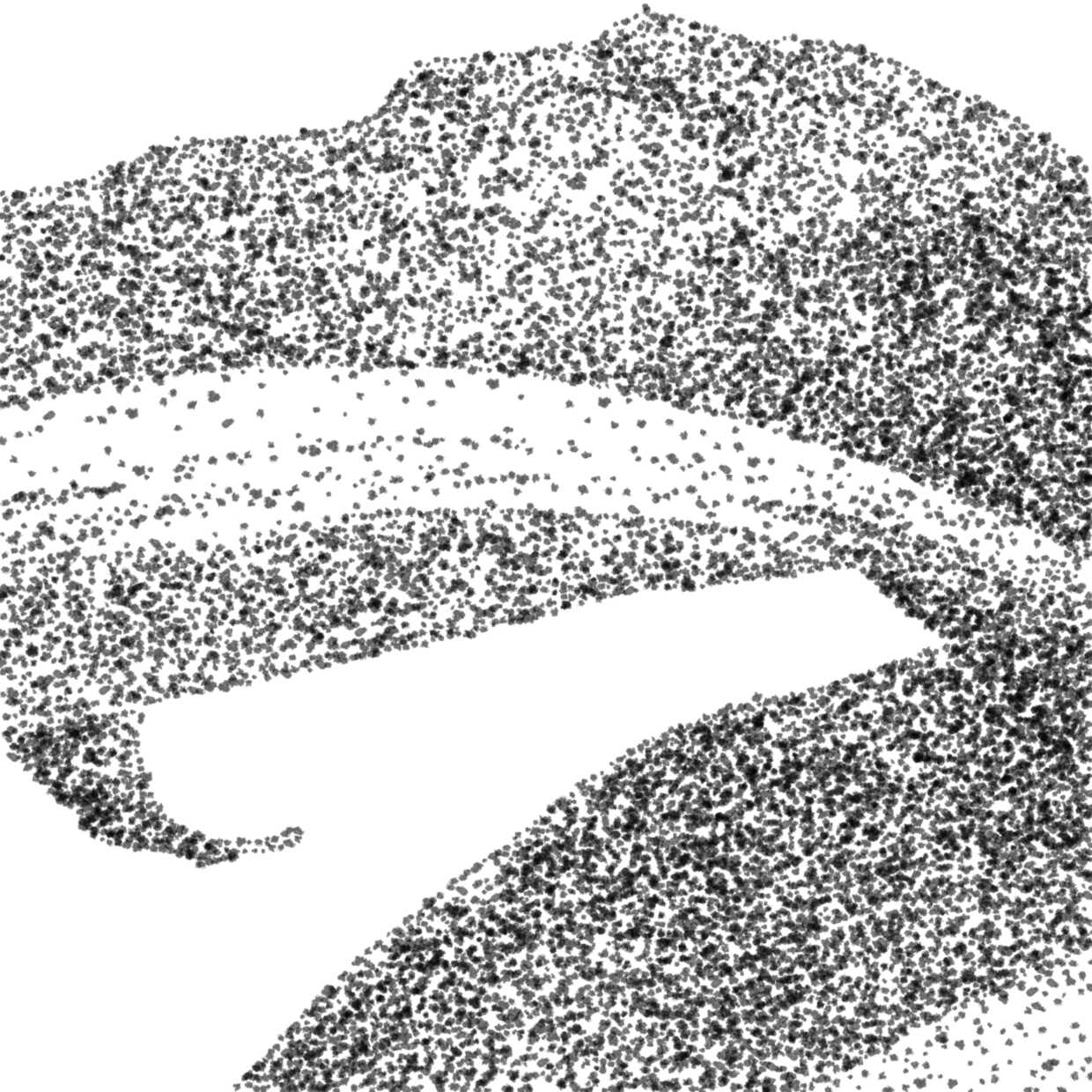}}\vspace{-0.8ex}
	\caption{Dot size range of 4--8; discrete, random, scanned dots at 1200\, ppi. First row WVS, second row SPS, third row EBG, and fourth row IPD.}\vspace{-1.5ex}
	\label{fig:results_random_ebg_scanned_dots}
\end{figure*}

\setlength{\picturewidth}{.245\textwidth}
\begin{figure*}[t]
	\centering
	\setlength{\subfigcapskip}{-3.5ex}%
	\subfigure[\hspace{\linewidth}]{\label{fig:filter_comparison:a}\includegraphics[width=\picturewidth]{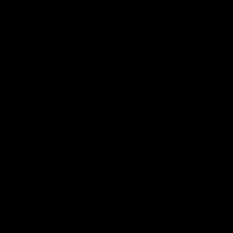}}\hfill%
	\subfigure[\hspace{\linewidth}]{\label{fig:filter_comparison:b}\includegraphics[width=\picturewidth]{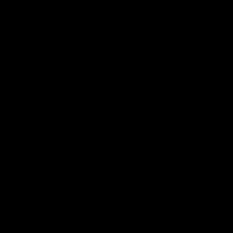}}\hfill%
	\subfigure[\hspace{\linewidth}]{\label{fig:filter_comparison:c}\includegraphics[width=\picturewidth]{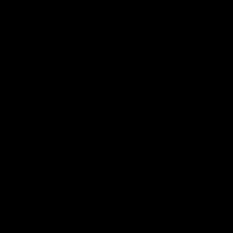}}\hfill%
	\subfigure[\hspace{\linewidth}]{\label{fig:filter_comparison:d}\includegraphics[width=\picturewidth]{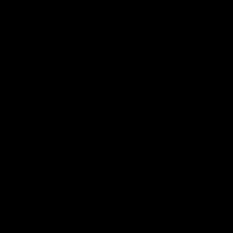}}\\[.5ex]
	\subfigure[\hspace{\linewidth}]{\label{fig:filter_comparison:e}\includegraphics[width=\picturewidth]{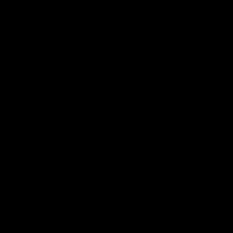}}\hfill%
	\subfigure[\hspace{\linewidth}]{\label{fig:filter_comparison:f}\includegraphics[width=\picturewidth]{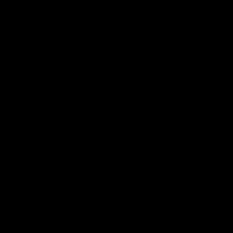}}\hfill%
	\subfigure[\hspace{\linewidth}]{\label{fig:filter_comparison:g}\includegraphics[width=\picturewidth]{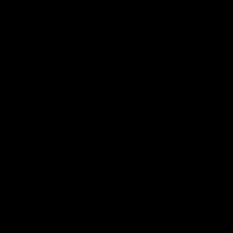}}\hfill%
	\subfigure[\hspace{\linewidth}]{\label{fig:filter_comparison:h}\includegraphics[width=\picturewidth]{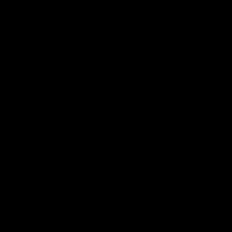}}\\[.5ex]
	\subfigure[\hspace{\linewidth}]{\label{fig:filter_comparison:i}\includegraphics[width=\picturewidth]{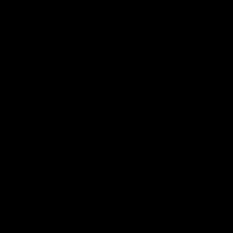}}\hfill%
	\subfigure[\hspace{\linewidth}]{\label{fig:filter_comparison:j}\includegraphics[width=\picturewidth]{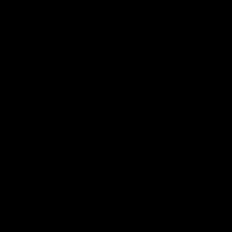}}\hfill%
	\subfigure[\hspace{\linewidth}]{\label{fig:filter_comparison:k}\includegraphics[width=\picturewidth]{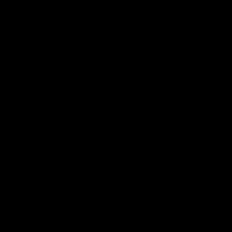}}\hfill%
	\subfigure[\hspace{\linewidth}]{\label{fig:filter_comparison:l}\includegraphics[width=\picturewidth]{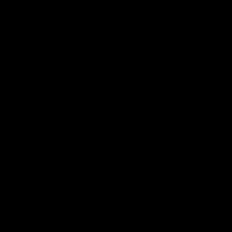}}\vspace{-1.0ex}
	\caption{Comparison of the use of different image filters for the IPD method. First row: Canny, second row: DoG, and third row: LoG. Also, we specifically highlighted the edge dots that originated from the WVD distribution. See the large versions in Figures~\ref{fig:filter_comparison_Canny}--\ref{fig:filter_comparison_LoG} in the additional material.}
	\label{fig:filter_comparison}
\end{figure*}

\setlength{\picturewidth}{.1225\textwidth}
\begin{figure*}[t]
	\centering
	\setlength{\subfigcapskip}{-3.5ex}%
	\subfigure[\hspace{\linewidth}]{\label{fig:filter_comparison_edges:a}\includegraphics[width=\picturewidth]{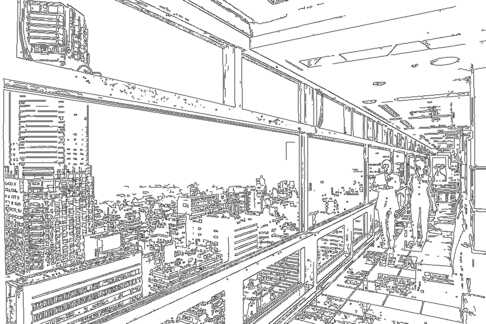}}\hfill%
	\subfigure[\hspace{\linewidth}]{\label{fig:filter_comparison_edges:b}\includegraphics[width=\picturewidth]{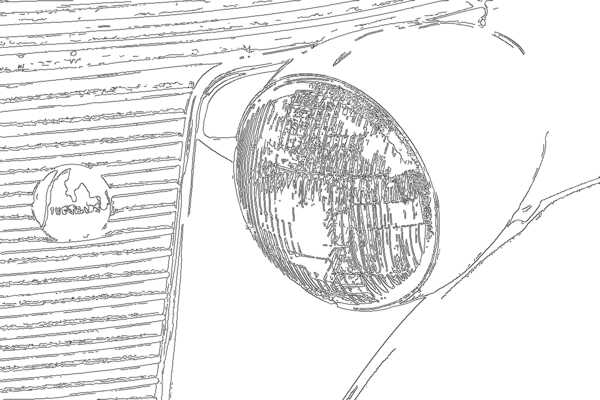}}\hfill%
	\subfigure[\hspace{\linewidth}]{\label{fig:filter_comparison_edges:c}\includegraphics[width=\picturewidth]{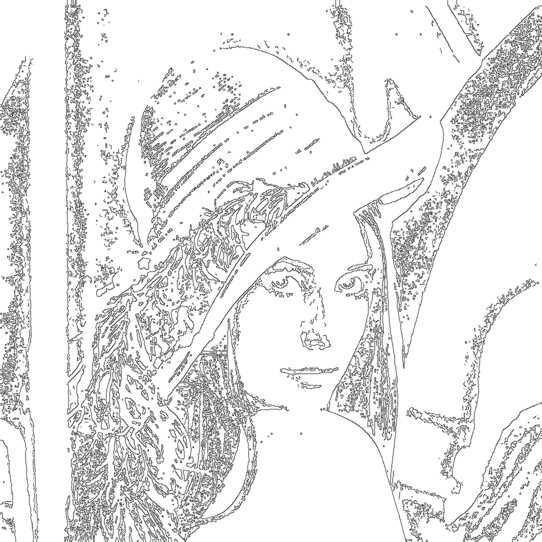}}\hfill%
	\subfigure[\hspace{\linewidth}]{\label{fig:filter_comparison_edges:d}\includegraphics[width=\picturewidth]{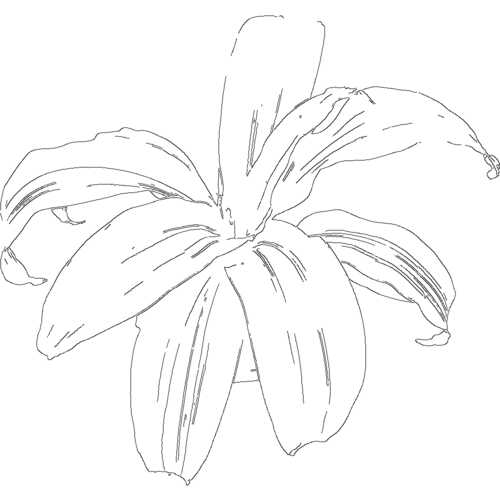}}\hfill%
	\subfigure[\hspace{\linewidth}]{\label{fig:filter_comparison_edges:e}\includegraphics[width=\picturewidth]{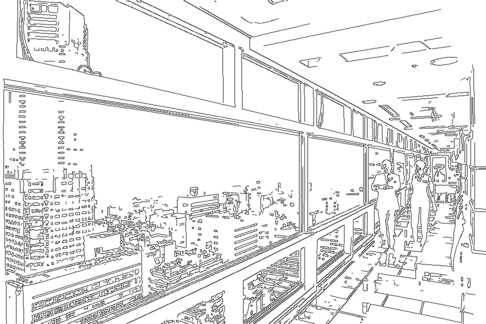}}\hfill%
	\subfigure[\hspace{\linewidth}]{\label{fig:filter_comparison_edges:f}\includegraphics[width=\picturewidth]{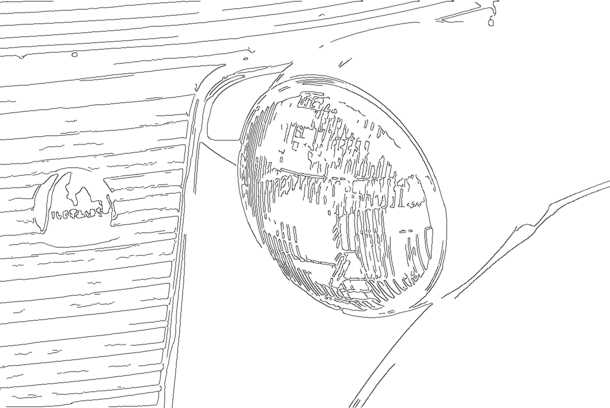}}\hfill%
	\subfigure[\hspace{\linewidth}]{\label{fig:filter_comparison_edges:g}\includegraphics[width=\picturewidth]{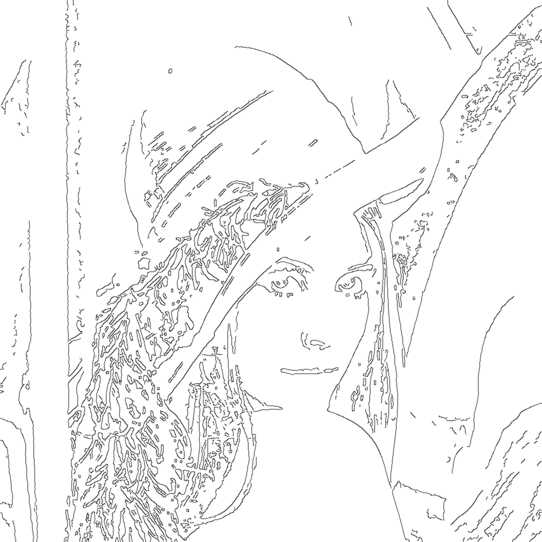}}\hfill%
	\subfigure[\hspace{\linewidth}]{\label{fig:filter_comparison_edges:h}\includegraphics[width=\picturewidth]{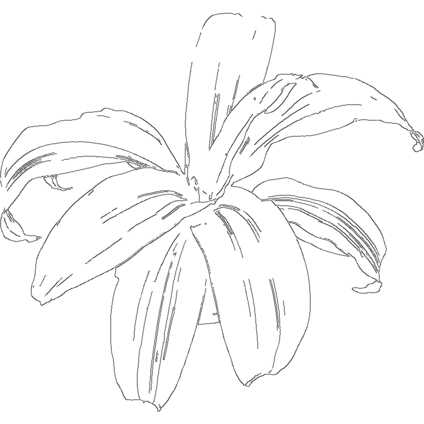}}\\[.25ex]%
	\subfigure[\hspace{\linewidth}]{\label{fig:filter_comparison_edges:i}\includegraphics[width=\picturewidth]{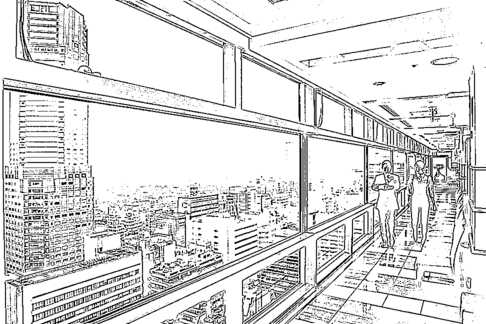}}\hfill%
	\subfigure[\hspace{\linewidth}]{\label{fig:filter_comparison_edges:j}\includegraphics[width=\picturewidth]{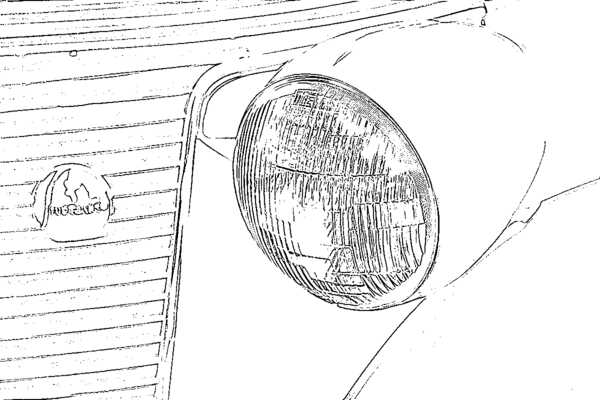}}\hfill%
	\subfigure[\hspace{\linewidth}]{\label{fig:filter_comparison_edges:k}\includegraphics[width=\picturewidth]{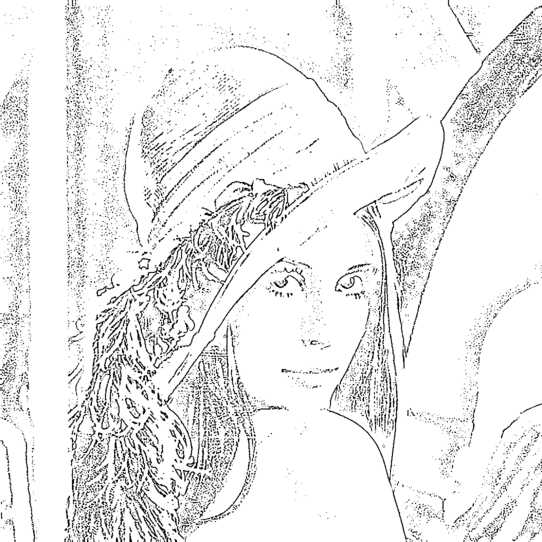}}\hfill%
	\subfigure[\hspace{\linewidth}]{\label{fig:filter_comparison_edges:l}\includegraphics[width=\picturewidth]{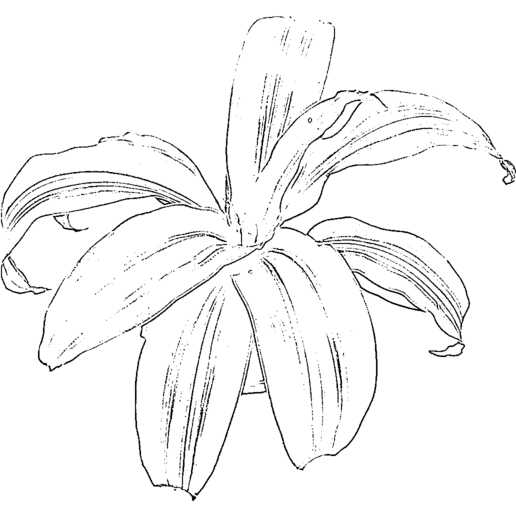}}\hfill%
	\subfigure[\hspace{\linewidth}]{\label{fig:filter_comparison_edges:m}\includegraphics[width=\picturewidth]{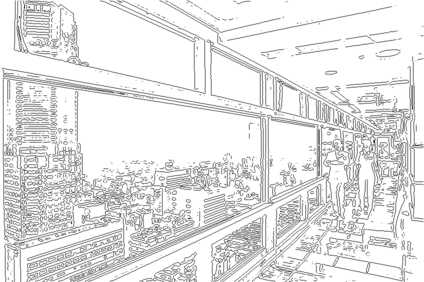}}\hfill%
	\subfigure[\hspace{\linewidth}]{\label{fig:filter_comparison_edges:n}\includegraphics[width=\picturewidth]{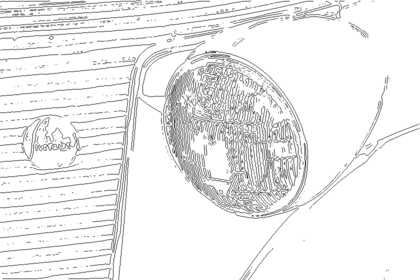}}\hfill%
	\subfigure[\hspace{\linewidth}]{\label{fig:filter_comparison_edges:o}\includegraphics[width=\picturewidth]{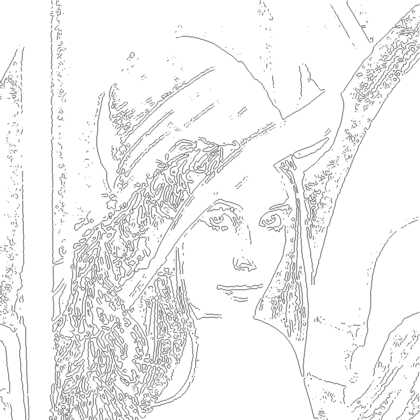}}\hfill%
	\subfigure[\hspace{\linewidth}]{\label{fig:filter_comparison_edges:p}\includegraphics[width=\picturewidth]{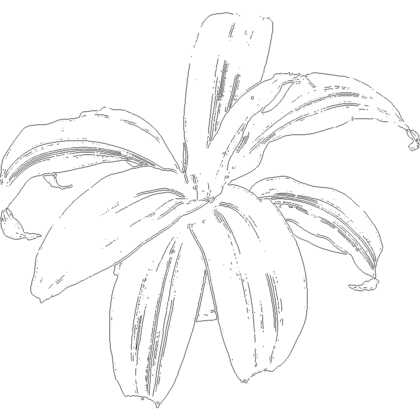}}\vspace{-1.0ex}%
	\caption{Edge image inputs for the results in Figures~\ref{fig:results_constant_bw}--\ref{fig:filter_comparison}. \subref{fig:filter_comparison_edges:a}--\subref{fig:filter_comparison_edges:d}: Canny-based edges for Figures~\ref{fig:results_constant_bw}--\ref{fig:results_random_ebg_scanned_dots}; 
	\subref{fig:filter_comparison_edges:e}--\subref{fig:filter_comparison_edges:h}: edges from filtered input with Canny for Figures~\ref{fig:filter_comparison:a}--\subref{fig:filter_comparison:d}; 
	\subref{fig:filter_comparison_edges:i}--\subref{fig:filter_comparison_edges:l}: edges from filtered input with DoG for Figures~\ref{fig:filter_comparison:e}--\subref{fig:filter_comparison:h}; 
	\subref{fig:filter_comparison_edges:m}--\subref{fig:filter_comparison_edges:p}: edges from filtered input with LoG for Figures~\ref{fig:filter_comparison:i}--\subref{fig:filter_comparison:l}.}
	\label{fig:filter_comparison_edges}
\end{figure*}

By using small circular dots with constant sizes, \autoref{fig:results_constant_bw} allows us to analyze the spatial character of the different dot distributions. We can observe, \eg, that SPS uses a regular grid alignment of the dots somewhat reminiscent of halftoning, which is not visible in its actual results as then the larger dot size leads to overlapping stippling (Figures~\ref{fig:results_modulated_ebg_scanned_dots} and~\ref{fig:results_random_ebg_scanned_dots}). WVS, in contrast, is not bound by a grid but exhibits chain artifacts, also in normally dense regions where dot size attenuation normally would hide this fact (\autoref{fig:results_modulated_ebg_scanned_dots}). Nonetheless, the chain artifacts also play to WVS's advantage for stipples that should line up, such as the example in \autoref{fig:results_constant_bw:b}. EBG shows less chain artifacts in areas, but its edges and borders are not as clear due to the halftoning-based placement and the added randomness. EBG' halftoning grid itself is not visible because, for \autoref{fig:results_constant_bw}, we used the EBG process as if with large dots and then used the final dot positions, without rounding them. Finally, we see that IPD combines the overall good distribution of EBG in areas with better edges from WVS, evident, \eg, in \autoref{fig:results_constant_bw:o}.

In a next comparison, we added variable, more realistic dot sizes that are also modulated based on the source image's brightness \cite{Secord:2002:RMP} (\autoref{fig:results_modulated_ebg_scanned_dots}). We computed the range of the actual dot sizes according to the considerations by EBG, which aimed to simulate realistic dot sizes. This change of dot sizes naturally makes all images much darker. The modulated dot sizes also make some artifacts such as the grid arrangement of SPS less apparent, but it does not completely hide them. Rather, in SPS it leads to completely black regions, while for WVS, EBG, and IPD we see more realistic stipple clusters forming. Note that this is not the same modulation as used by \citet{Secord:2002:RMP} (and which we show in \autoref{fig:results_secord_modulation}); \citeauthor{Secord:2002:RMP} did not only rely on the local image graylevel for deciding the modulation and also uses a range of dot sizes for a given graylevel. For the size of the line features in IPD we used a dedicated way to specify the dot size because we could not apply modulation---for linear features extracted by some edge detection mechanism there is no meaningful brightness to use. To avoid having to use a constant size, we instead chose the mean value of the dot size range used for the areas, with an additional random scaling of $\pm$\,25\% to provide some variability. In comparing the line features, we see that they come out well for WVS and SPS, and, in contrast to EBG, also IPD shows much better aligned features. Unfortunately, however, IPD has problems with single lines of dots such as the highlight on the car headlamp as can be seen in \autoref{fig:results_modulated_ebg_scanned_dots:n}. The root cause for these double lines is our use of the Canny filter to control the interpolation. We discuss this effect and how to address it below in \autoref{sec:alternative-results}.

Finally, \autoref{fig:results_random_ebg_scanned_dots} is based on the same modulated dot sizes as \autoref{fig:results_modulated_ebg_scanned_dots}, but with realistic stipple dot textures applied. For this figure, we used the normal scale-dependent process for EBG, and for WVS, SPS, and IPD we placed dot textures to the vector positions of \autoref{fig:results_modulated_ebg_scanned_dots}, with rounding of the resulting scale-dependent texels to avoid texture resampling. We enlarged the modulated dot sizes of \autoref{fig:results_modulated_ebg_scanned_dots} by doubling their size to accomodate the fact that a texture of an irreguarly shaped dot is not perceived as a black circle of the same size, but smaller due the intensity change toward the center of the dot and the overall irregular shape.\footnote{We determined the scaling factor of 2\texttimes\ empirically. In fact, the sizes of the dots in \autoref{fig:results_random_ebg_scanned_dots} are mandated by the EBG process. In practice we thus adjusted the dot sizes in \autoref{fig:results_modulated_ebg_scanned_dots} to be half of the sizes of the textured EBG dots in \autoref{fig:results_random_ebg_scanned_dots}.}
As we can see from the comparison of the images in \autoref{fig:results_random_ebg_scanned_dots}, the irregular shapes of the real dots even further hide the regularity artifacts of SPS, yet the areas of SPS still appear very dark and the images appear to have less tonal variety than the other techniques. Moreover, we see that the interpolation of IPD allowed us to create edges of similar quality as those used in SPS---both for isolated lines and for edges of areas. This result is visible, in particular, in \autoref{fig:results_random_ebg_scanned_dots:p} vs.~\subref{fig:results_random_ebg_scanned_dots:h} vs.~\subref{fig:results_random_ebg_scanned_dots:l}. Some specific edges, however, are still best portrayed by WVS, such as the lines on top of the headlight of the car---neither SPS nor IPD capture them quite as well as WVS, yet both are about the same and much better than WVS. With respect to the area portrayal, WVS shows some unnatural clusters, and EBG and IPD seem both better than the other two---with IPD showing some more tonal variation than EBG. The reason for this effect is that IPD distributes slightly differently and places more points in borders than EBG.

These examples show that our new IPD approach indeed combines high-quality stippled areas with a good reproduction of edges, addressing the problem of EBG's fuzzy lines. As noted before, however, the issue of isolated lines remains, as we discuss next. Moreover, our comparisons used \emph{two specific} source techniques with a \emph{specific parametrization}, while IPD can combine the results from \emph{any two} techniques and with a \emph{flexible parametrization}, as we also show next.

\setlength{\picturewidth}{.245\textwidth}
\begin{figure*}[t]
	\centering
	\setlength{\subfigcapskip}{-3.5ex}%
	\subfigure[\hspace{\linewidth}]{\label{fig:effect_results:a}\includegraphics[width=\picturewidth]{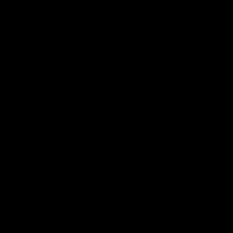}}\hfill%
	\subfigure[\hspace{\linewidth}]{\label{fig:effect_results:b}\includegraphics[width=\picturewidth]{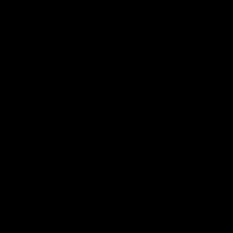}}\hfill%
	\subfigure[\hspace{\linewidth}]{\label{fig:effect_results:c}\includegraphics[width=\picturewidth]{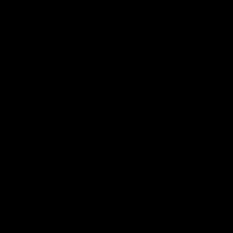}}\hfill%
	\subfigure[\hspace{\linewidth}]{\label{fig:effect_results:d}\includegraphics[width=\picturewidth]{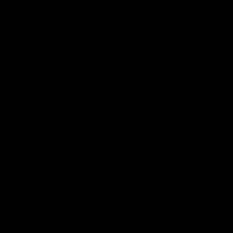}}\\[.5ex]
	\subfigure[\hspace{\linewidth}]{\label{fig:effect_results:e}\includegraphics[width=\picturewidth]{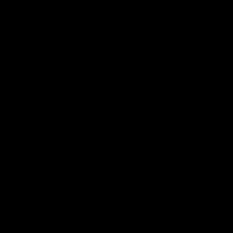}}\hfill%
	\subfigure[\hspace{\linewidth}]{\label{fig:effect_results:f}\includegraphics[width=\picturewidth]{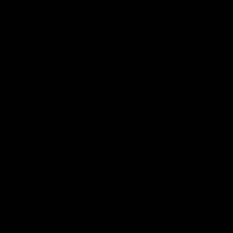}}\hfill%
	\subfigure[\hspace{\linewidth}]{\label{fig:effect_results:g}\includegraphics[width=\picturewidth]{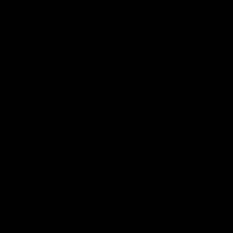}}\hfill%
	\subfigure[\hspace{\linewidth}]{\label{fig:effect_results:h}\includegraphics[width=\picturewidth]{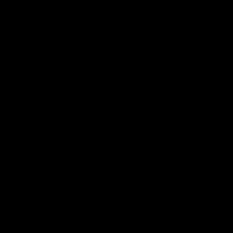}}\vspace{-1.0ex}
	\caption{Interpolation effects that can be achieved with our IPD method. See the large versions in Figures~\ref{fig:effect_results_wb}--\ref{fig:effect_results_wide} in the additional material.}
	\label{fig:effect_results}
\end{figure*}

\setlength{\picturewidth}{.245\textwidth}
\setlength{\pictureheight}{.242\textwidth}
\begin{figure*}[t]
	\centering
	\setlength{\subfigcapskip}{-3.5ex}%
	\subfigure[\hspace{\linewidth}]{\label{fig:other_effect_results:a}\includegraphics[height=\pictureheight]{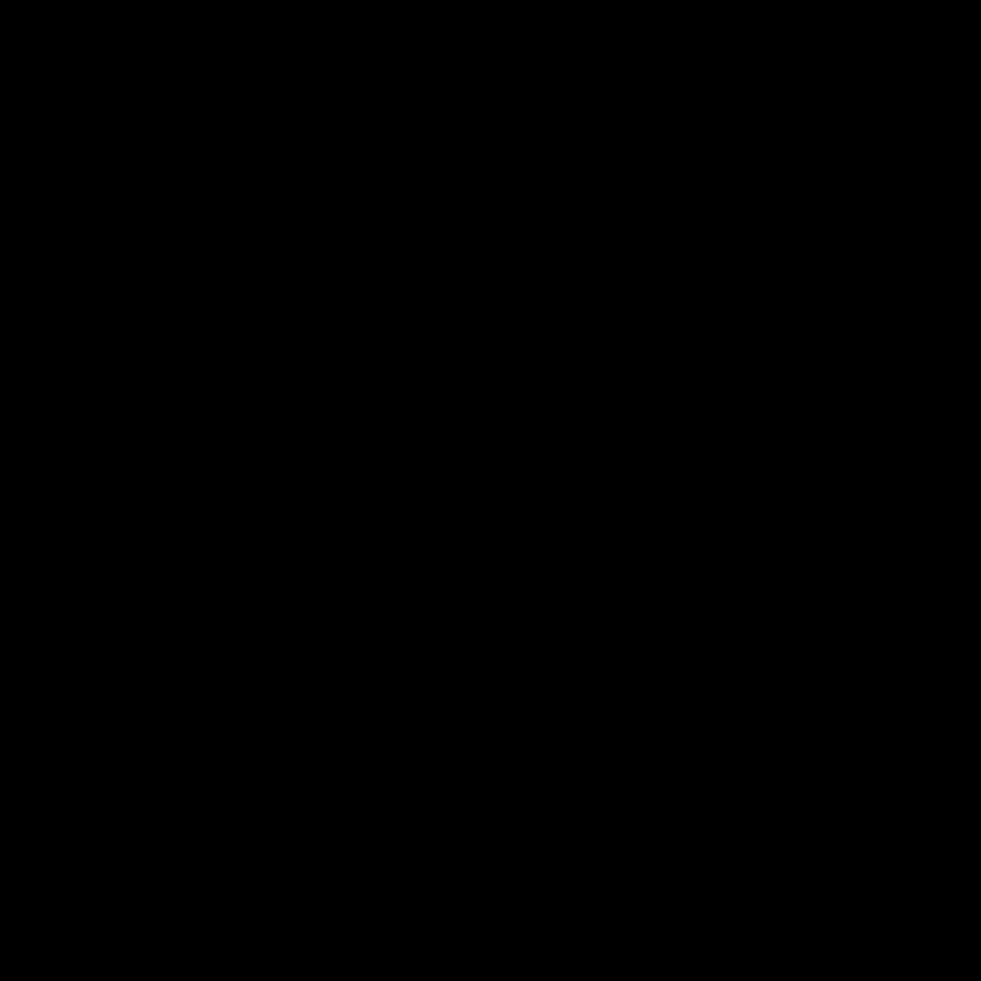}}\hfill
	\subfigure[\hspace{\linewidth}]{\label{fig:other_effect_results:b}\includegraphics[height=\pictureheight]{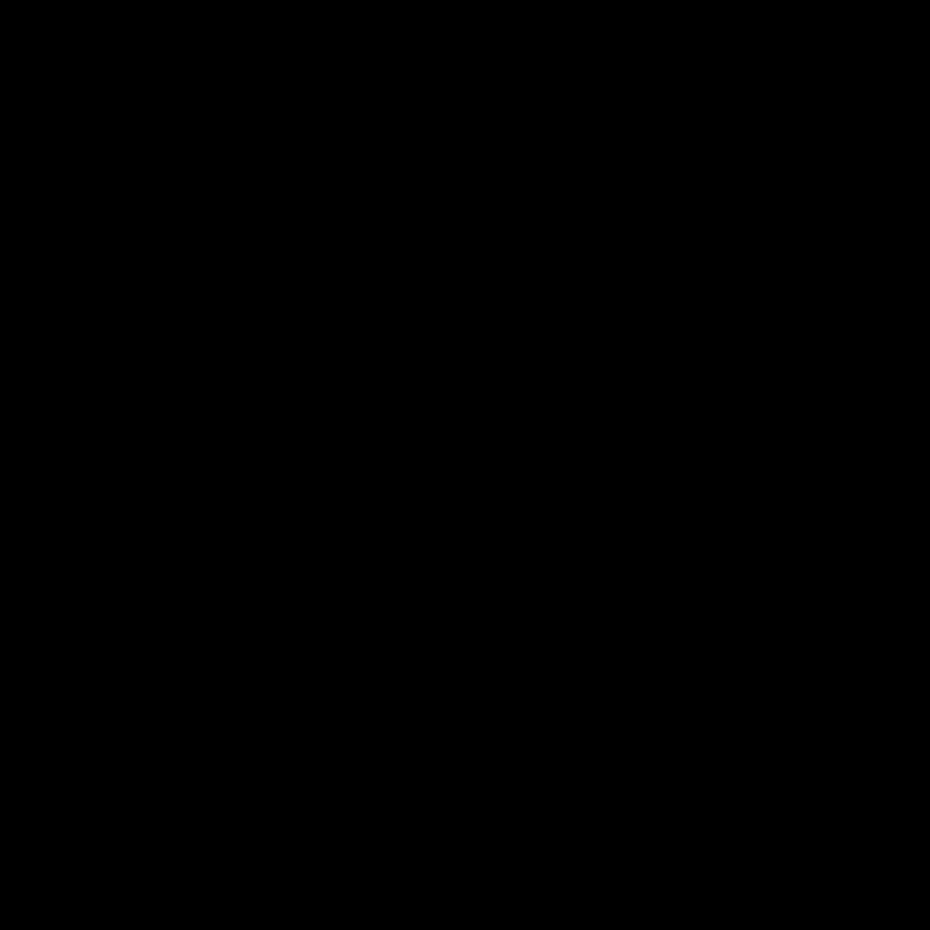}}\hfill
	\subfigure[\hspace{\linewidth}]{\label{fig:other_effect_results:c}\includegraphics[height=\pictureheight]{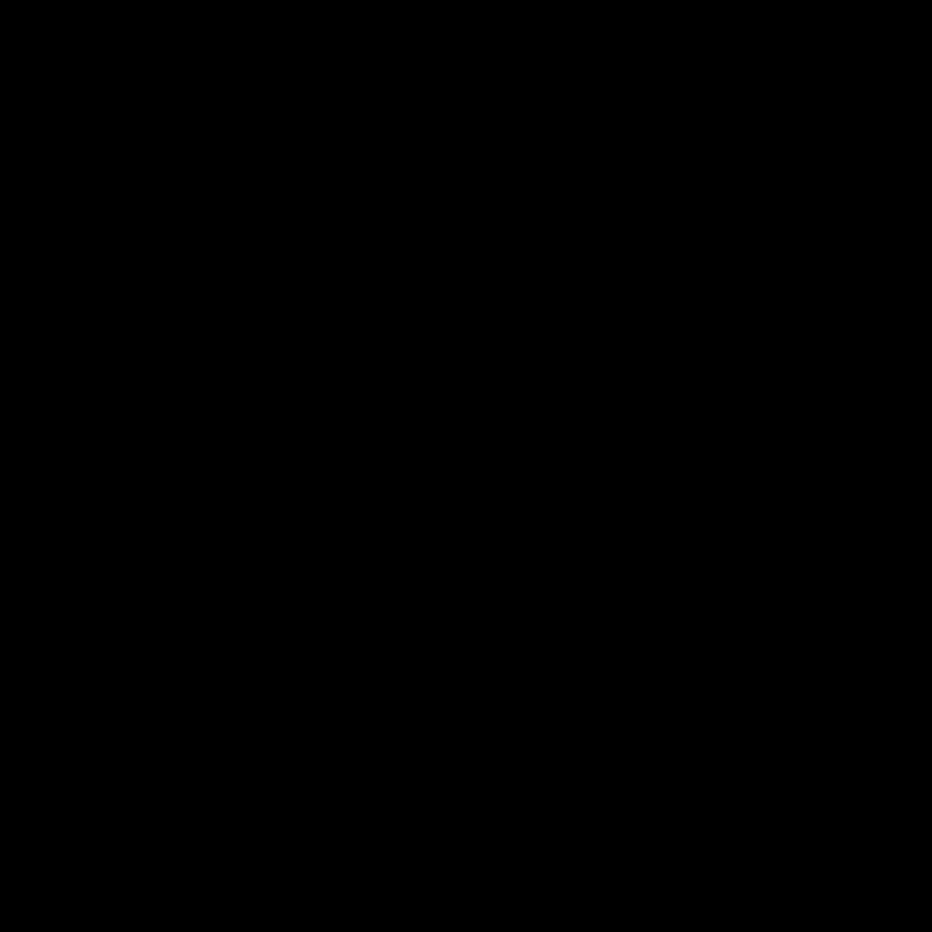}}
	\vspace{-1.0ex}
	\caption{Examples for effects that make use of masks to adjust the distance function $\Delta$, to control the use of stipples from the input distributions.}
	\label{fig:other_effect_results}
\end{figure*}

\subsection{Alternative results}
\label{sec:alternative-results}

\vspace{-0.5ex}For other results we can drop the limitations that we applied to the previous comparison, such as the fixed number of stipple dots used in Figures~\ref{fig:results_constant_bw}--\ref{fig:results_random_ebg_scanned_dots}. These led to certain artifacts for IPD such as the duplicated rows of dots for isolated lines and the removal of dots around the linear isolated features of the plant example, which may or may not be desired. We thus first explore the use of alternative edge detection filters for controlling the interpolation as well as use different parameterizations for the dots, depending on from which of the two input distributions they arose.

\autoref{fig:filter_comparison} shows the result for the use of Canny, Difference of Gaussian (DoG), and Laplace of Gaussian (LoG), with the additional use of filters to address the issue of double edges for single lines. To better highlight the properties of the generated edge stipples---and to produce an alternative visual result---, in \autoref{fig:filter_comparison} we use a constant small dot size of 2 for the stipples generated for areas, while showing edge stipples larger with size 4. For this purpose we manually used a combination of Gaussian blur as well as contrast and brightness adjustments, before applying the respective edge filters. We show the resulting edge images in Figures~\ref{fig:filter_comparison_edges:e}--\ref{fig:filter_comparison_edges:p}. With this process we can see that we can remove the double edges even for the Canny filter as shown in Figures~\ref{fig:filter_comparison:a}--\subref{fig:filter_comparison:d}. The DoG filter is able to create single edges from single lines even without the use of other filters, but has the problem that its edges are somewhat noisier and may lead to zig-zags in places as can be seen in Figures~\ref{fig:filter_comparison:e}--\subref{fig:filter_comparison:h}. Finally, the results for LoG (Figures~\ref{fig:filter_comparison:i}--\subref{fig:filter_comparison:l}) show a slightly different portrayal of the edges, with some cleaner (\eg, Secord's \emph{Plant}) and some noisier (\eg, \emph{Headlight}) than DoG. Ultimately, it is thus upon the artist to select the most suitable edge detection process and to potentially apply additional filters to get the best results for a given visual goal, depending on the chosen input image.

We can also adjust the interpolation function to increase the distance around the borders such that area dots have less possibility of being drawn, as shown in \autoref{fig:effect_results:a}--\subref{fig:effect_results:d}. This leads to a white border around edges, which may be desired in some cases. Alternatively, we can also change the borders filter to use, for example, a DoG and remove the block that tries to align the strokes, producing that every black pixels is drawn as a dot as shown in \autoref{fig:effect_results:e}--\subref{fig:effect_results:h}. This effect emphasizes the linear edges particularly well through dark clusters of stipple dots, as can be seen particularly well in the full-size versions in \autoref{fig:effect_results_wide} in the additional material. If the process results in large, essentially black regions then a process to replace the stipples with solid polygons could be added \cite{Azami:2019:SRE}.

Finally, we can combine our process also with additional masks as shown in \autoref{fig:other_effect_results} to adjust the way the distance function is computed in \autoref{eq:final-interp}. For example, we can use the original edge image (for a chosen edge detection process) to generate offset lines similar to \citeauthor{Kim:2008:FGI}'s [\citeyear{Kim:2008:FGI}] hedcut stippling, and then use this offset line image as the input to compute the distance function in \autoref{eq:final-interp}. This process generates an effect as shown in \autoref{fig:other_effect_results:a}. We can also use dedicated masks to emphasize specific parts of the image as shown in \autoref{fig:other_effect_results:b}. 
Here, we masked the region of the lamp in the image in white, leave the rest black, and then use this image as the input to compute the distance $\Delta$. By then adjusting $b$ such that edge stipples are always drawn we achieve the emphasis effect in \autoref{fig:other_effect_results:b} because, in the white region of the mask (for the area distribution), we define $\Delta$ to be 0 and thus ensure that the region's area stipples are always shown.
Finally, with the same approach we can also use image-independent masks such as wavy lines to generate the background effects as we show in \autoref{fig:other_effect_results:c}.\vspace{-1ex}

\section{Conclusion}
\label{sec:conclusion}

In summary, we described a method to smoothly interpolate between two point distributions. The fundamental contribution of our approach is that we do not interpolate the resulting images, but the underlying dot distributions by means of Probability Density Functions (PDFs). Our approach is general: it can deal with any input distribution by relying on the discrete version of the PDFs such that it can also deal with input distributions that are not expressed as analytic PDFs but as discrete sets of dots. Moreover, the discrete approach also allows us to take advantage of GPU-based implementations and raster image filters.

This new distribution interpolation allows us to combine traditional stippling techniques for different regions in an input image---those that excel in shaded areas with those that focus on structure preservation. We thus can now generate stippling images that express the best of these two worlds and smoothly interpolate both point distributions---without the artifacts that would arise from a purely pixel-based interpolation of final stipple image results. For this interpolation between the distributions we made use of a rasterized distance field, based on an edge image that expresses the features that one wants to preserve. Our overall process provides a lot of flexibility in that individual stages can be adjusted, to create additional visual effects and feature emphasis.

The main limitation of our method is its inherent use of discrete raster representations during the interpolation process. For example, this approach can produce ``stair'' effects, in particular, for linear structures. Typical ways to address these problems is to increase the resolution of the image or to apply anti-aliasing, the former of which we can also use in our case. The discrete approach also leads to another limitation, which is that our results are probabilistic---meaning that we only reproduce the input distribution up to the resolution of the interpolation raster of the discrete probability function. With a sufficiently fine interpolation raster, however, we can address this limitation because the resolution of the interpolation raster is largely independent from that of the input images, provided that we can compute respective distances from edge features.

In the future we thus want to focus on ways to reduce rasterization artifacts and to better control the interpolation output. For instance, we plan to investigate the mentioned higher-resolution interpolation processing (\eg, using vectorization of the edge image), explore better edge detection for isolated linear features, and ultimately come up with a fully analytic way of interpolation. Independent of these efforts, we also want to explore ways in which our tool can best be given to illustrators, to provide them with detailed control yet without the need to deal with complex parameter settings.

\begin{acks}
We thank the author of WVS for his tool and the authors of SPS for their comparison images. This work was partially funded by the Spanish Ministry of Economy and Competitiveness with FEDER funds through project TIN\,2017-85259-R.
\end{acks}

\bibliographystyle{ACM-Reference-Format}
\bibliography{survey_stippling,survey_stippling_extra}


\cleardoublepage
\appendix
\section{Additional material}

Below we provide additional material to support our discussion.

\subsection{Original images}
Figures~\ref{fig:originals}--\ref{fig:originals-grayscale} show the input to our algorithm comparison in \autoref{sec:results}. Figures~\ref{fig:results_secord_constant}--\ref{fig:results_mould_constant} show the original results produced by the WVS and SPS techniques as they were produced by the original tools.

\setlength{\picturewidth}{.245\textwidth}
\setlength{\wd0}{\picturewidth}
\setlength{\ht0}{\picturewidth}
\begin{figure}[!b]
	\begin{minipage}{\textwidth}
	\centering
	\setlength{\subfigcapskip}{-3.5ex}%
	\textcolor{white}{\subfigure[\hspace{0.22\linewidth}]{\label{fig:originals:a}\includegraphics[width=\picturewidth]{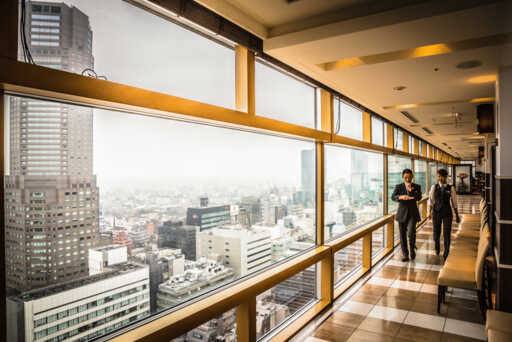}}}\hfill
	\textcolor{white}{\subfigure[\hspace{0.22\linewidth}]{\label{fig:originals:b}\includegraphics[width=\picturewidth]{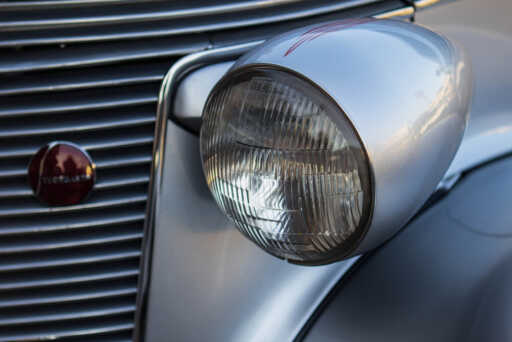}}}\hfill
	\textcolor{white}{\subfigure[\hspace{0.22\linewidth}]{\label{fig:originals:c}\includegraphics[width=\picturewidth]{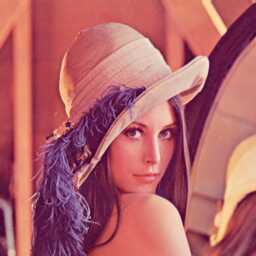}}}\hfill
	\subfigure[\hspace{\linewidth}]{\label{fig:originals:d}\includegraphics[width=\picturewidth]{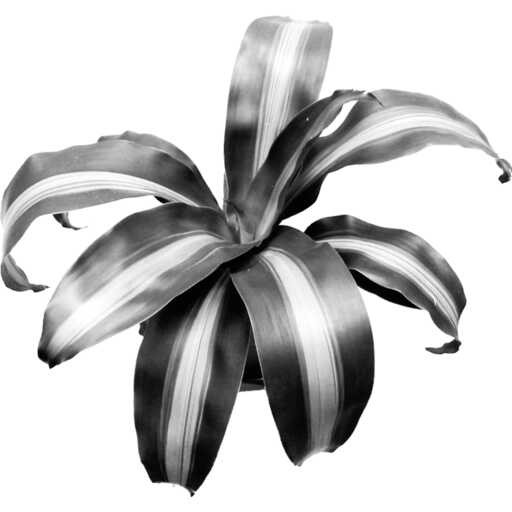}}\vspace{-1ex}%
	\caption{Original versions of the images we used as the input for comparing the algorithms.
	Images~\subref{fig:originals:a} and~\subref{fig:originals:b} are part of \citeauthor{Mould:2017:DAB}'s [\citeyear{Mould:2017:DAB}] benchmark set for non-photorealistic rendering.
	Image~\subref{fig:originals:a} \href{http://www.flickr.com/photo.gne?id=13897692457}{(Flickr image 13897692457)} is by Richard Schneider, \href{https://creativecommons.org/licenses/by-nc/2.0/}{\ccLogo\,\ccAttribution\,\ccNonCommercial~CC BY-NC 2.0}.
	Image~\subref{fig:originals:b} \href{http://www.flickr.com/photo.gne?id=16453416352}{(Flickr image 16453416352)} is by Photos By Clark, \href{https://creativecommons.org/licenses/by-nc/2.0/}{\ccLogo\,\ccAttribution\,\ccNonCommercial~CC BY-NC 2.0}.
	Image~\subref{fig:originals:c} shows Lena (Soderberg)---a frequently used benchmark image from image processing, \href{https://web.archive.org/web/20160301041549/http://sipi.usc.edu/database/database.php?volume=misc&image=12}{formerly part of the USC-SIPI Image Database}, used under the fair-use clause.
	Image~\subref{fig:originals:d} is \textcopyright~Adrien Secord, permission will be requested.}
	\label{fig:originals}
	\end{minipage}
\end{figure}


\begin{figure}[!b]
	\begin{minipage}{\textwidth}
	\centering
	\setlength{\subfigcapskip}{-3.5ex}%
	\textcolor{white}{\subfigure[\hspace{0.22\linewidth}]{\includegraphics[width=\picturewidth]{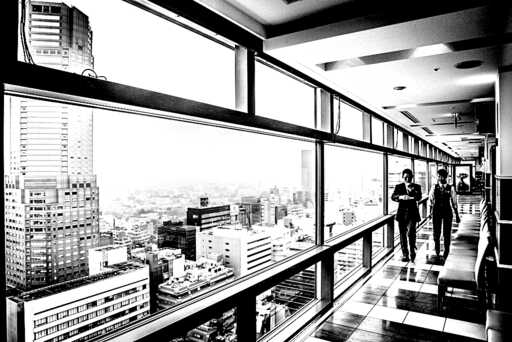}}}\hfill
	\textcolor{white}{\subfigure[\hspace{0.22\linewidth}]{\includegraphics[width=\picturewidth]{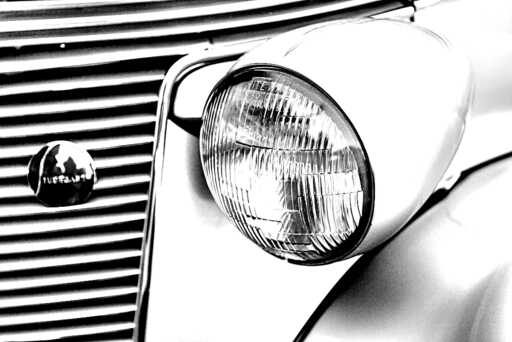}}}\hfill
	\subfigure[\hspace{0.22\linewidth}]{\includegraphics[width=\picturewidth]{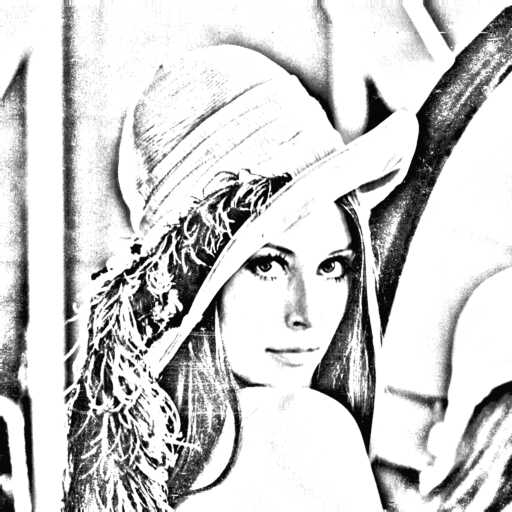}}\hfill
	\subfigure[\hspace{\linewidth}]{\includegraphics[width=\picturewidth]{plant_1024}}\vspace{-1ex}%
	\caption{Contrast-enhanced grayscale versions of \autoref{fig:originals} (width of 1024 pixels) that we used as the input to our algorithm comparison in \autoref{sec:results}.}
	\label{fig:originals-grayscale}
	\end{minipage}
\end{figure}

\setlength{\picturewidth}{84mm}\newcommand{\imageproportion}{We show each image at \nicefrac{2}{5} of its intended target size of A5 (\ie, with base width of 84\,mm instead of 210\,mm, to conserve space).}
\begin{figure*}[!p]
	\centering
	\subfigure[]{\label{fig:results_secord_constant:a}\includegraphics[width=\picturewidth]{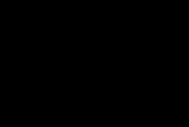}}\hfill
	\subfigure[]{\label{fig:results_secord_constant:b}\includegraphics[width=\picturewidth]{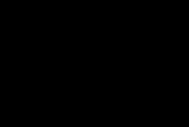}}\\[2ex]
	\subfigure[]{\label{fig:results_secord_constant:c}\includegraphics[width=\picturewidth]{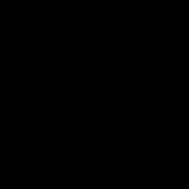}}\hspace{10mm}%
	\subfigure[]{\label{fig:results_secord_constant:d}\includegraphics[width=\picturewidth]{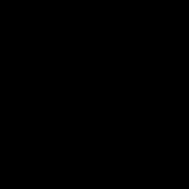}}
	\caption{Original results of \citeauthor{Secord:2002:WVS}'s [\citeyear{Secord:2002:WVS}] WVS method with dots of constant size. \imageproportion}
	\label{fig:results_secord_constant}
\end{figure*}

\begin{figure*}[!p]
	\centering
	\subfigure[]{\label{fig:results_secord_modulation:a}\includegraphics[width=\picturewidth]{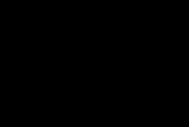}}\hfill
	\subfigure[]{\label{fig:results_secord_modulation:b}\includegraphics[width=\picturewidth]{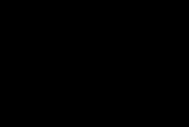}}\\[2ex]
	\subfigure[]{\label{fig:results_secord_modulation:c}\includegraphics[width=\picturewidth]{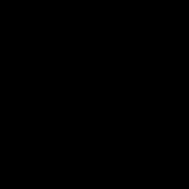}}\hfill
	\subfigure[]{\label{fig:results_secord_modulation:d}\includegraphics[width=\picturewidth]{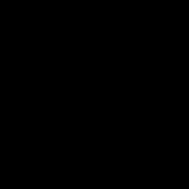}}	
	\caption{Original results of \citeauthor{Secord:2002:WVS}'s [\citeyear{Secord:2002:WVS}] WVS method with its own modulation approach \cite{Secord:2002:RMP}. \imageproportion\ The one unusually large stipple per image is an artifact (bug) of \citeauthor{Secord:2002:WVS}'s own original program which we used to create these images, we decided not to remove or adjust them---even if this would be possible.}
	\label{fig:results_secord_modulation}
\end{figure*}

\begin{figure*}[!p]
	\centering
	\subfigure[]{\label{fig:results_mould_constant:e}\includegraphics[width=\picturewidth]{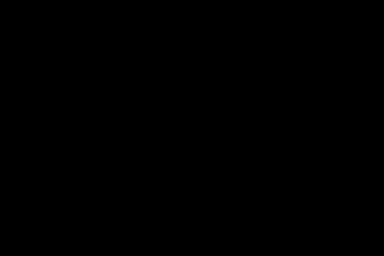}}\hfill
	\subfigure[]{\label{fig:results_mould_constant:f}\includegraphics[width=\picturewidth]{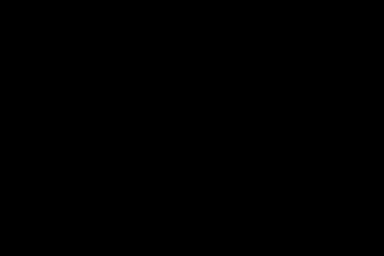}}\\[2ex]
	\subfigure[]{\label{fig:results_mould_constant:g}\includegraphics[width=\picturewidth]{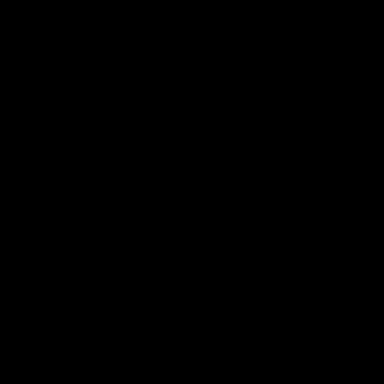}}\hfill
	\subfigure[]{\label{fig:results_mould_constant:h}\includegraphics[width=\picturewidth]{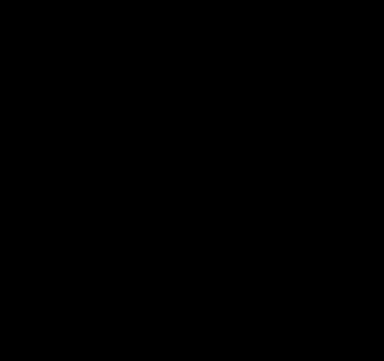}}
	\caption{Original results of \citeauthor{Li:2011:SPS}'s [\citeyear{Li:2011:SPS,Li:2017:PSS}] SPS method with dots of constant size. \imageproportion}
	\label{fig:results_mould_constant}
\end{figure*}

\subsection{Large, complete results}

\begin{figure*}[!p]
	\centering
	\subfigure[]{\label{fig:results_constant_bw_WVS:a}\includegraphics[width=\picturewidth]{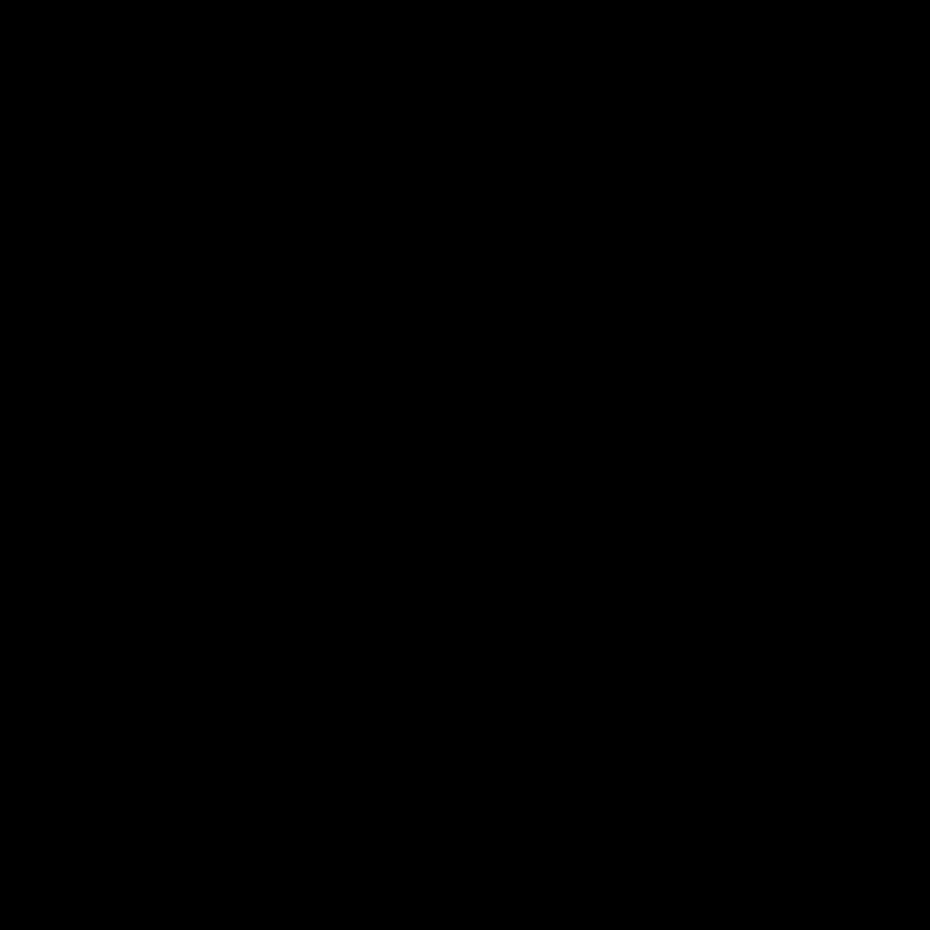}}\hfill
	\subfigure[]{\label{fig:results_constant_bw_WVS:b}\includegraphics[width=\picturewidth]{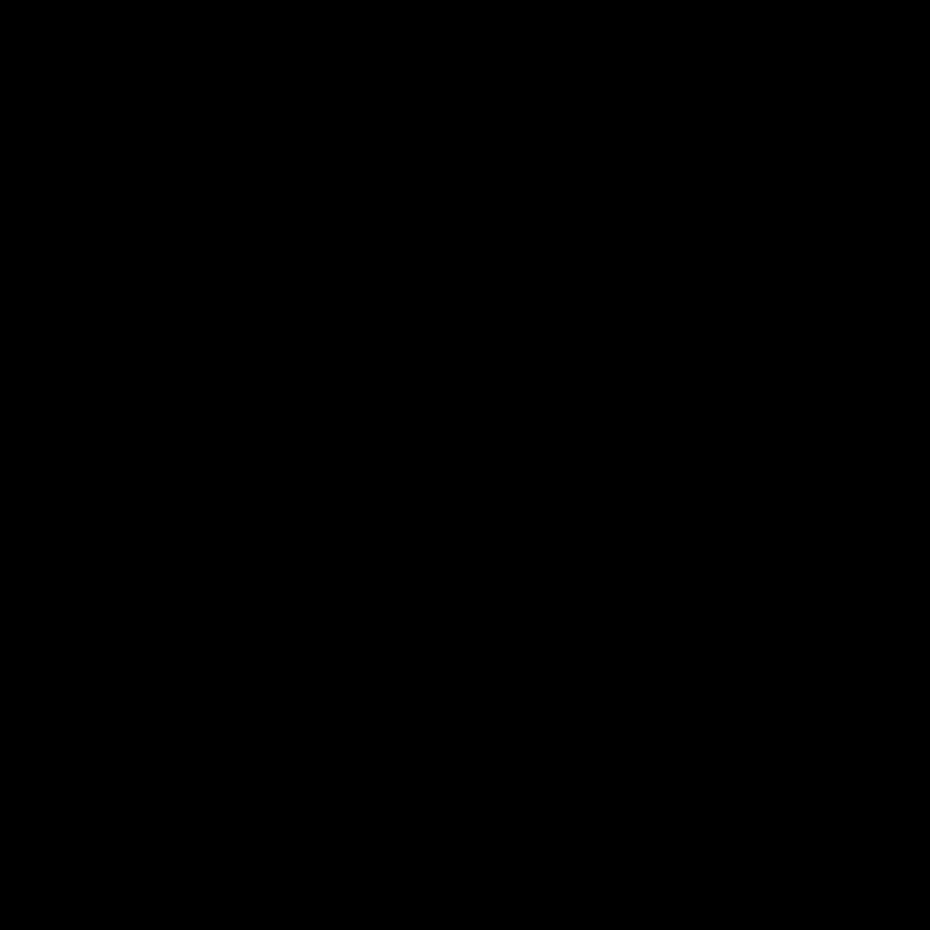}}\\[2ex]
	\subfigure[]{\label{fig:results_constant_bw_WVS:c}\includegraphics[width=\picturewidth]{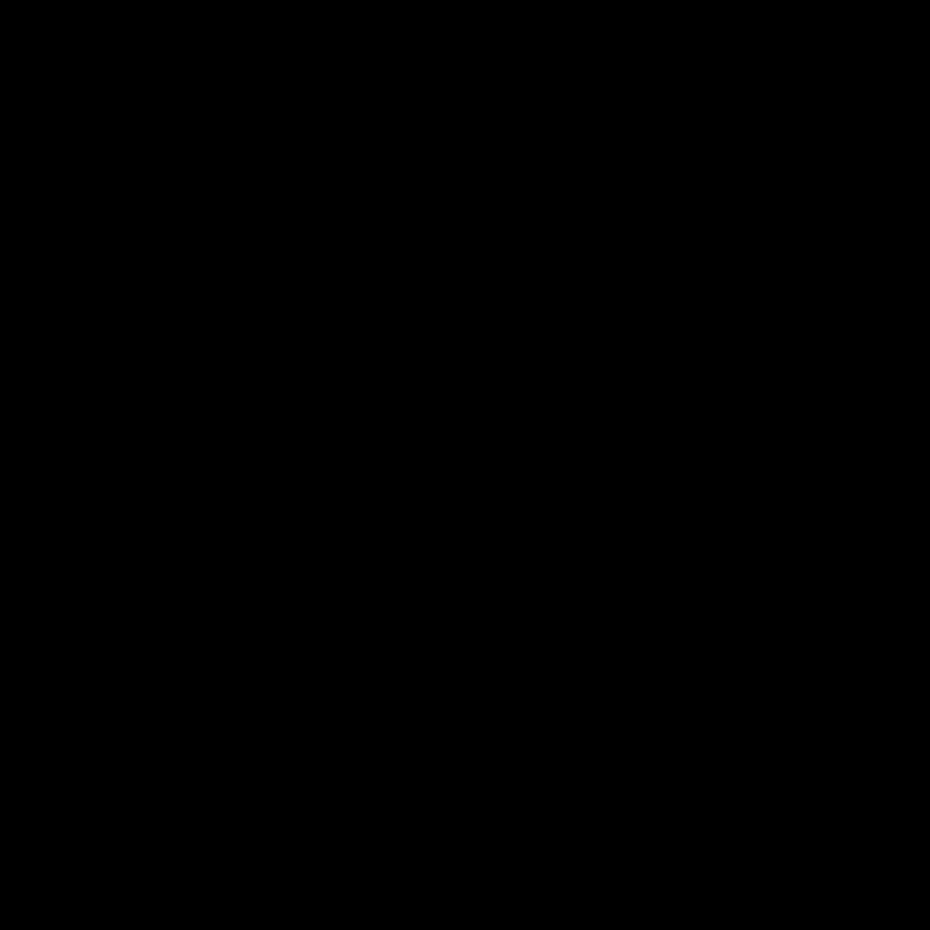}}\hfill
	\subfigure[]{\label{fig:results_constant_bw_WVS:d}\includegraphics[width=\picturewidth]{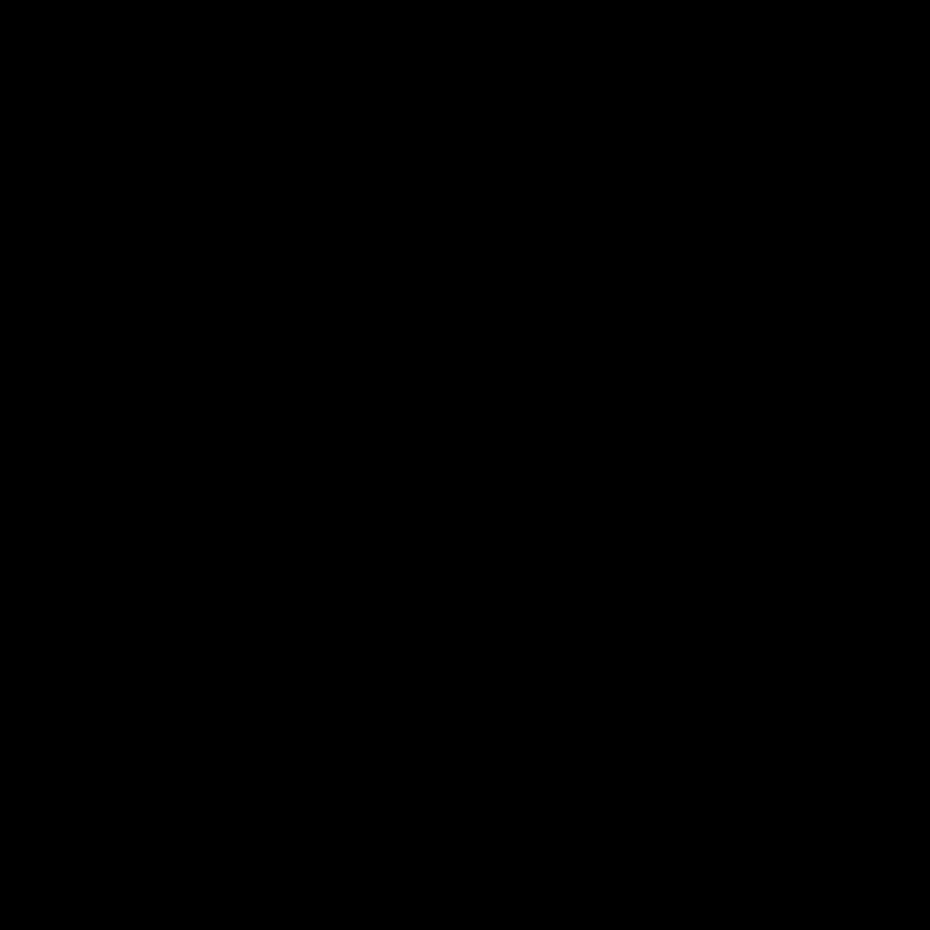}}	
	\caption{Large and complete version of the detail sections in \autoref{fig:results_constant_bw:a}--\subref{fig:results_constant_bw:d}:  Constant size dots, radius 0.5, B\&W for WVS. \imageproportion}
	\label{fig:results_constant_bw_WVS}
\end{figure*}

\begin{figure*}[!p]
	\centering
	\subfigure[]{\label{fig:results_constant_bw_SPS:a}\includegraphics[width=\picturewidth]{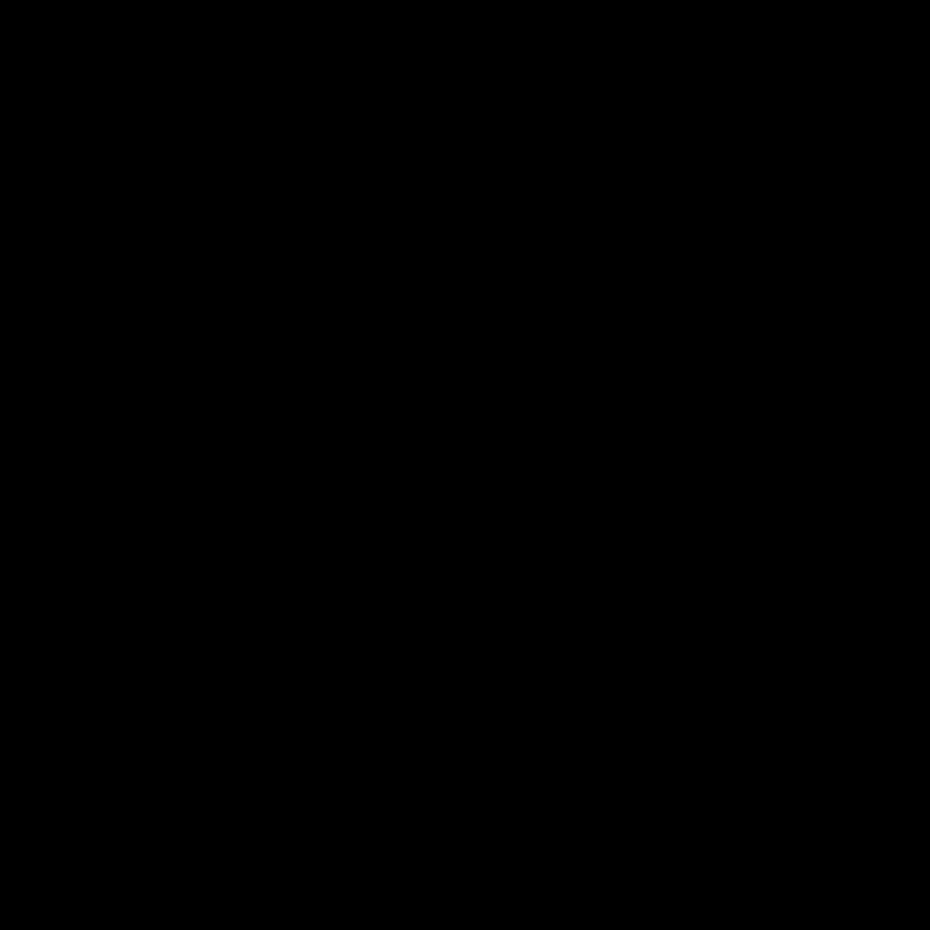}}\hfill
	\subfigure[]{\label{fig:results_constant_bw_SPS:b}\includegraphics[width=\picturewidth]{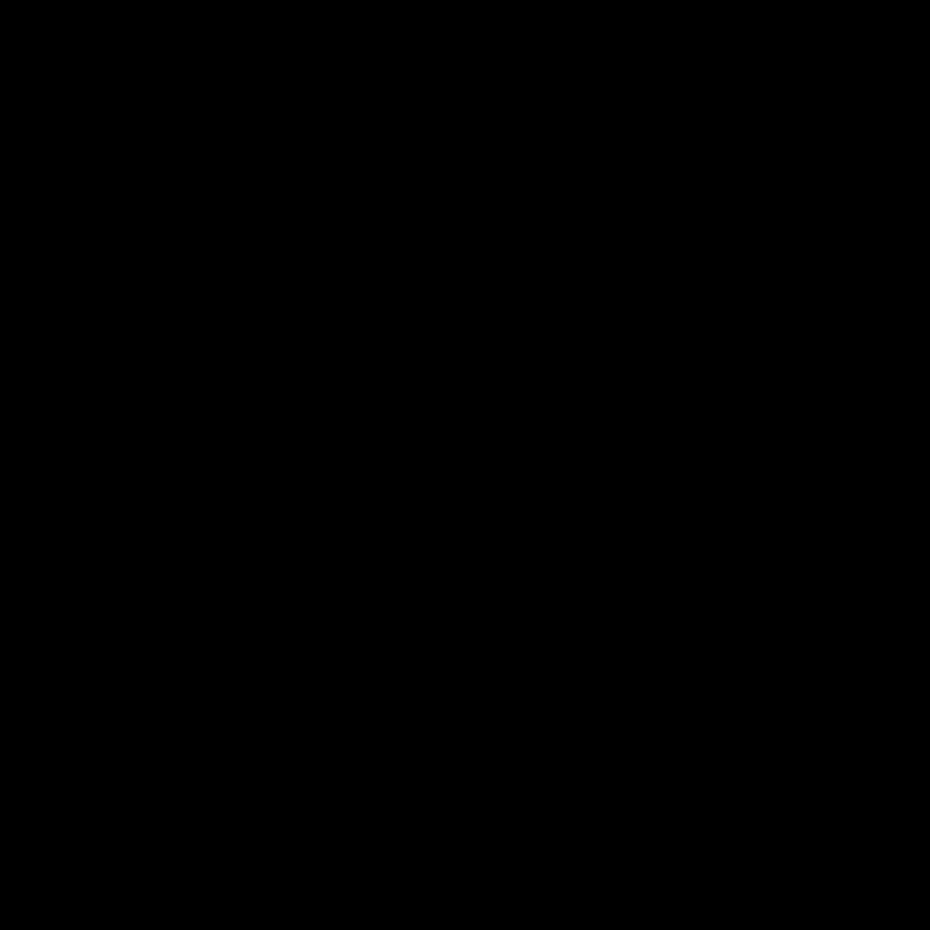}}\\[2ex]
	\subfigure[]{\label{fig:results_constant_bw_SPS:c}\includegraphics[width=\picturewidth]{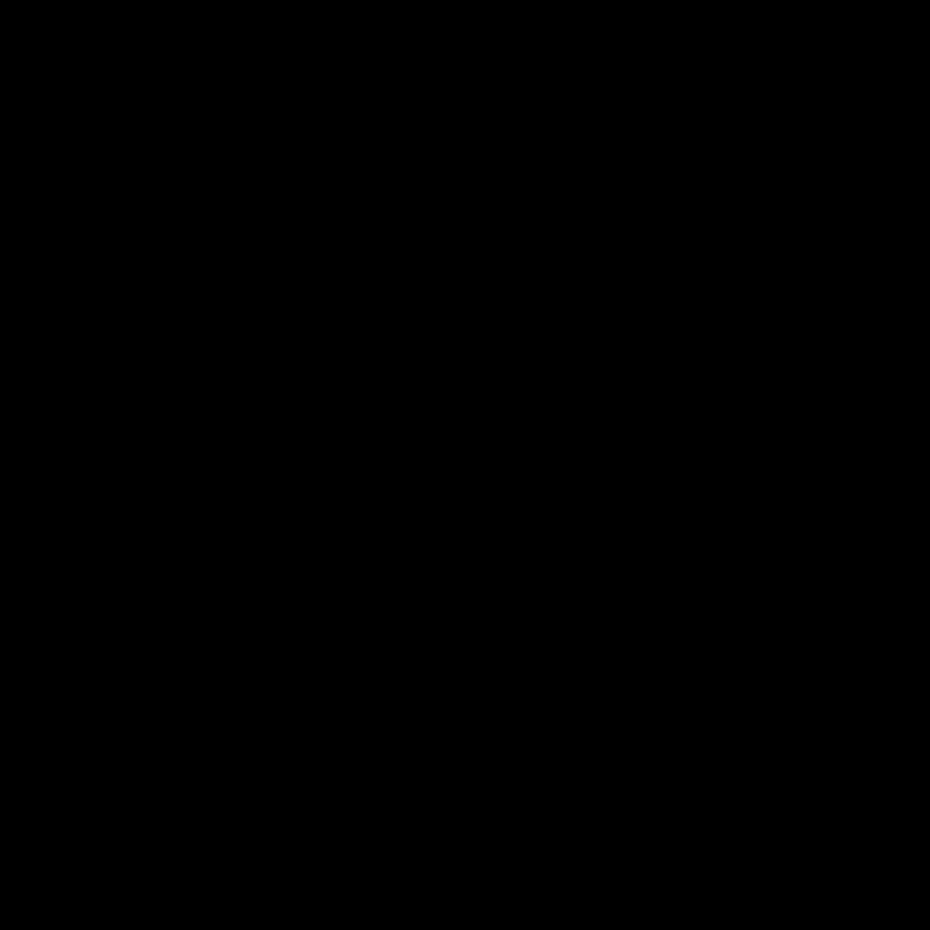}}\hfill
	\subfigure[]{\label{fig:results_constant_bw_SPS:d}\includegraphics[width=\picturewidth]{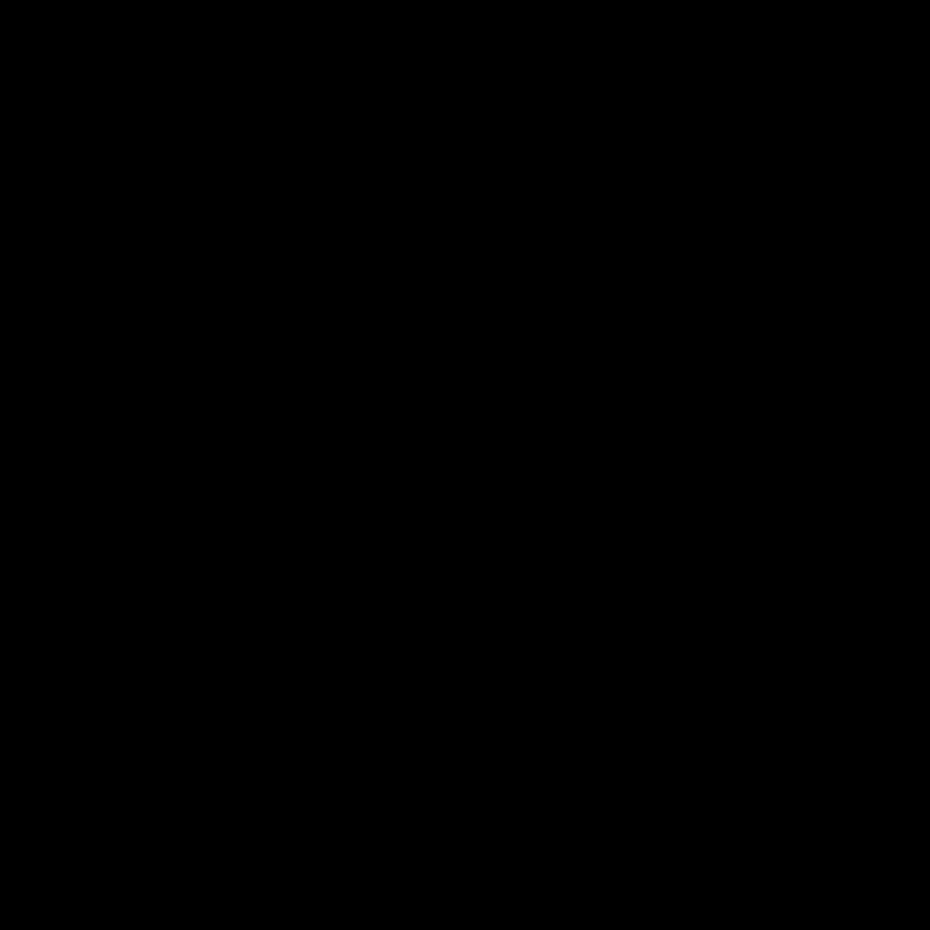}}	
	\caption{Large and complete version of the detail sections in \autoref{fig:results_constant_bw:e}--\subref{fig:results_constant_bw:h}:  Constant size dots, radius 0.5, B\&W for SPS. \imageproportion}
	\label{fig:results_constant_bw_SPS}
\end{figure*}

\begin{figure*}[!p]
	\centering
	\subfigure[]{\label{fig:results_constant_bw_EBG:a}\includegraphics[width=\picturewidth]{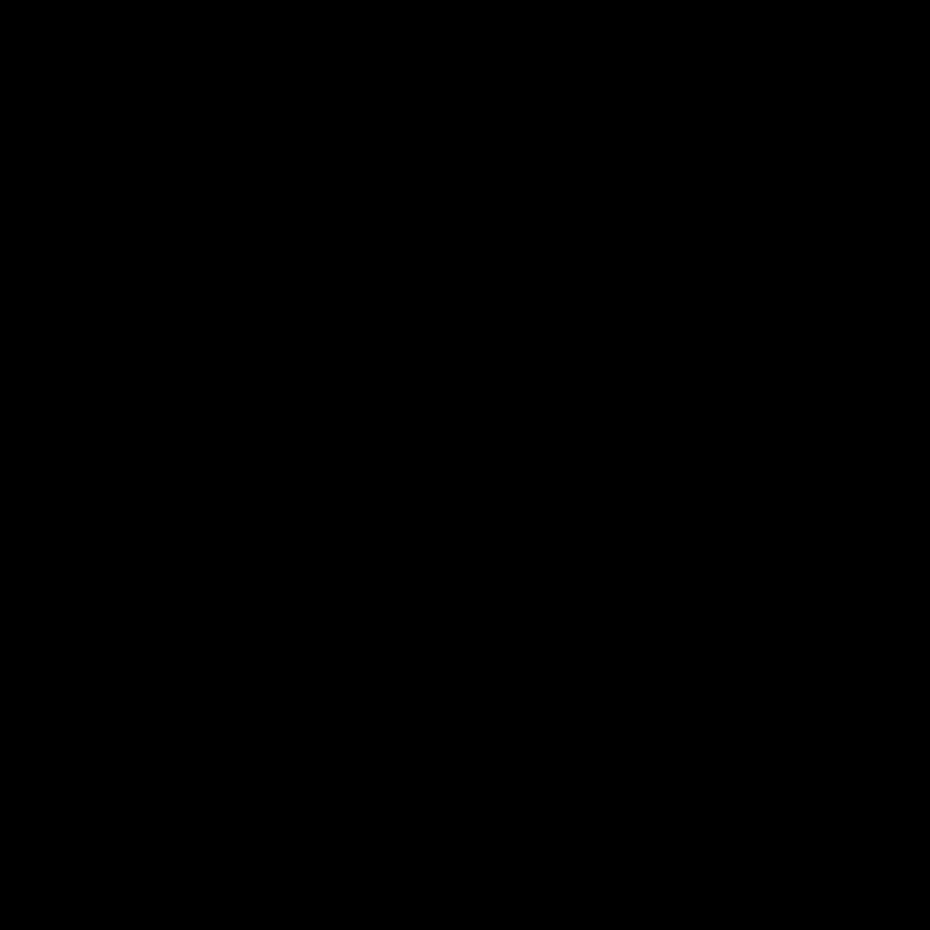}}\hfill
	\subfigure[]{\label{fig:results_constant_bw_EBG:b}\includegraphics[width=\picturewidth]{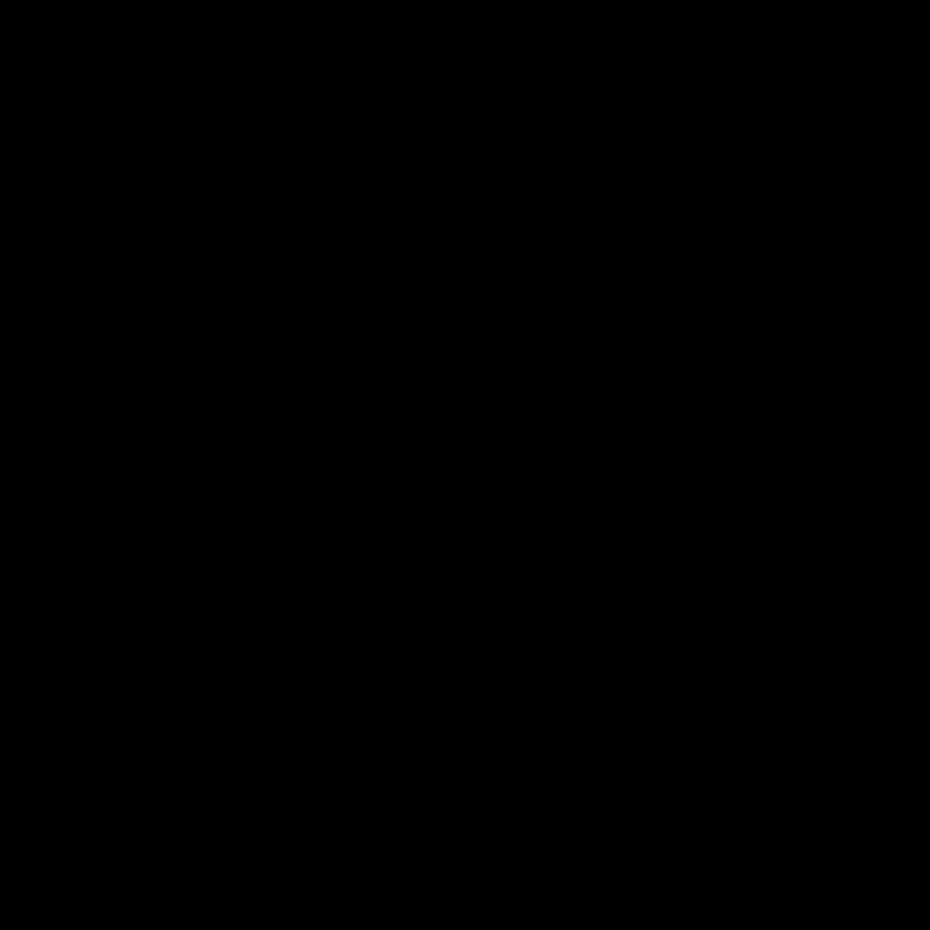}}\\[2ex]
	\subfigure[]{\label{fig:results_constant_bw_EBG:c}\includegraphics[width=\picturewidth]{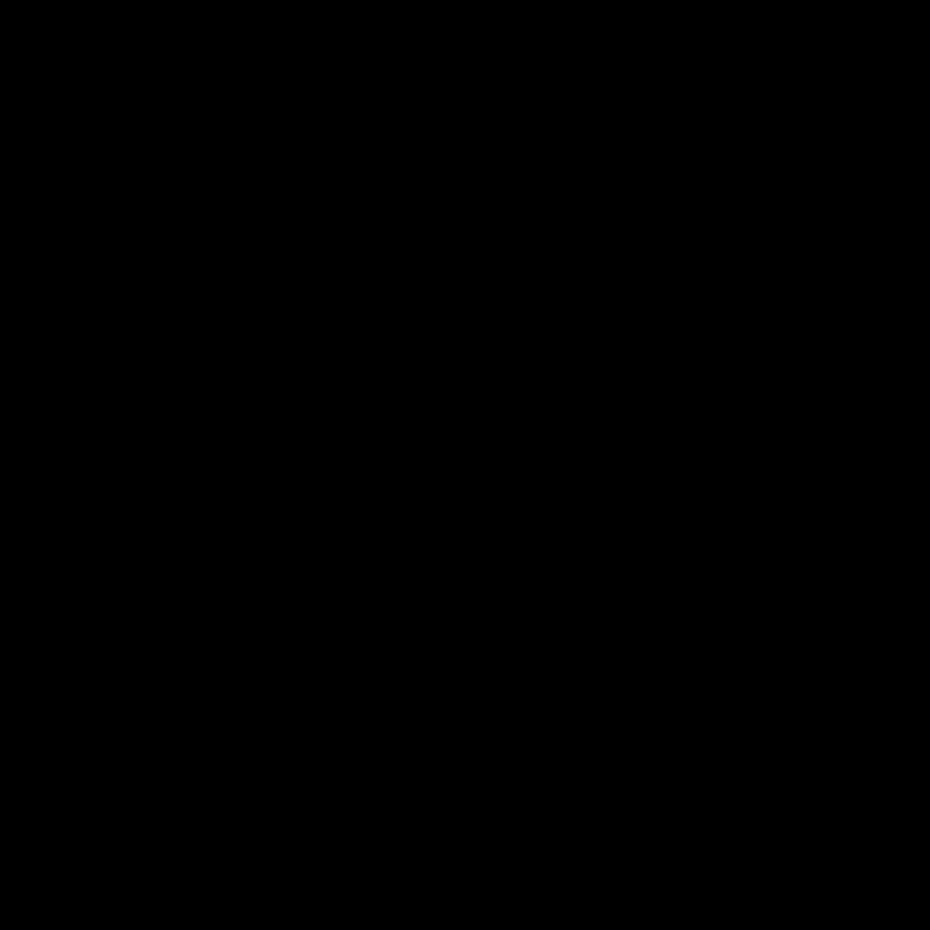}}\hfill
	\subfigure[]{\label{fig:results_constant_bw_EBG:d}\includegraphics[width=\picturewidth]{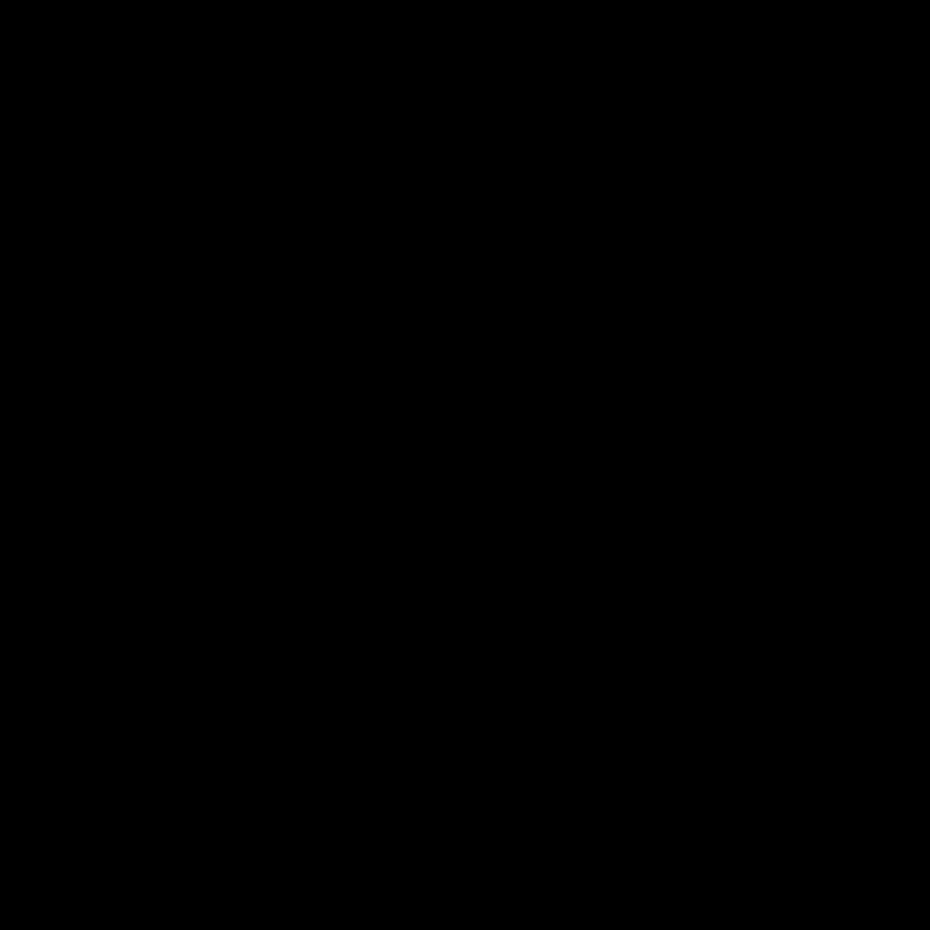}}	
	\caption{Large and complete version of the detail sections in \autoref{fig:results_constant_bw:i}--\subref{fig:results_constant_bw:l}:  Constant size dots, radius 0.5, B\&W for EBG. \imageproportion}
	\label{fig:results_constant_bw_EBG}
\end{figure*}

\begin{figure*}[!p]
	\centering
	\subfigure[]{\label{fig:results_constant_bw_IPD:a}\includegraphics[width=\picturewidth]{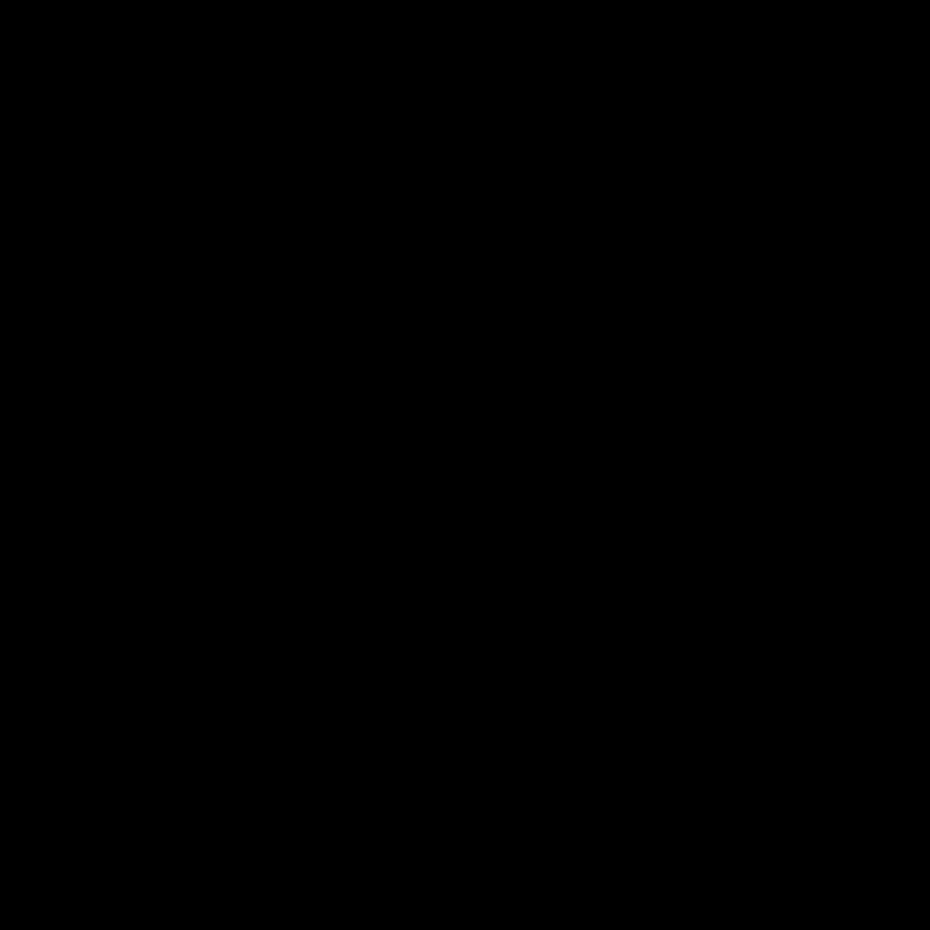}}\hfill
	\subfigure[]{\label{fig:results_constant_bw_IPD:b}\includegraphics[width=\picturewidth]{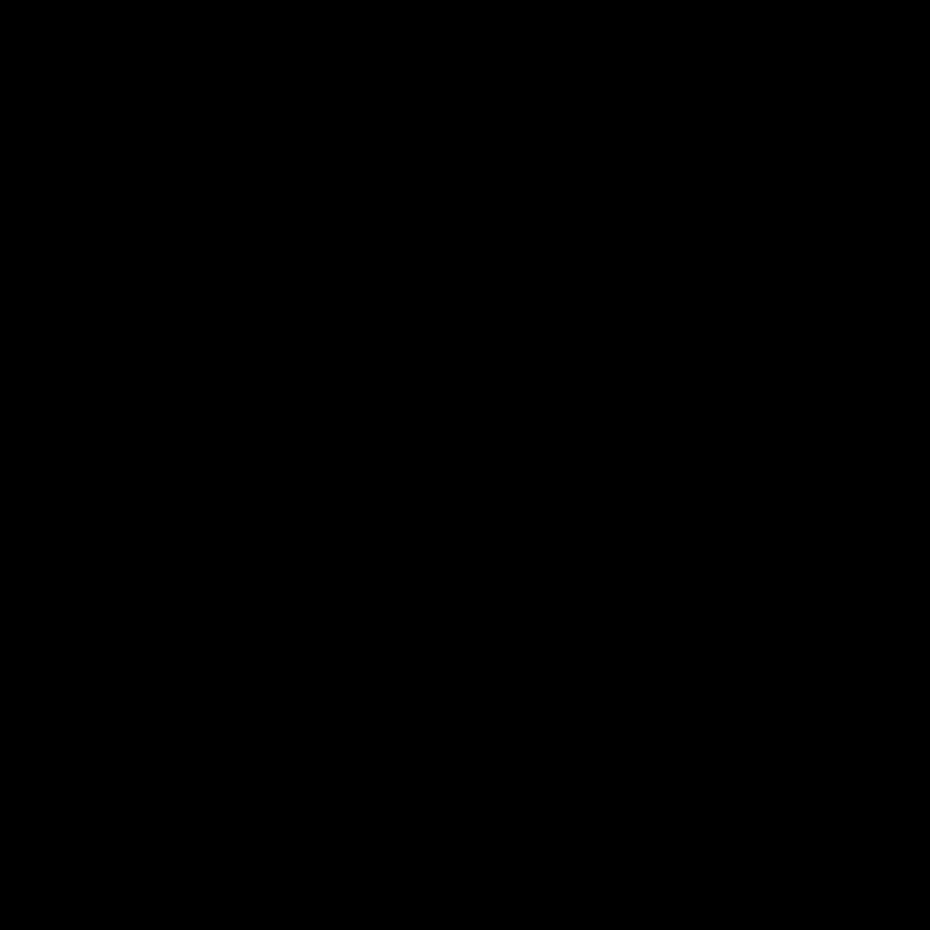}}\\[2ex]
	\subfigure[]{\label{fig:results_constant_bw_IPD:c}\includegraphics[width=\picturewidth]{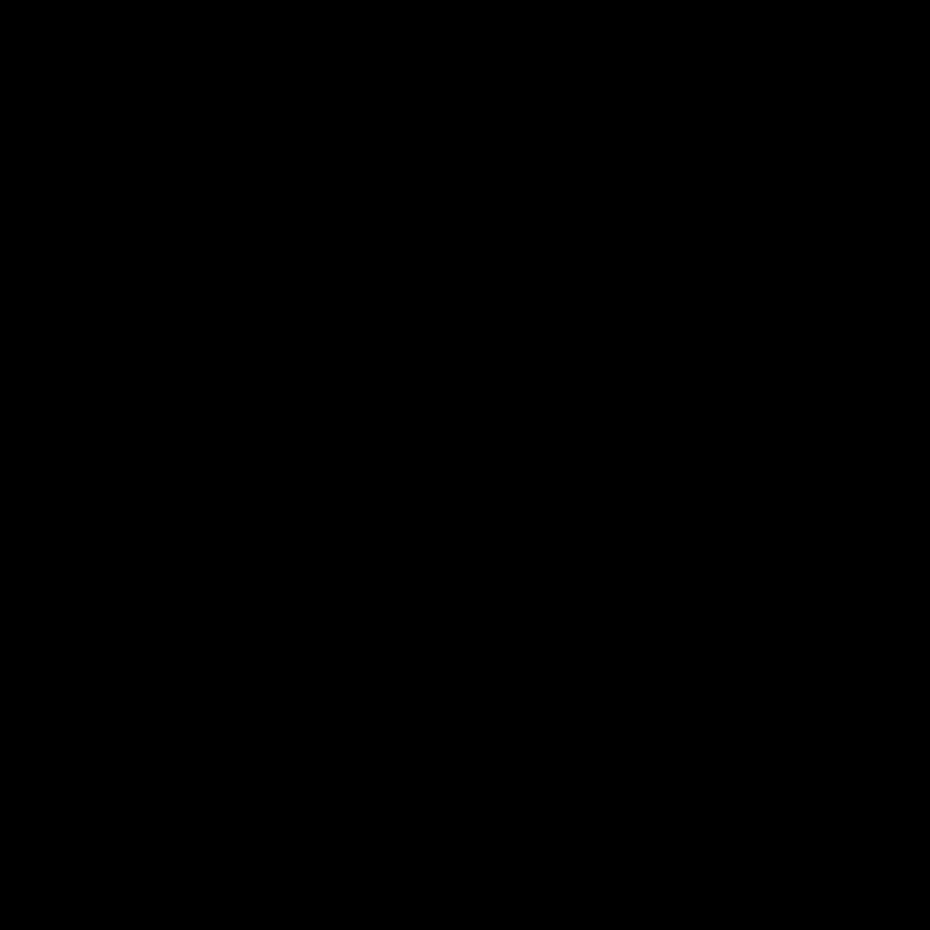}}\hfill
	\subfigure[]{\label{fig:results_constant_bw_IPD:d}\includegraphics[width=\picturewidth]{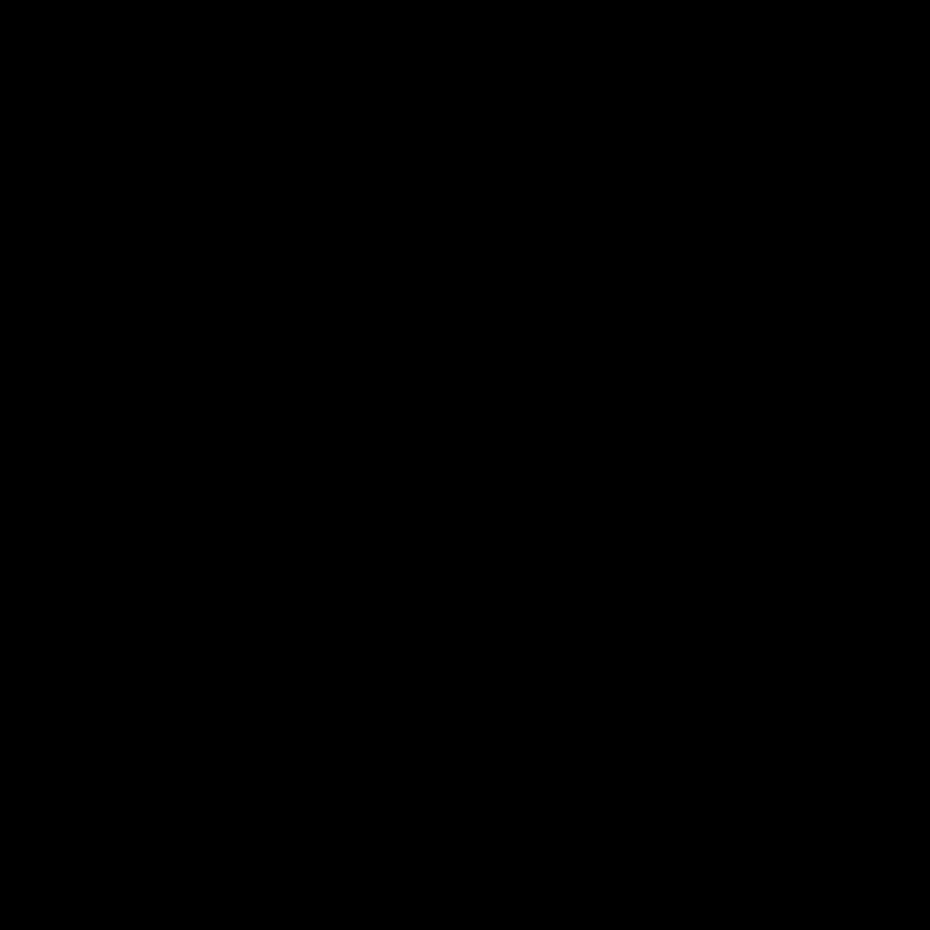}}	
	\caption{Large and complete version of the detail sections in \autoref{fig:results_constant_bw:m}--\subref{fig:results_constant_bw:p}:  Constant size dots, radius 0.5, B\&W for IPD. \imageproportion}
	\label{fig:results_constant_bw_IPD}
\end{figure*}

\begin{figure*}[!p]
	\centering
	\subfigure[]{\label{fig:results_modulated_bw_WVS:a}\includegraphics[width=\picturewidth]{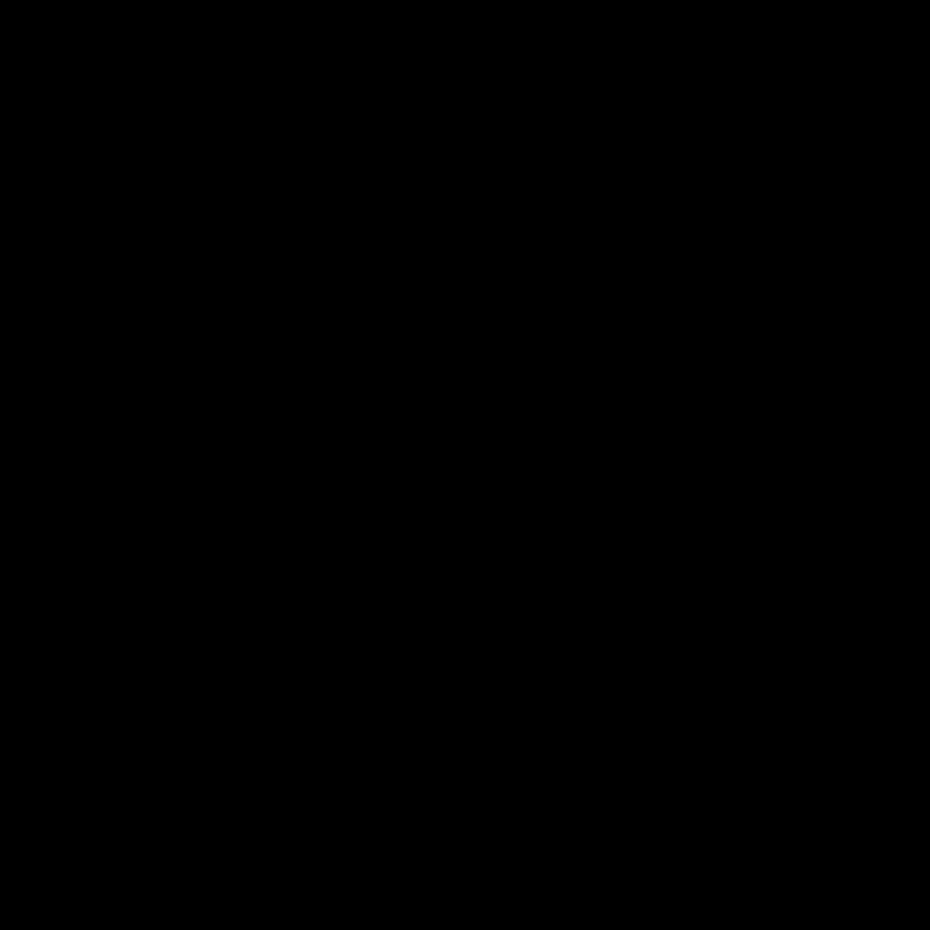}}\hfill
	\subfigure[]{\label{fig:results_modulated_bw_WVS:b}\includegraphics[width=\picturewidth]{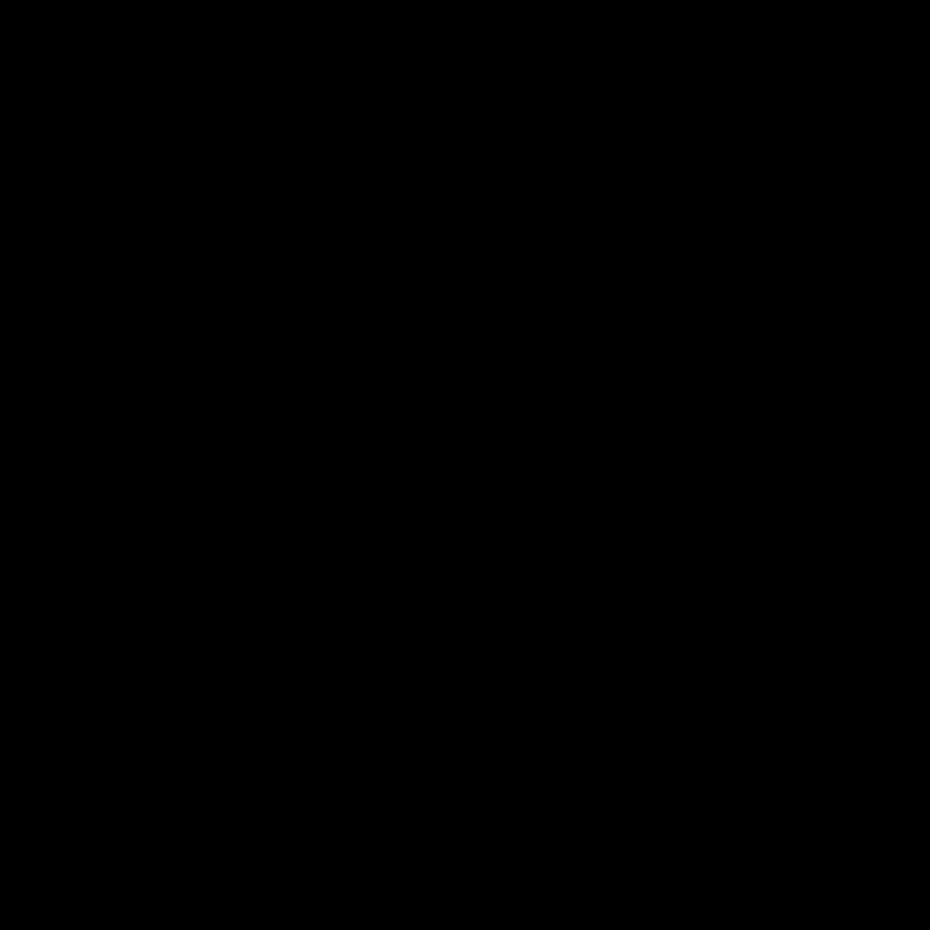}}\\[2ex]
	\subfigure[]{\label{fig:results_modulated_bw_WVS:c}\includegraphics[width=\picturewidth]{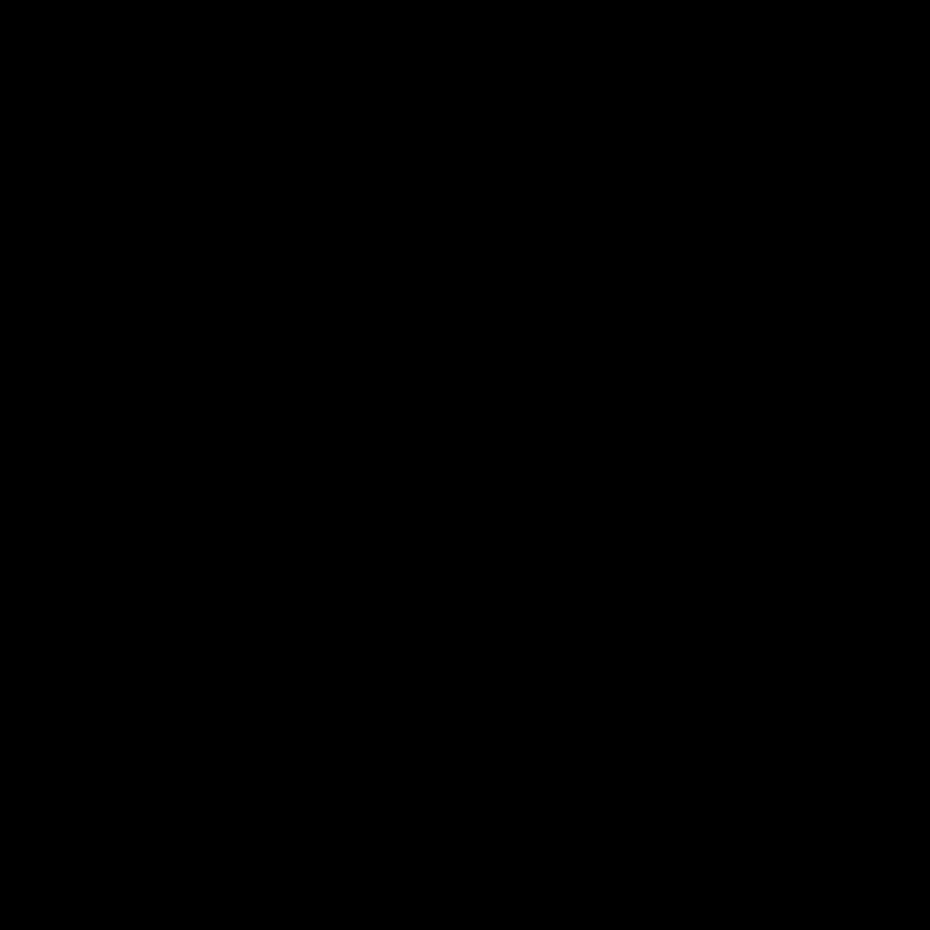}}\hfill
	\subfigure[]{\label{fig:results_modulated_bw_WVS:d}\includegraphics[width=\picturewidth]{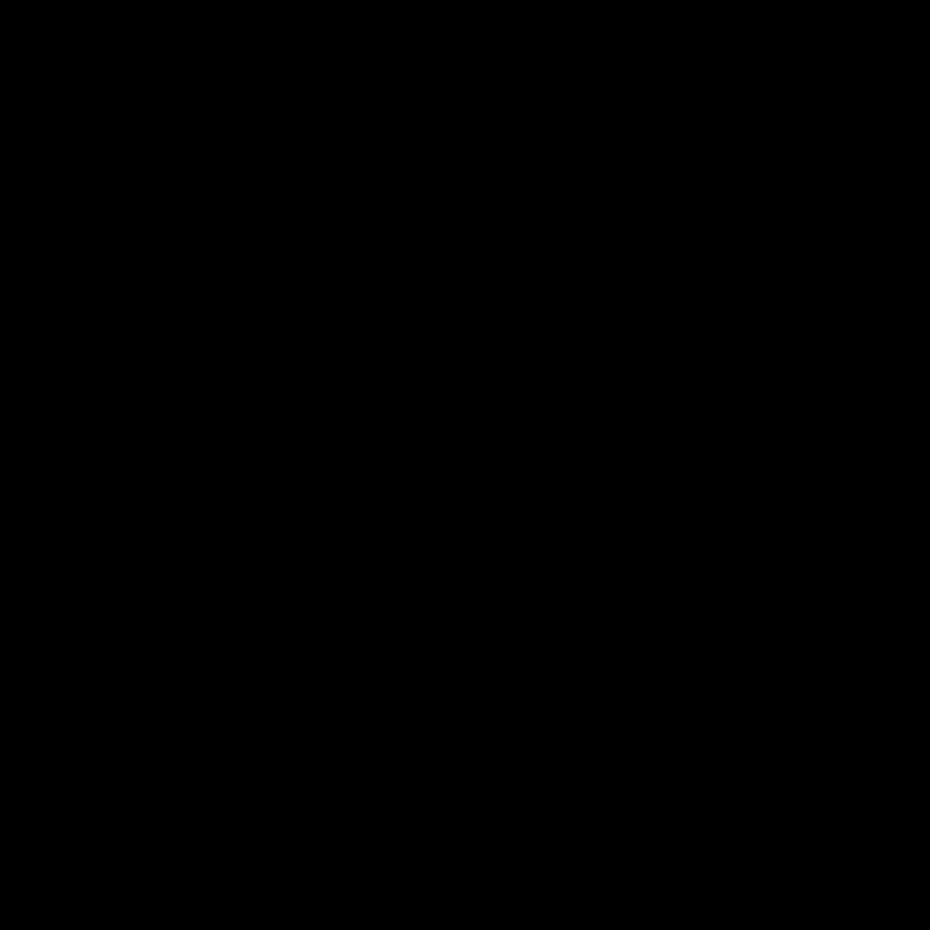}}	
	\caption{Large and complete version of the detail sections in \autoref{fig:results_modulated_ebg_scanned_dots:a}--\subref{fig:results_modulated_ebg_scanned_dots:d}:  Modulated size dots between 2 and 4, B\&W for WVS. \imageproportion}
	\label{fig:results_modulated_bw_WVS}
\end{figure*}

\begin{figure*}[!p]
	\centering
	\subfigure[]{\label{fig:results_modulated_bw_SPS:a}\includegraphics[width=\picturewidth]{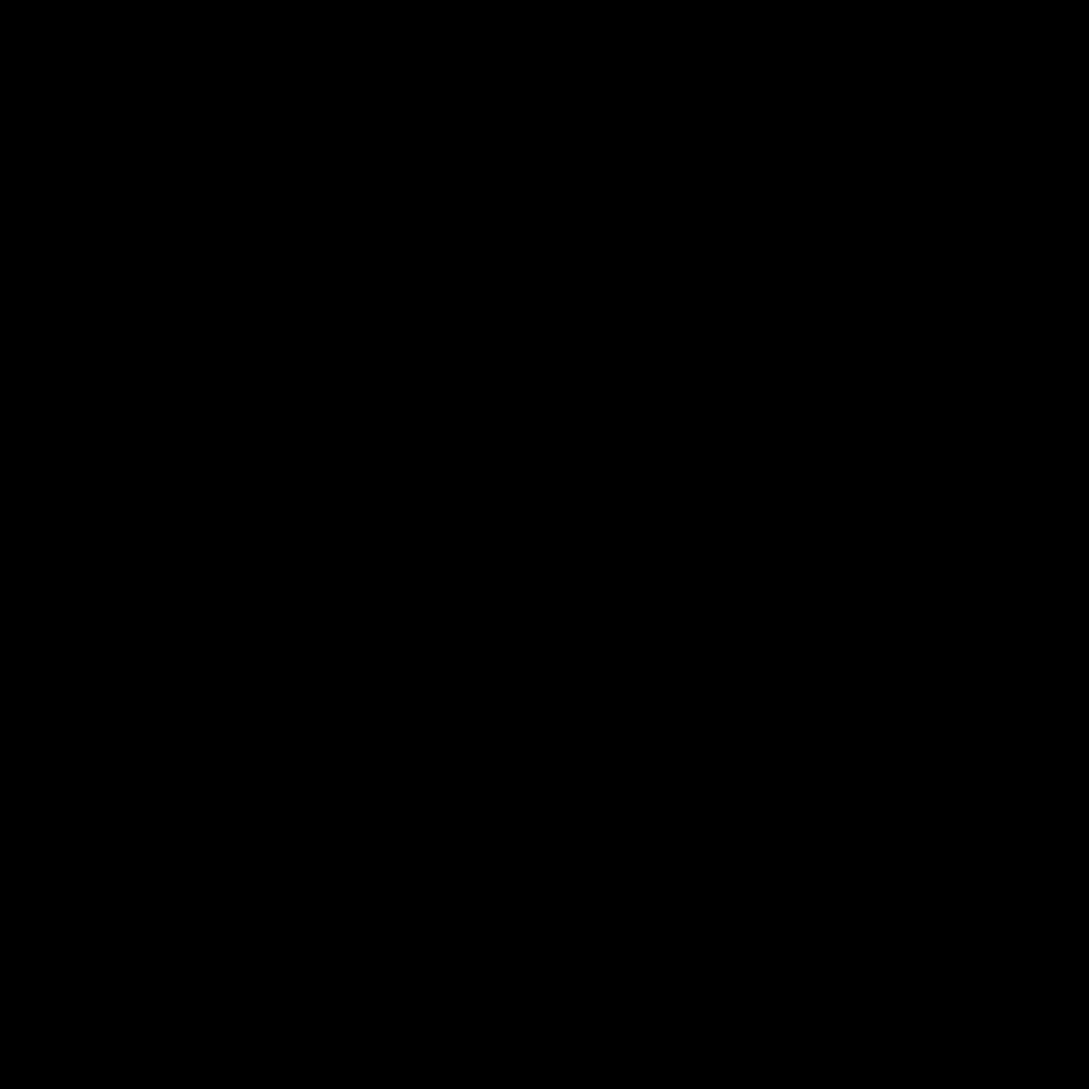}}\hfill
	\subfigure[]{\label{fig:results_modulated_bw_SPS:b}\includegraphics[width=\picturewidth]{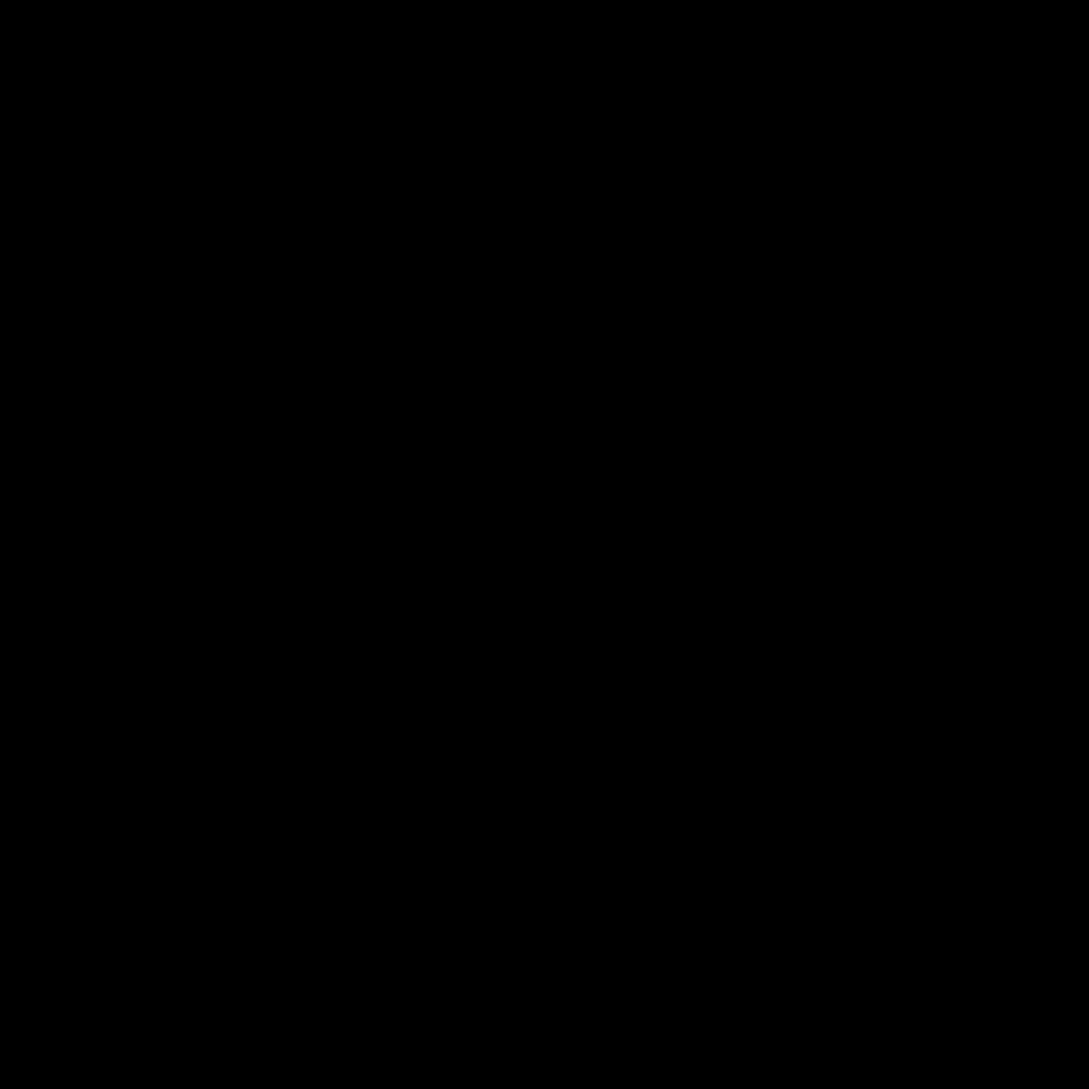}}\\[2ex]
	\subfigure[]{\label{fig:results_modulated_bw_SPS:c}\includegraphics[width=\picturewidth]{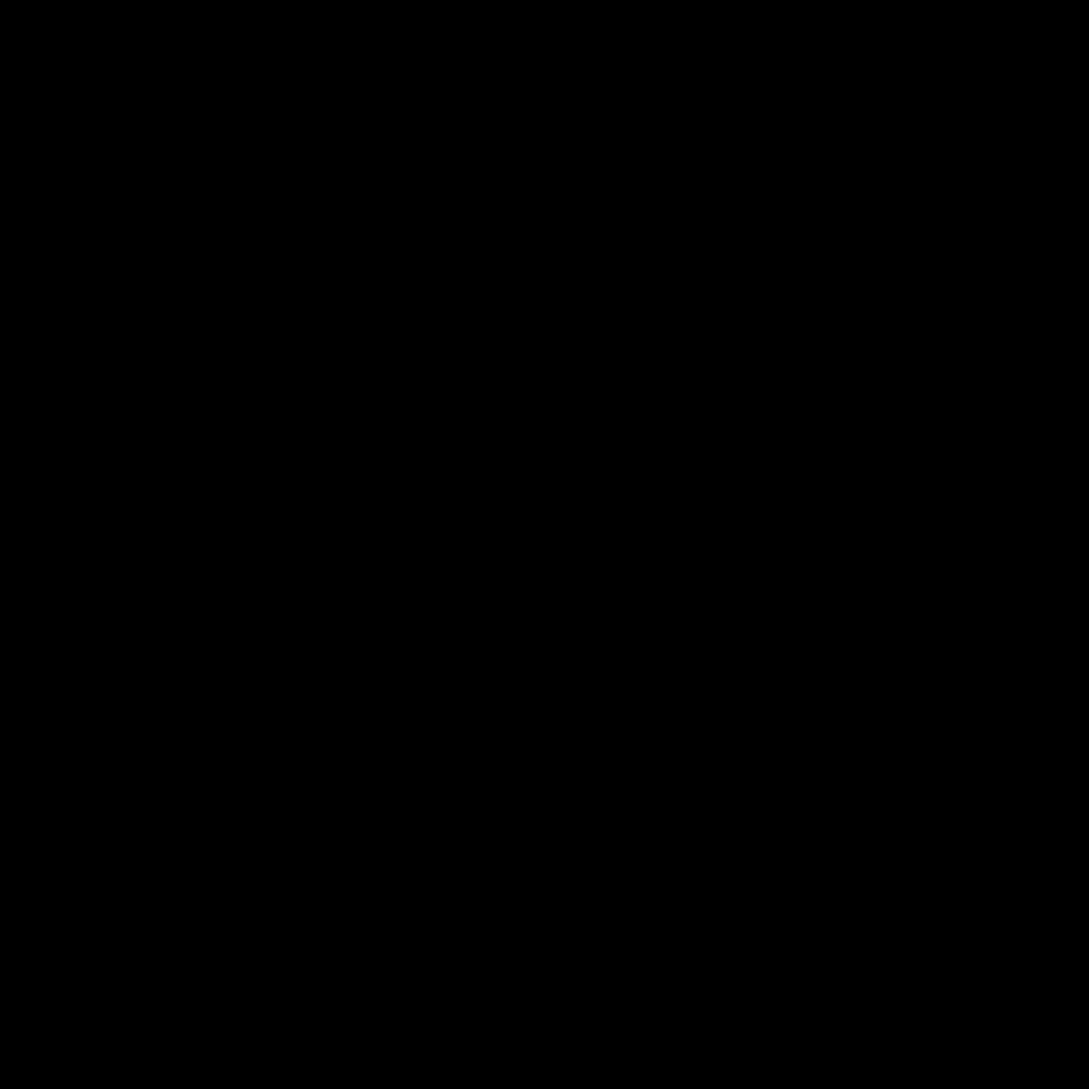}}\hfill
	\subfigure[]{\label{fig:results_modulated_bw_SPS:d}\includegraphics[width=\picturewidth]{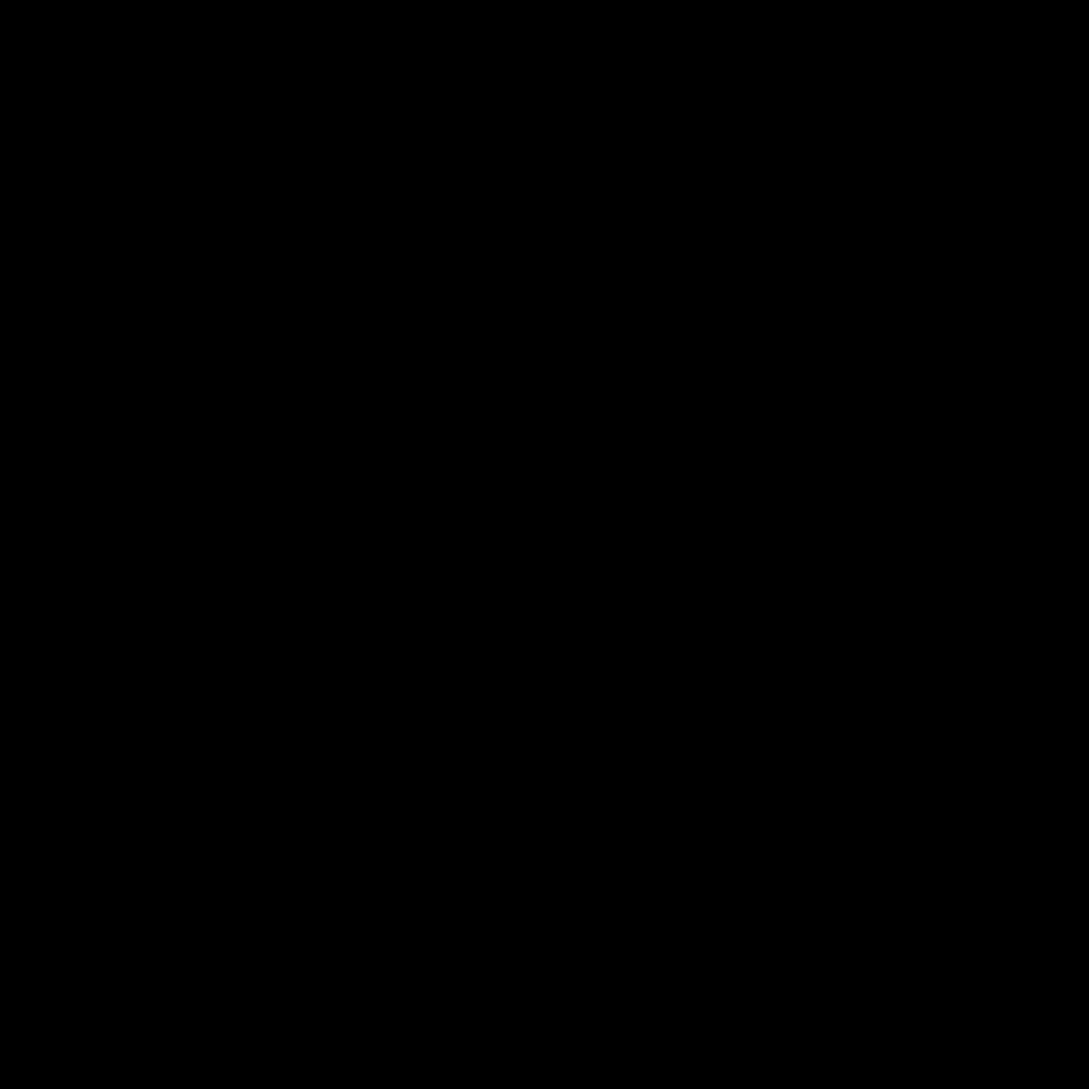}}	
	\caption{Large and complete version of the detail sections in \autoref{fig:results_modulated_ebg_scanned_dots:e}--\subref{fig:results_modulated_ebg_scanned_dots:h}:  Modulated size dots between 2 and 4, B\&W for SPS. \imageproportion}
	\label{fig:results_modulated_bw_SPS}
\end{figure*}

\begin{figure*}[!p]
	\centering
	\subfigure[]{\label{fig:results_modulated_bw_EBG:a}\includegraphics[width=\picturewidth]{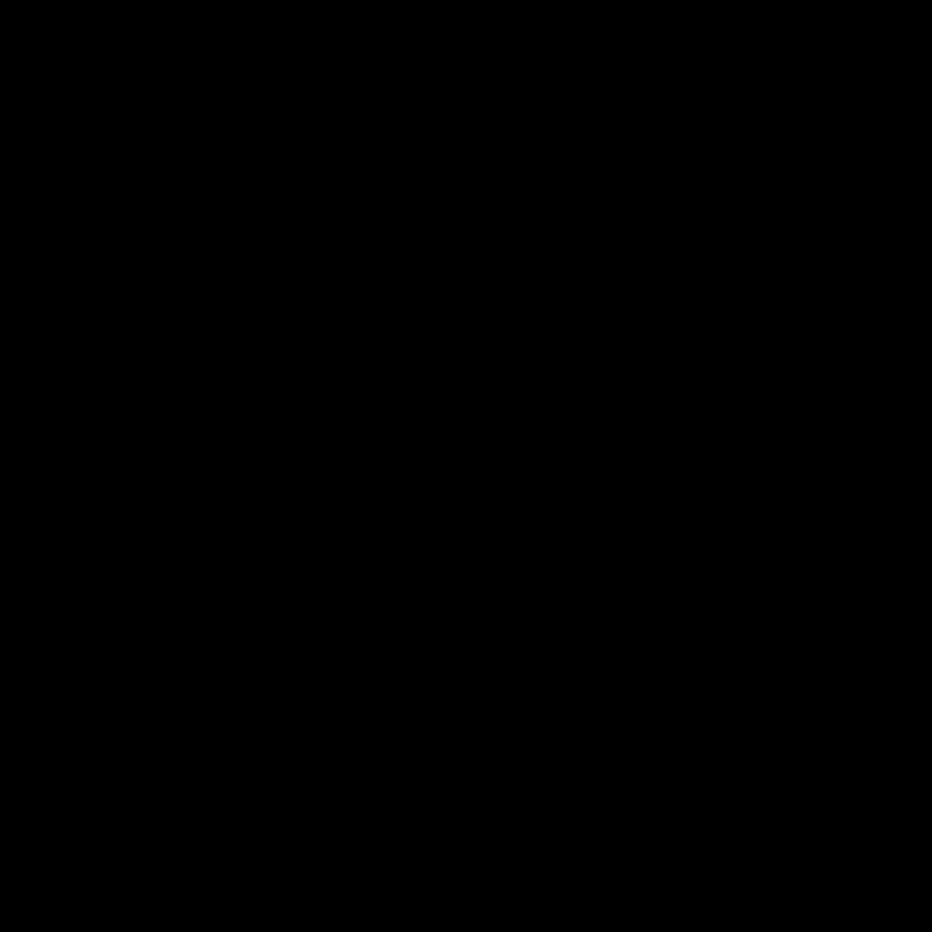}}\hfill
	\subfigure[]{\label{fig:results_modulated_bw_EBG:b}\includegraphics[width=\picturewidth]{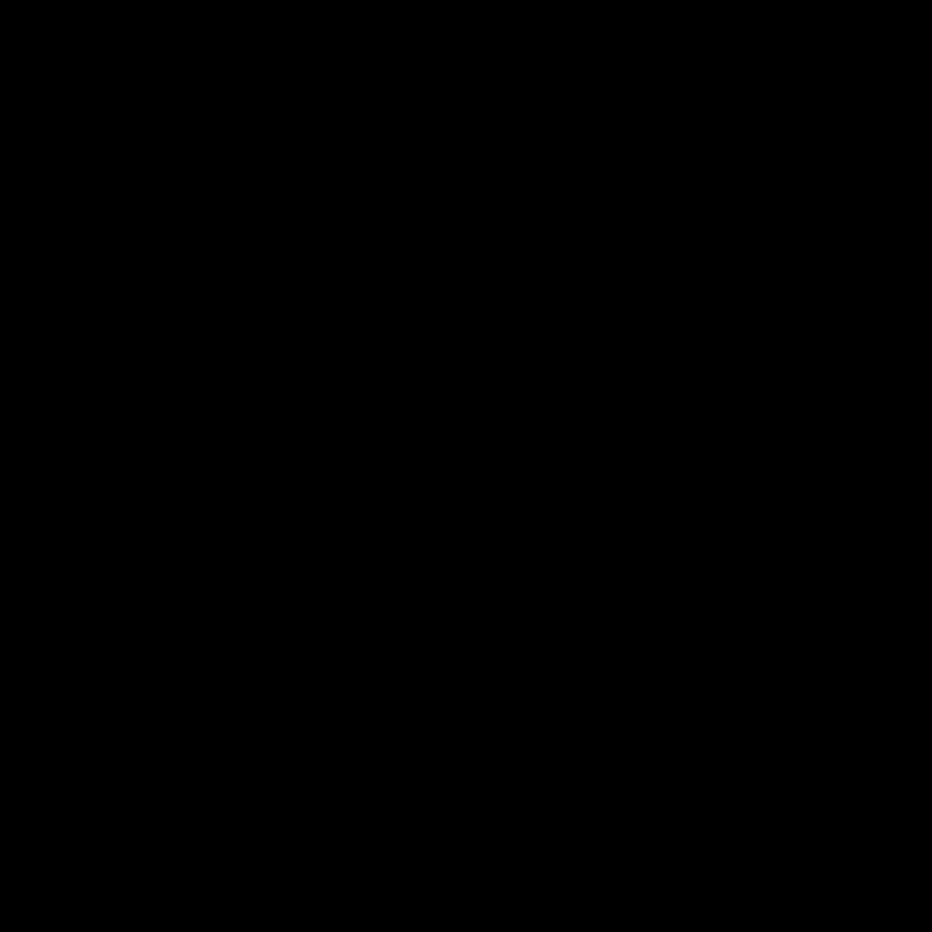}}\\[2ex]
	\subfigure[]{\label{fig:results_modulated_bw_EBG:c}\includegraphics[width=\picturewidth]{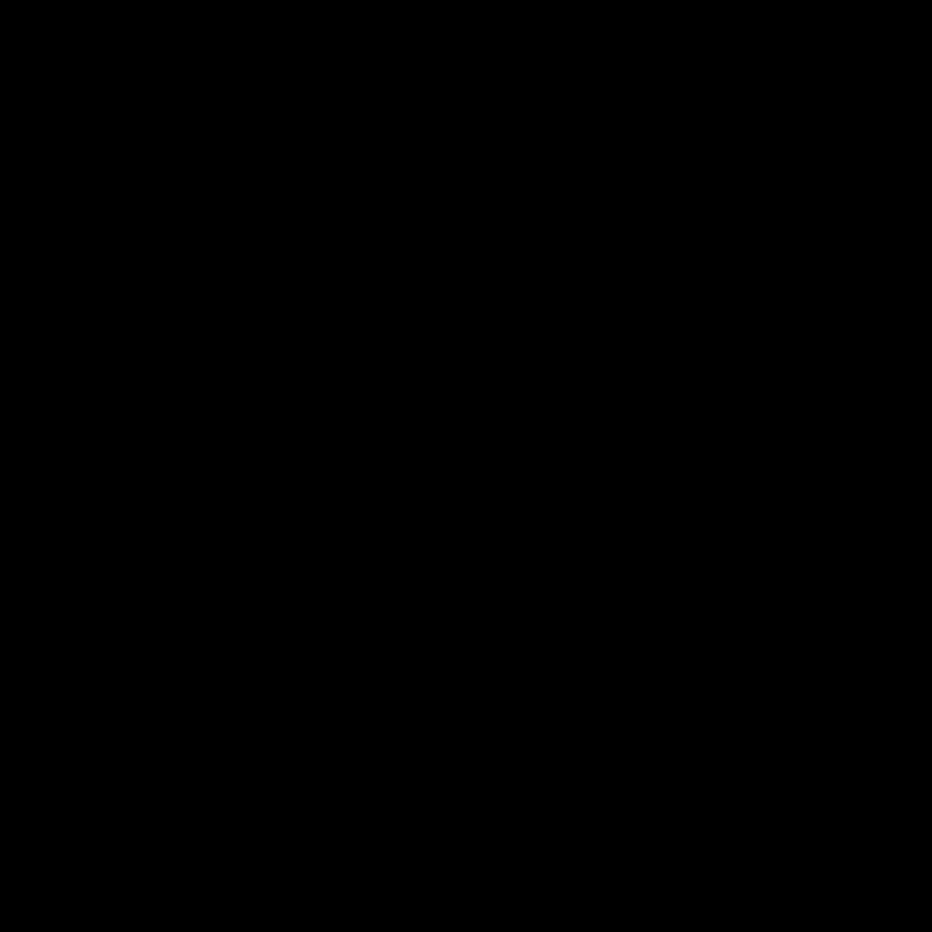}}\hfill
	\subfigure[]{\label{fig:results_modulated_bw_EBG:d}\includegraphics[width=\picturewidth]{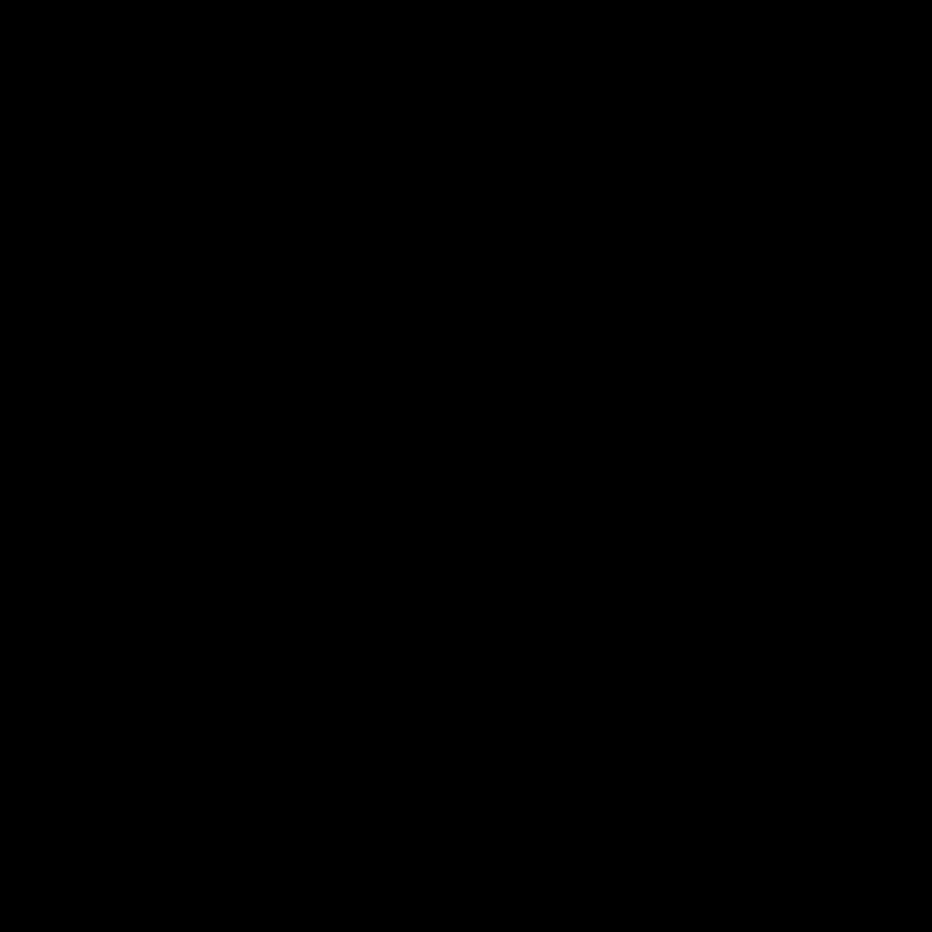}}	
	\caption{Large and complete version of the detail sections in \autoref{fig:results_modulated_ebg_scanned_dots:i}--\subref{fig:results_modulated_ebg_scanned_dots:l}:  Modulated size dots between 2 and 4, B\&W for EBG. \imageproportion}
	\label{fig:results_modulated_bw_EBG}
\end{figure*}

\begin{figure*}[!p]
	\centering
	\subfigure[]{\label{fig:results_modulated_bw_IPD:a}\includegraphics[width=\picturewidth]{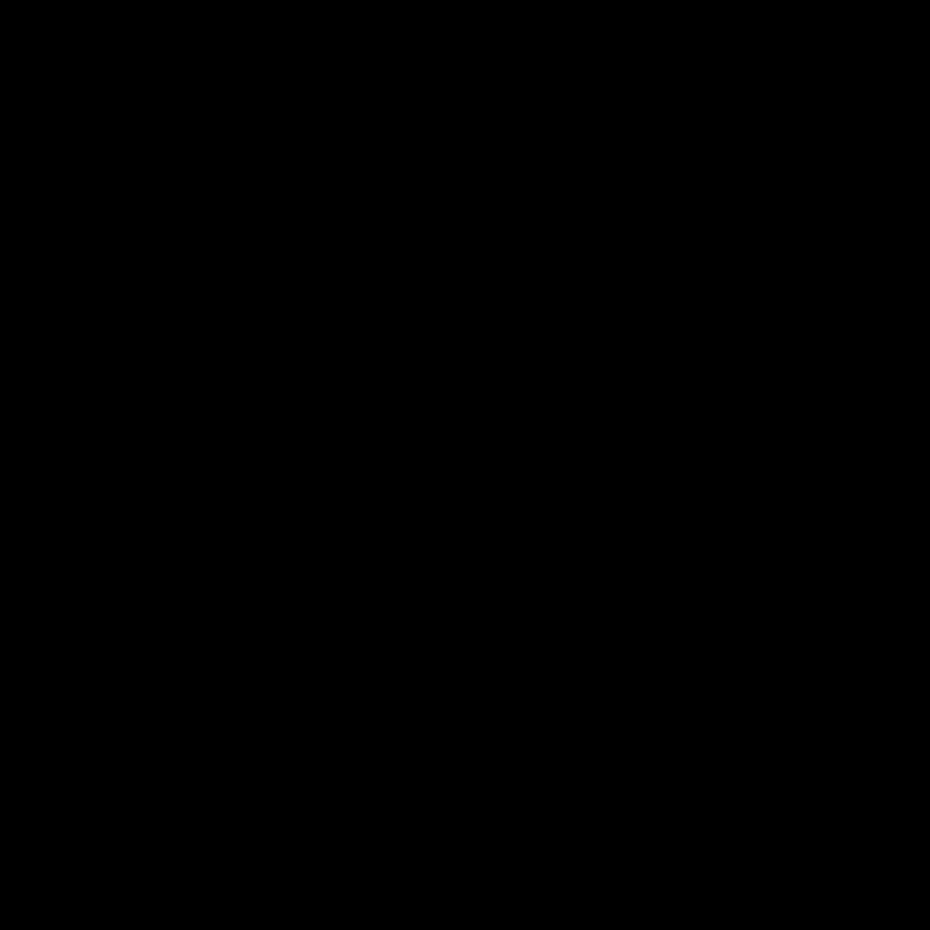}}\hfill
	\subfigure[]{\label{fig:results_modulated_bw_IPD:b}\includegraphics[width=\picturewidth]{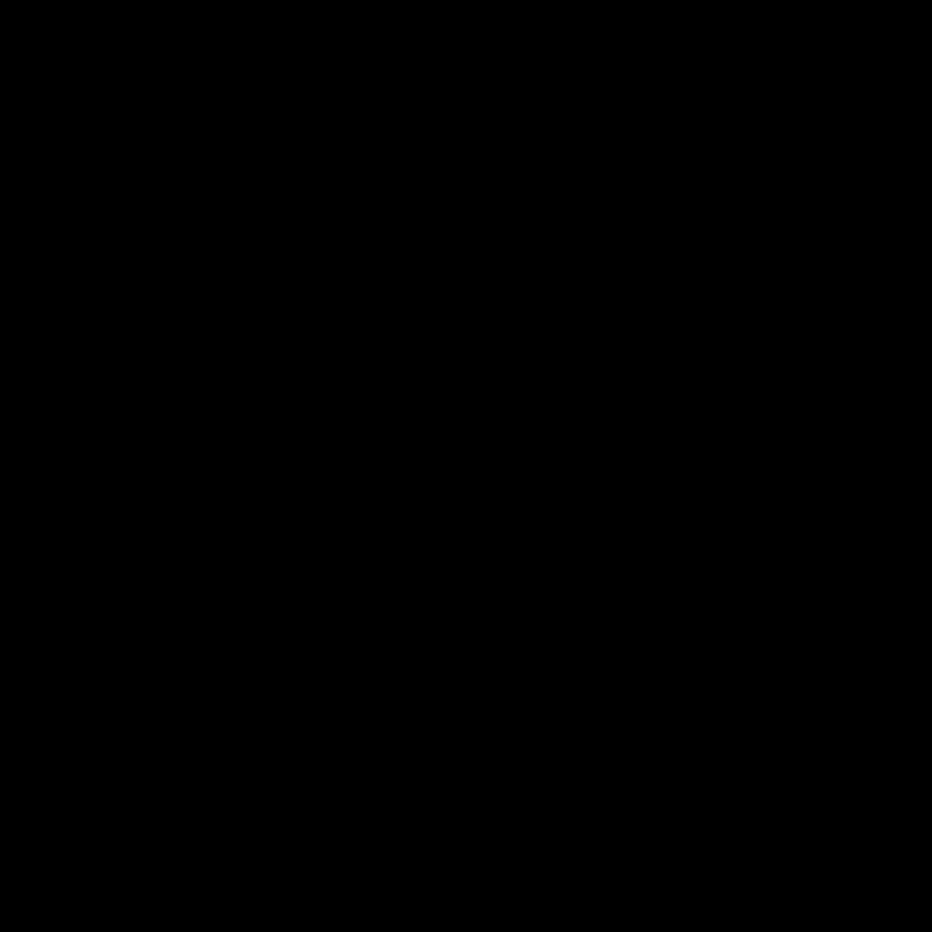}}\\[2ex]
	\subfigure[]{\label{fig:results_modulated_bw_IPD:c}\includegraphics[width=\picturewidth]{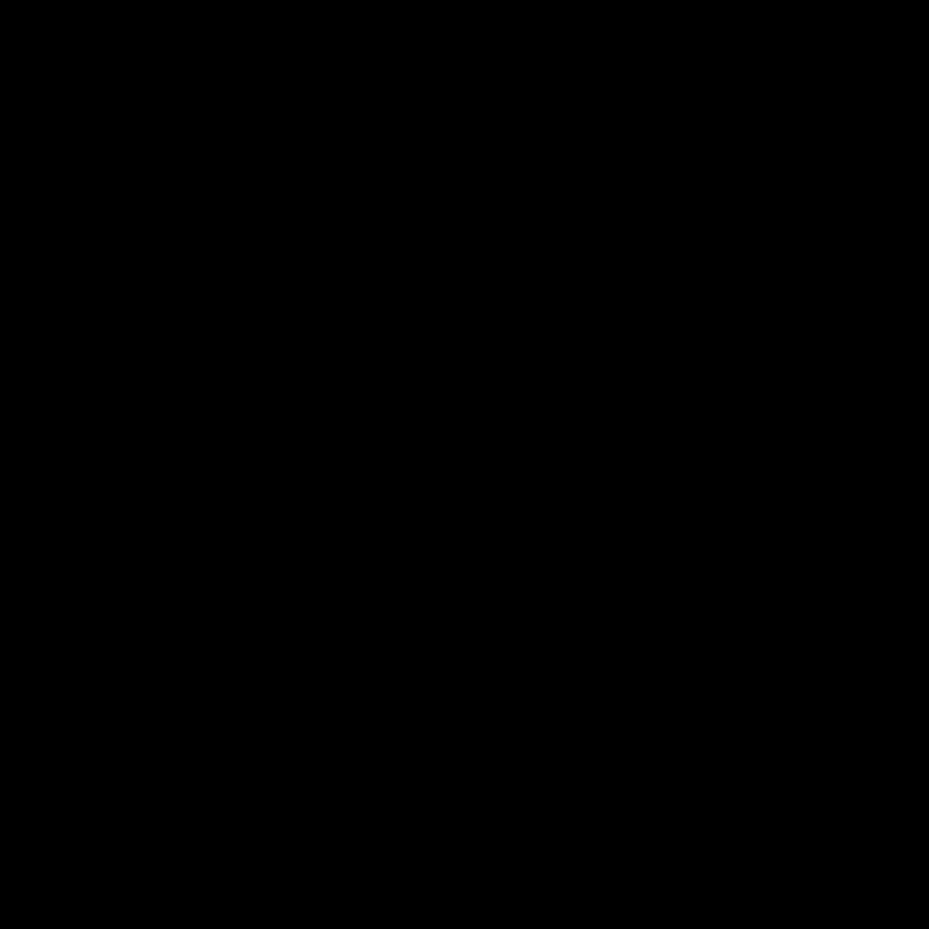}}\hfill
	\subfigure[]{\label{fig:results_modulated_bw_IPD:d}\includegraphics[width=\picturewidth]{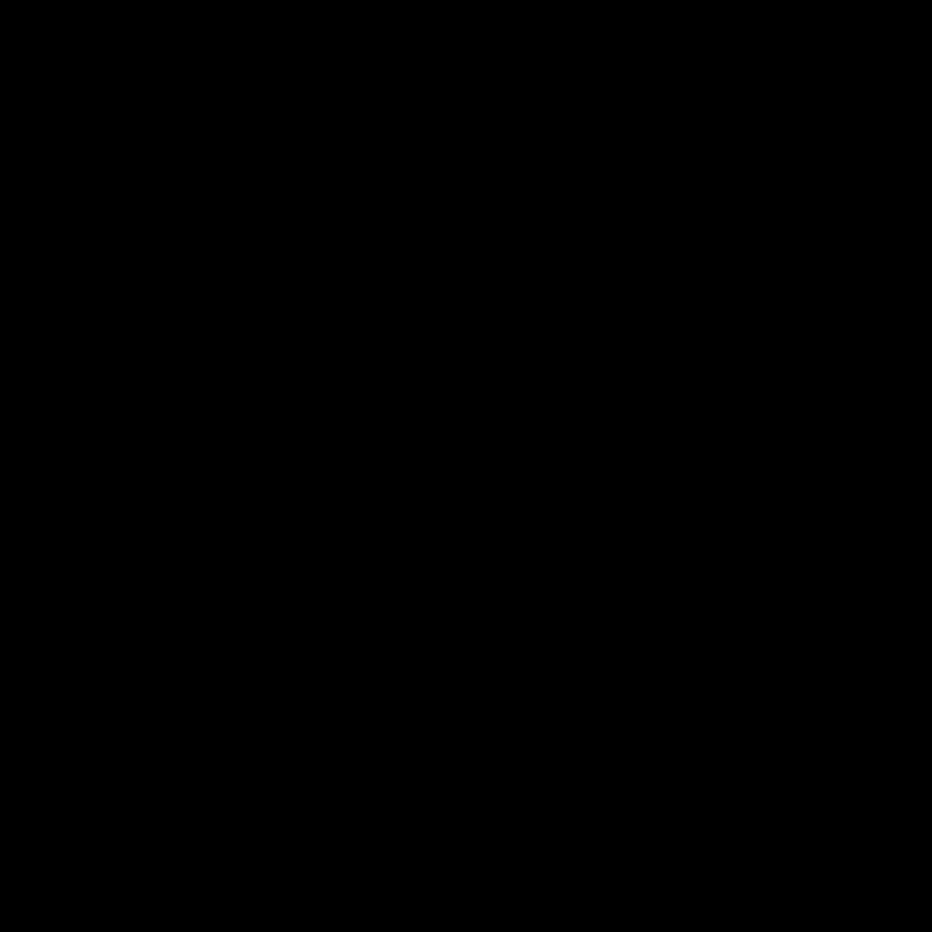}}	
	\caption{Large and complete version of the detail sections in \autoref{fig:results_modulated_ebg_scanned_dots:m}--\subref{fig:results_modulated_ebg_scanned_dots:p}:  Modulated size dots between 2 and 4, B\&W for IPD. \imageproportion}
	\label{fig:results_modulated_bw_IPD}
\end{figure*}

\begin{figure*}[!p]
	\centering
	\subfigure[]{\label{fig:results_scanned_bw_WVS:a}\includegraphics[width=\picturewidth]{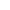}}\hfill
	\subfigure[]{\label{fig:results_scanned_bw_WVS:b}\includegraphics[width=\picturewidth]{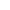}}\\[2ex]
	\subfigure[]{\label{fig:results_scanned_bw_WVS:c}\includegraphics[width=\picturewidth]{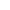}}\hfill
	\subfigure[]{\label{fig:results_scanned_bw_WVS:d}\includegraphics[width=\picturewidth]{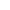}}	
	\caption{Large and complete version of the detail sections in \autoref{fig:results_random_ebg_scanned_dots:a}--\subref{fig:results_random_ebg_scanned_dots:d}:  Scanned dots between 4 and 8, 1200 ppi, gray scale for WVS. \imageproportion}
	\label{fig:results_random_ebg_scanned_dots_WVS}
\end{figure*}

\begin{figure*}[!p]
	\centering
	\subfigure[]{\label{fig:results_scanned_bw_SPS:a}\includegraphics[width=\picturewidth]{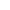}}\hfill
	\subfigure[]{\label{fig:results_scanned_bw_SPS:b}\includegraphics[width=\picturewidth]{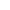}}\\[2ex]
	\subfigure[]{\label{fig:results_scanned_bw_SPS:c}\includegraphics[width=\picturewidth]{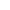}}\hfill
	\subfigure[]{\label{fig:results_scanned_bw_SPS:d}\includegraphics[width=\picturewidth]{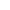}}	
	\caption{Large and complete version of the detail sections in \autoref{fig:results_random_ebg_scanned_dots:e}--\subref{fig:results_random_ebg_scanned_dots:h}:  Scanned dots between 4 and 8, 1200 ppi, gray scale for SPS. \imageproportion}
	\label{fig:results_random_ebg_scanned_dots_SPS}
\end{figure*}

\begin{figure*}[!p]
	\centering
	\subfigure[]{\label{fig:results_scanned_bw_EBG:a}\includegraphics[width=\picturewidth]{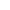}}\hfill
	\subfigure[]{\label{fig:results_scanned_bw_EBG:b}\includegraphics[width=\picturewidth]{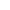}}\\[2ex]
	\subfigure[]{\label{fig:results_scanned_bw_EBG:c}\includegraphics[width=\picturewidth]{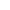}}\hfill
	\subfigure[]{\label{fig:results_scanned_bw_EBG:d}\includegraphics[width=\picturewidth]{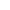}}	
	\caption{Large and complete version of the detail sections in \autoref{fig:results_random_ebg_scanned_dots:i}--\subref{fig:results_random_ebg_scanned_dots:l}:  Scanned dots between 4 and 8, 1200 ppi, gray scale for EBG. \imageproportion}
	\label{fig:results_random_ebg_scanned_dots_EBG}
\end{figure*}

\begin{figure*}[!p]
	\centering
	\subfigure[]{\label{fig:results_scanned_bw_IPD:a}\includegraphics[width=\picturewidth]{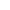}}\hfill
	\subfigure[]{\label{fig:results_scanned_bw_IPD:b}\includegraphics[width=\picturewidth]{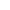}}\\[2ex]
	\subfigure[]{\label{fig:results_scanned_bw_IPD:c}\includegraphics[width=\picturewidth]{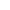}}\hfill
	\subfigure[]{\label{fig:results_scanned_bw_IPD:d}\includegraphics[width=\picturewidth]{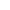}}	
	\caption{Large and complete version of the detail sections in \autoref{fig:results_random_ebg_scanned_dots:m}--\subref{fig:results_random_ebg_scanned_dots:p}:  Scanned dots between 4 and 8, 1200 ppi, gray scale for IPD. \imageproportion}
	\label{fig:results_random_ebg_scanned_dots_IPD}
\end{figure*}

\begin{figure*}[!p]
	\centering
	\subfigure[]{\label{fig:filter_comparison_Canny:a}\includegraphics[width=\picturewidth]{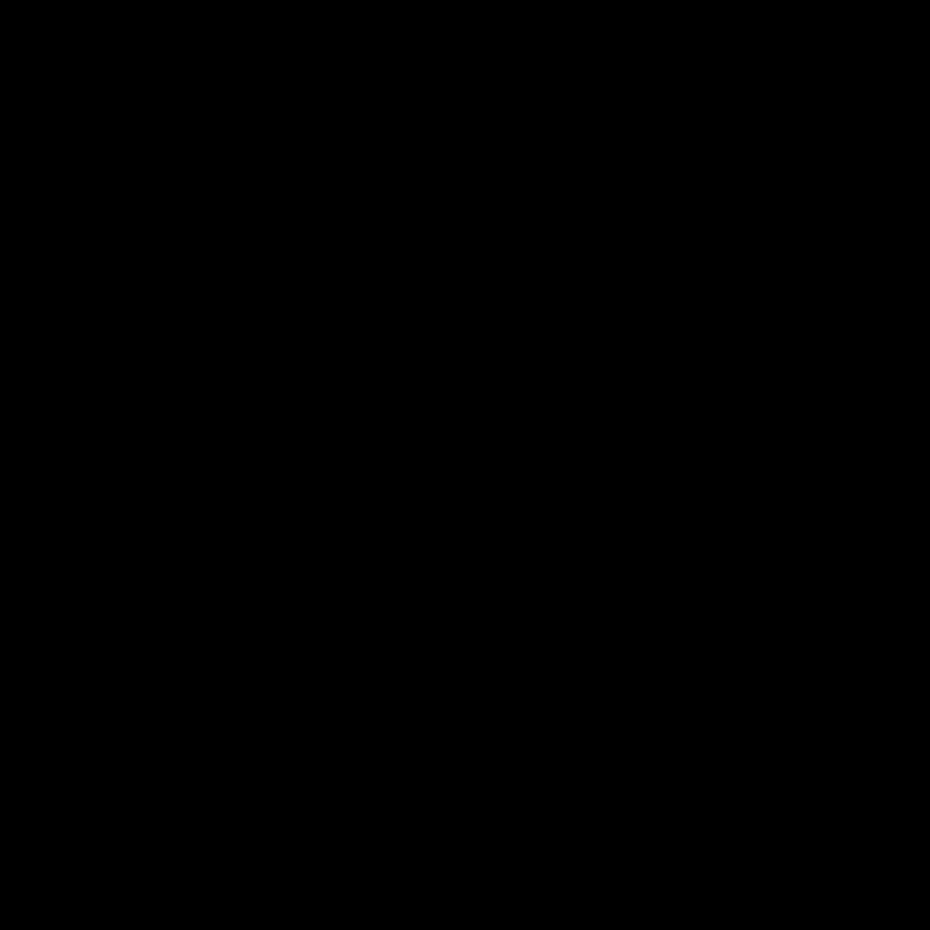}}\hfill
	\subfigure[]{\label{fig:filter_comparison_Canny:b}\includegraphics[width=\picturewidth]{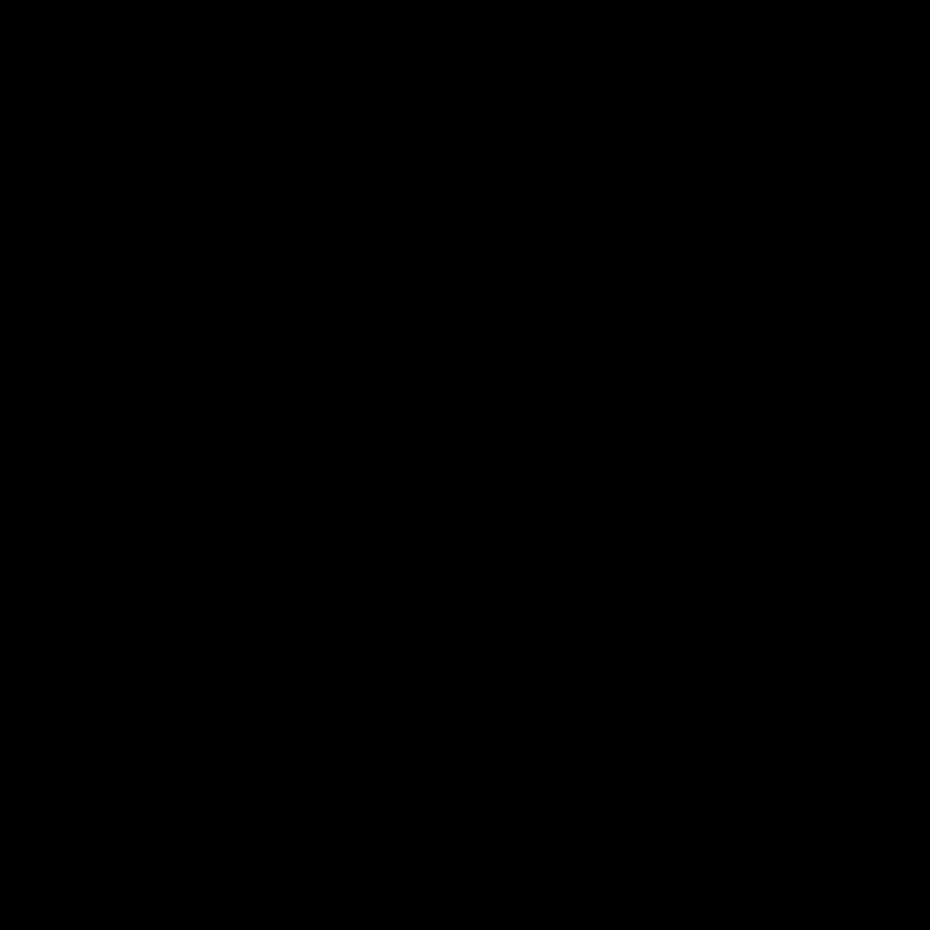}}\\[2ex]
	\subfigure[]{\label{fig:filter_comparison_Canny:c}\includegraphics[width=\picturewidth]{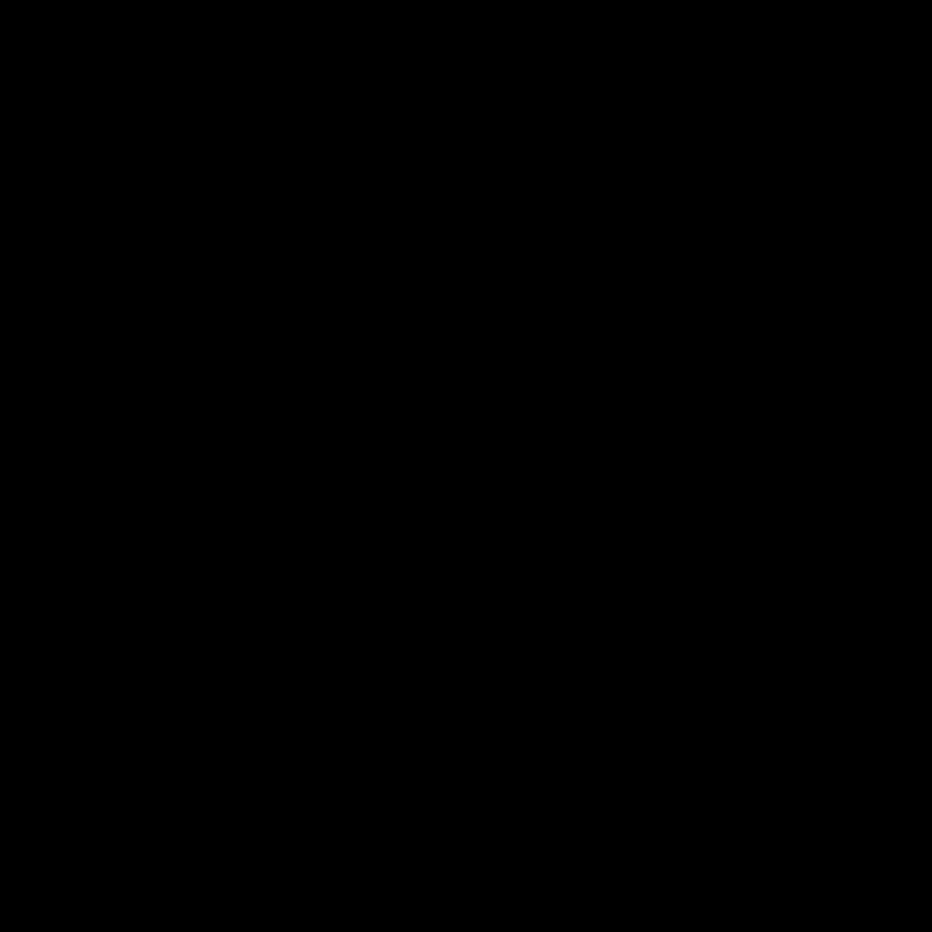}}\hfill%
	\subfigure[]{\label{fig:filter_comparison_Canny:d}\includegraphics[width=\picturewidth]{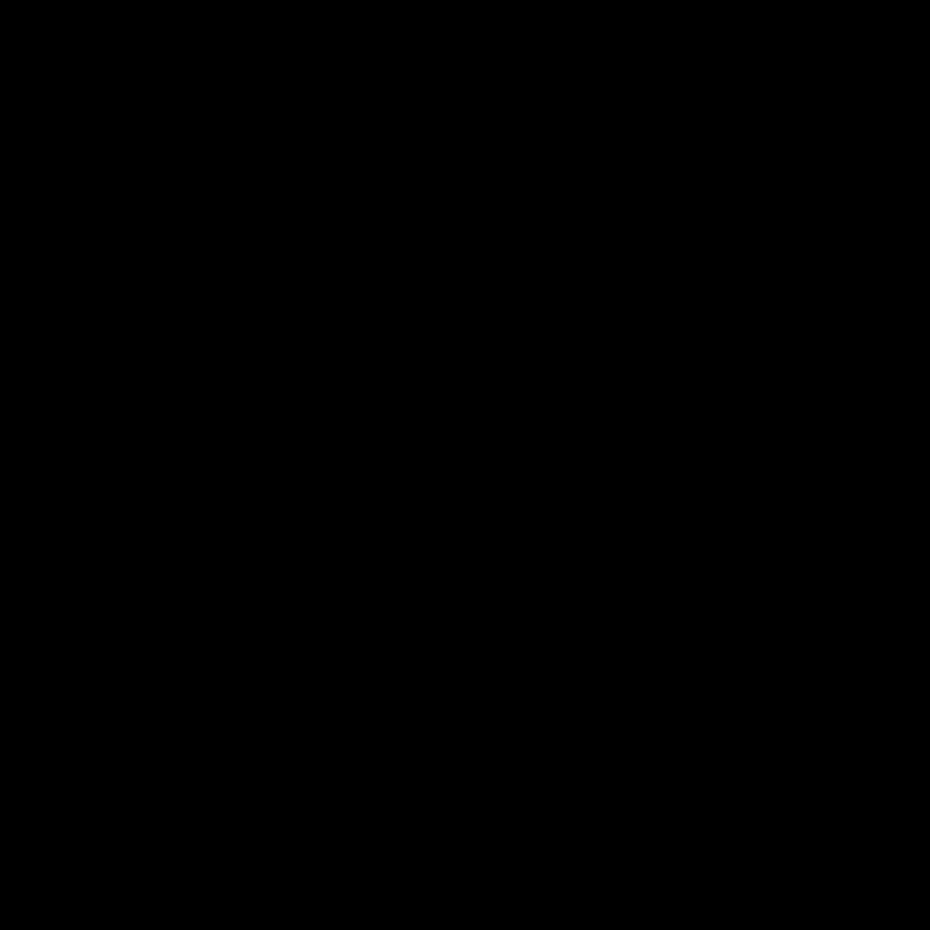}}%
	\caption{Large and complete version of the detail sections in \autoref{fig:filter_comparison:a}--\subref{fig:filter_comparison:d}:  Use of the Canny filter for for the IPD method. \imageproportion}
	\label{fig:filter_comparison_Canny}
\end{figure*}

\begin{figure*}[!p]
	\centering
	\subfigure[]{\label{fig:filter_comparison_DoG:a}\includegraphics[width=\picturewidth]{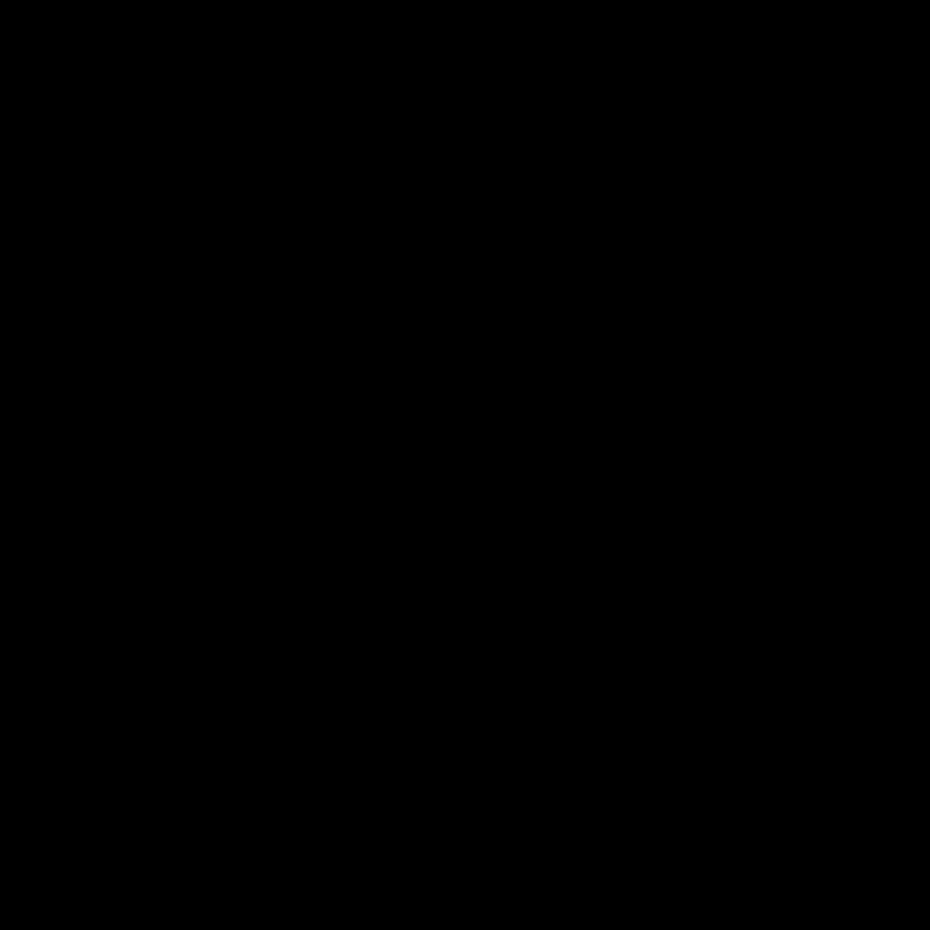}}\hfill
	\subfigure[]{\label{fig:filter_comparison_DoG:b}\includegraphics[width=\picturewidth]{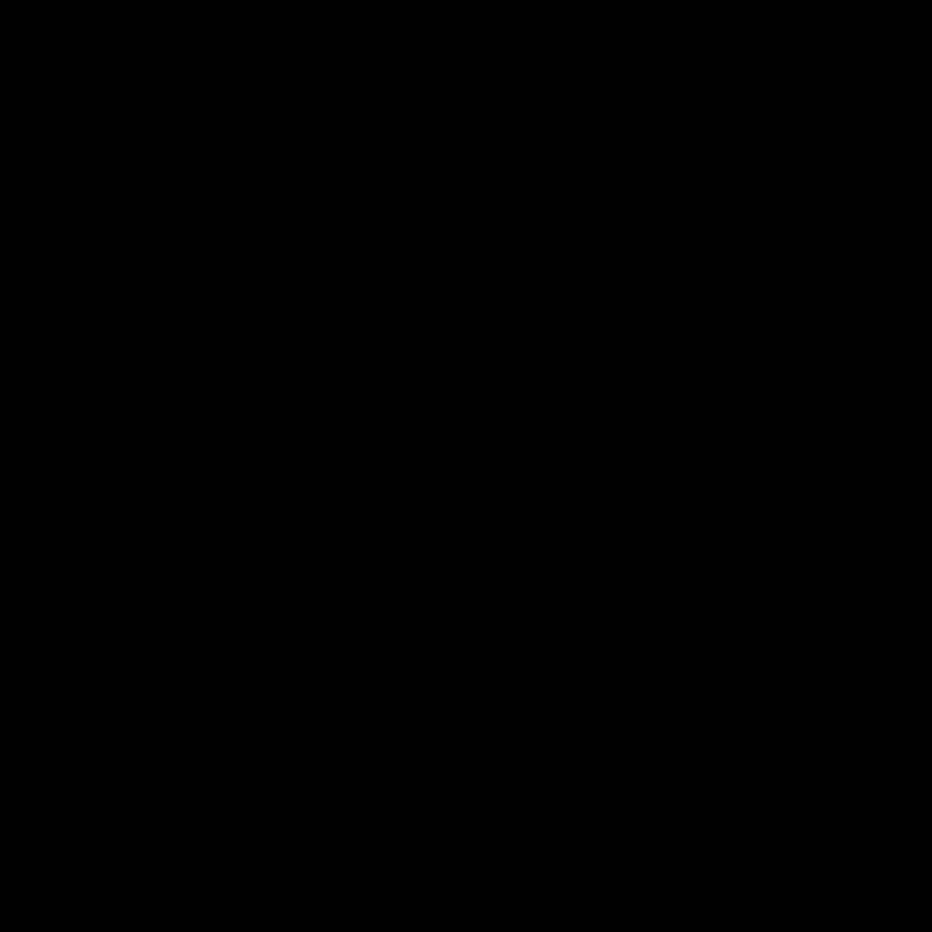}}\\[2ex]
	\subfigure[]{\label{fig:filter_comparison_DoG:c}\includegraphics[width=\picturewidth]{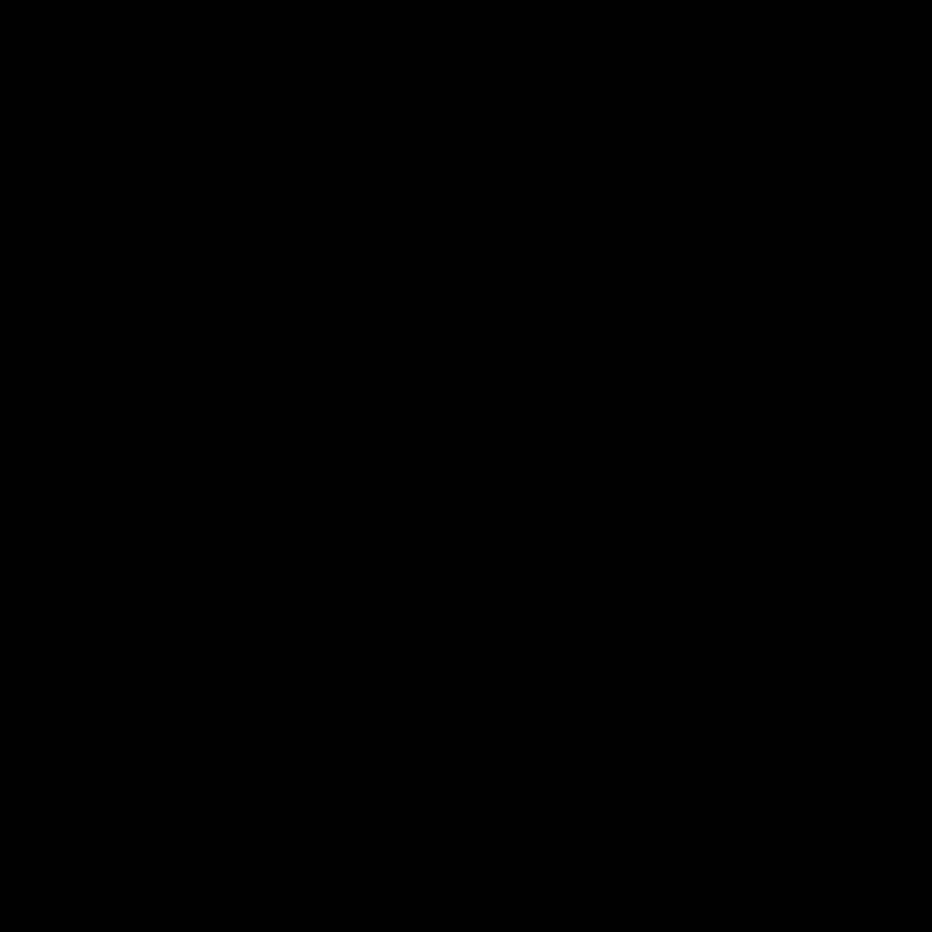}}\hfill%
	\subfigure[]{\label{fig:filter_comparison_DoG:d}\includegraphics[width=\picturewidth]{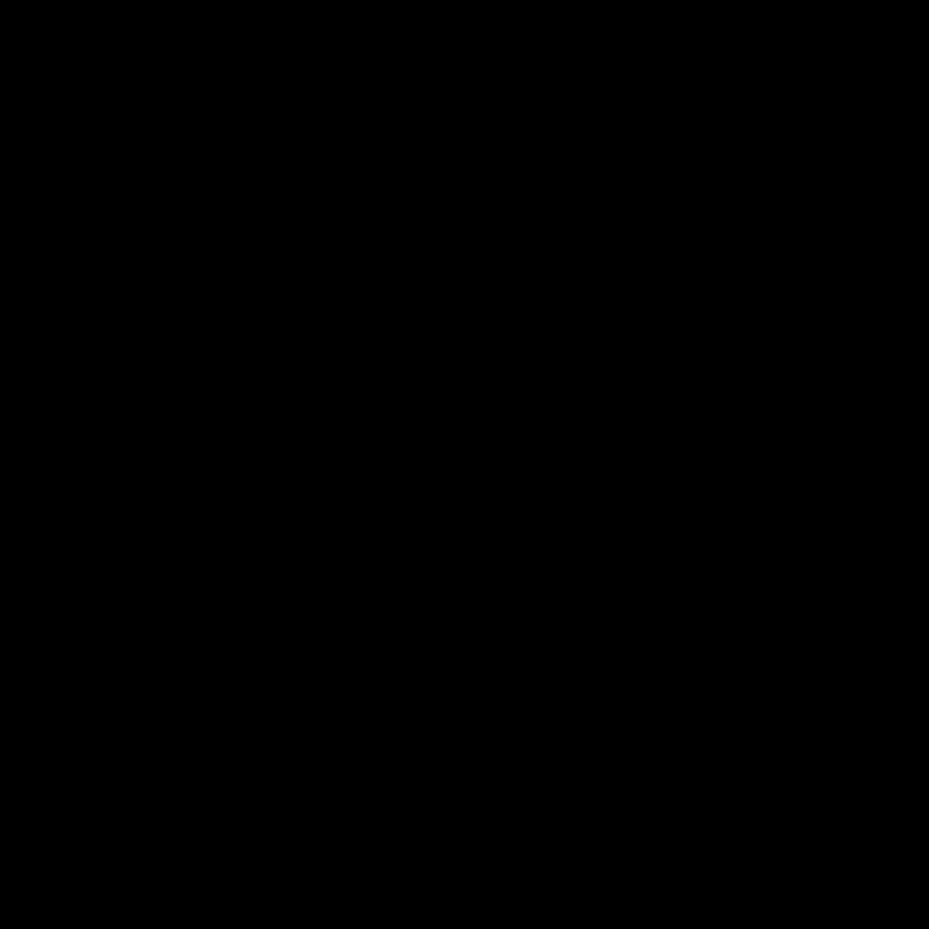}}%
	\caption{Large and complete version of the detail sections in \autoref{fig:filter_comparison:e}--\subref{fig:filter_comparison:h}:  Use of the DoG filter for for the IPD method. \imageproportion}
	\label{fig:filter_comparison_DoG}
\end{figure*}

\begin{figure*}[!p]
	\centering
	\subfigure[\hspace{\linewidth}]{\label{fig:filter_comparison_LoG:a}\includegraphics[width=\picturewidth]{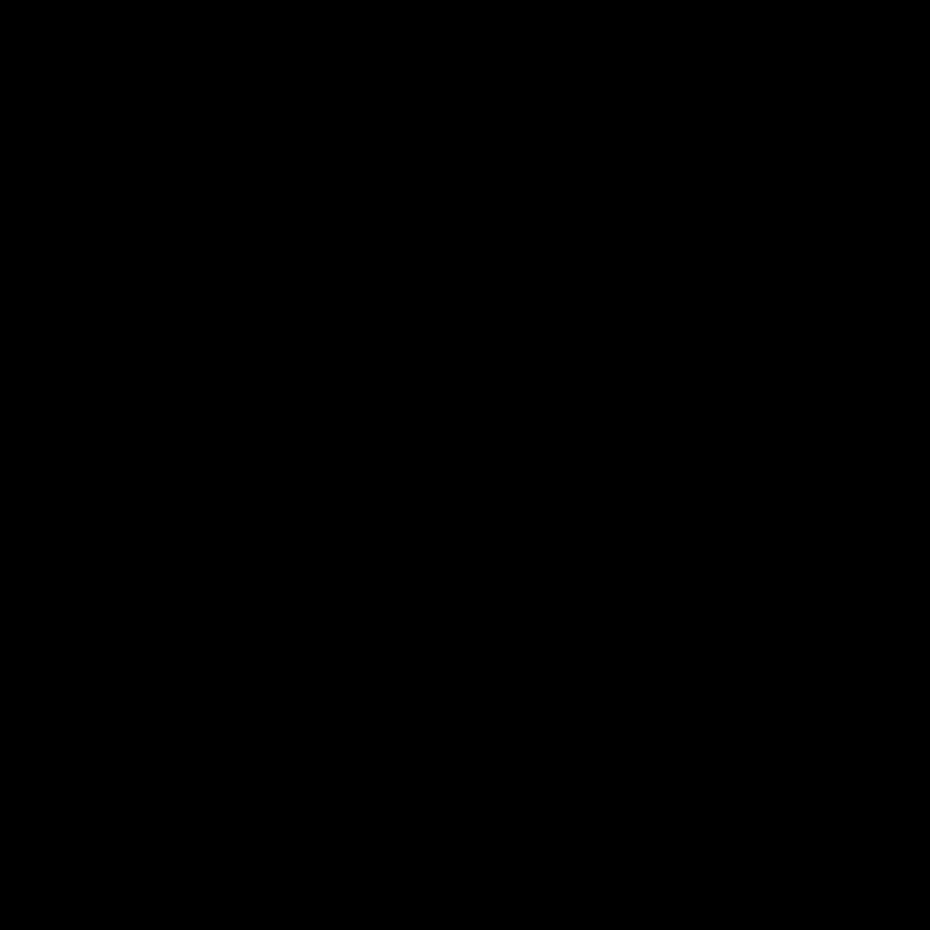}}\hfill
	\subfigure[\hspace{\linewidth}]{\label{fig:filter_comparison_LoG:b}\includegraphics[width=\picturewidth]{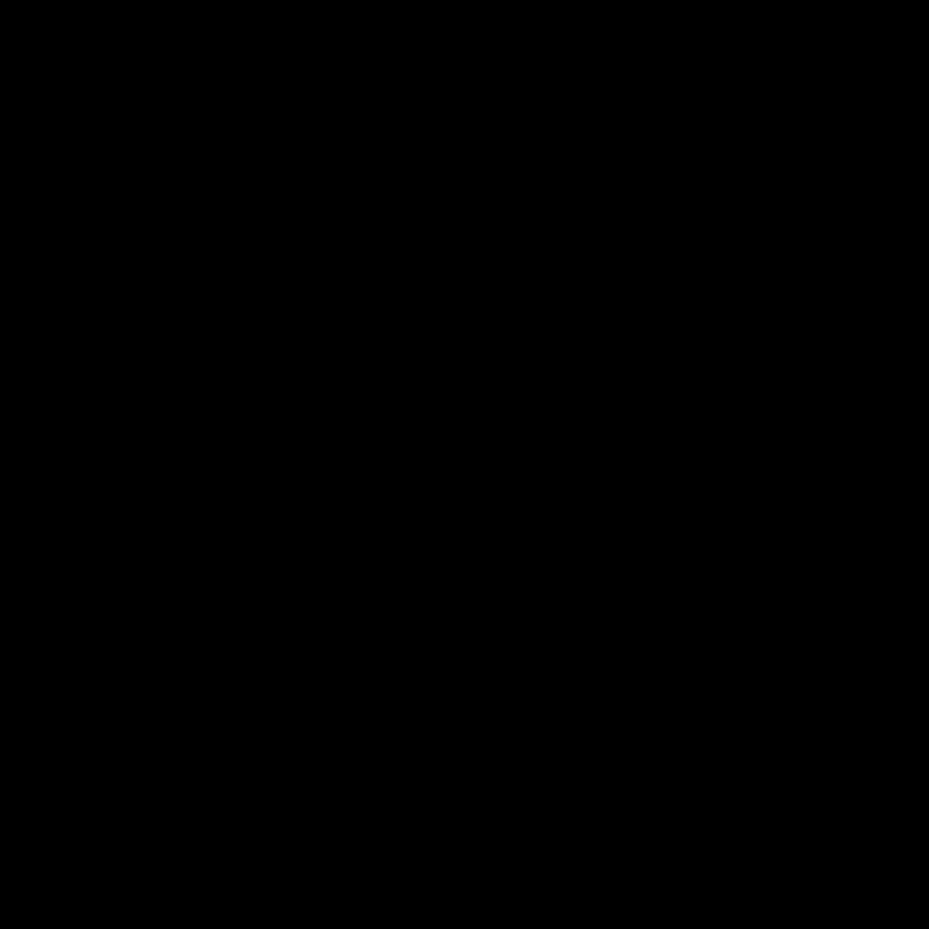}}\\[2ex]
	\subfigure[\hspace{\linewidth}]{\label{fig:filter_comparison_LoG:c}\includegraphics[width=\picturewidth]{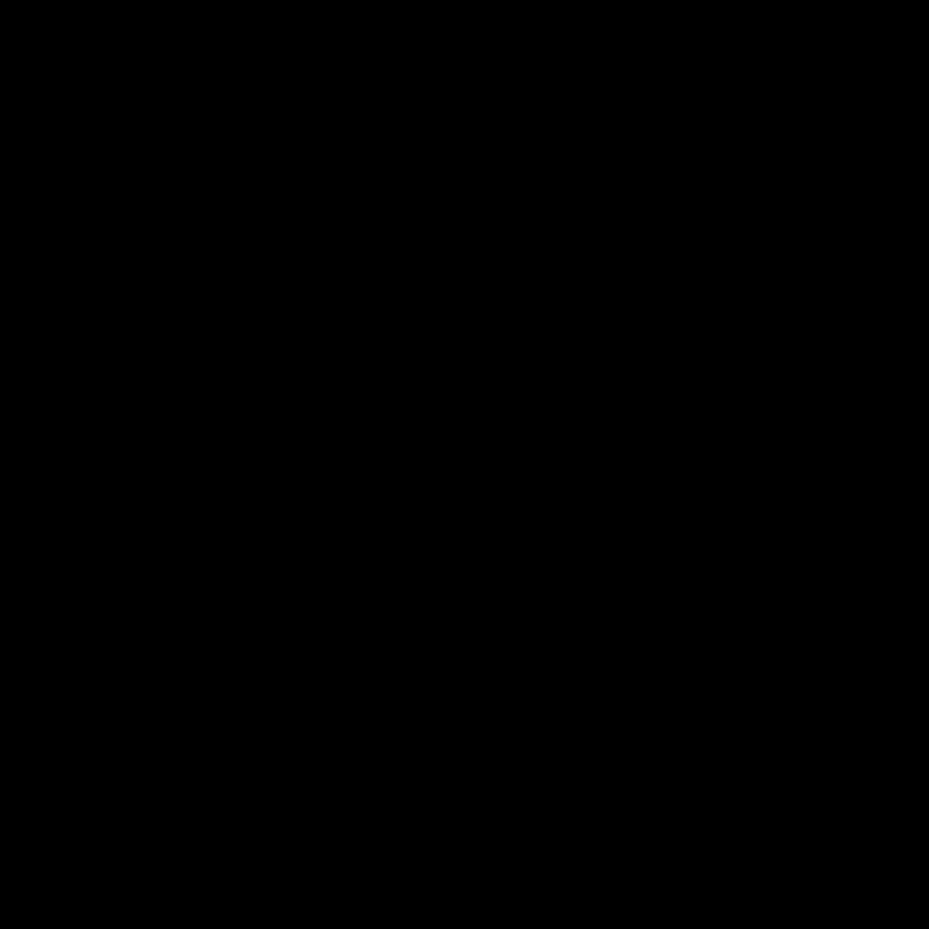}}\hfill%
	\subfigure[\hspace{\linewidth}]{\label{fig:filter_comparison_LoG:d}\includegraphics[width=\picturewidth]{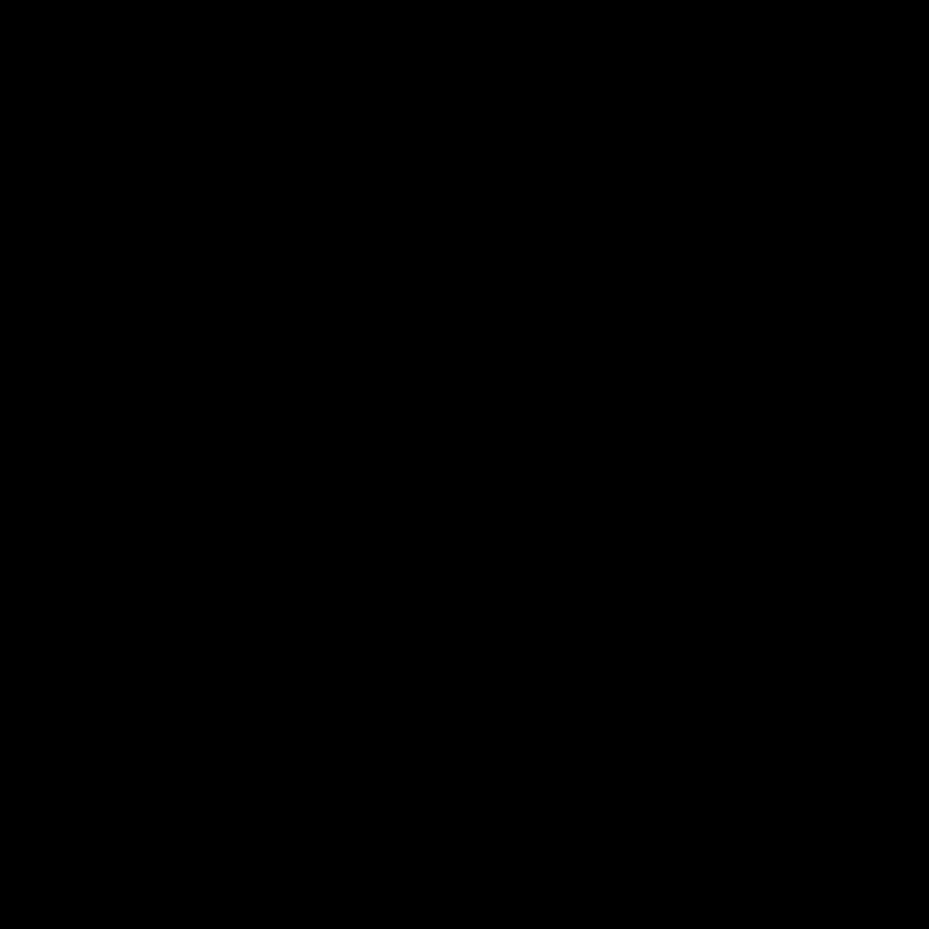}}%
	\caption{Large and complete version of the detail sections in \autoref{fig:filter_comparison:i}--\subref{fig:filter_comparison:l}:  Use of the LoG filter for for the IPD method. \imageproportion}
	\label{fig:filter_comparison_LoG}
\end{figure*}

\begin{figure*}[!p]
	\centering
	\subfigure[]{\label{fig:effect_results_wb:a}\includegraphics[width=\picturewidth]{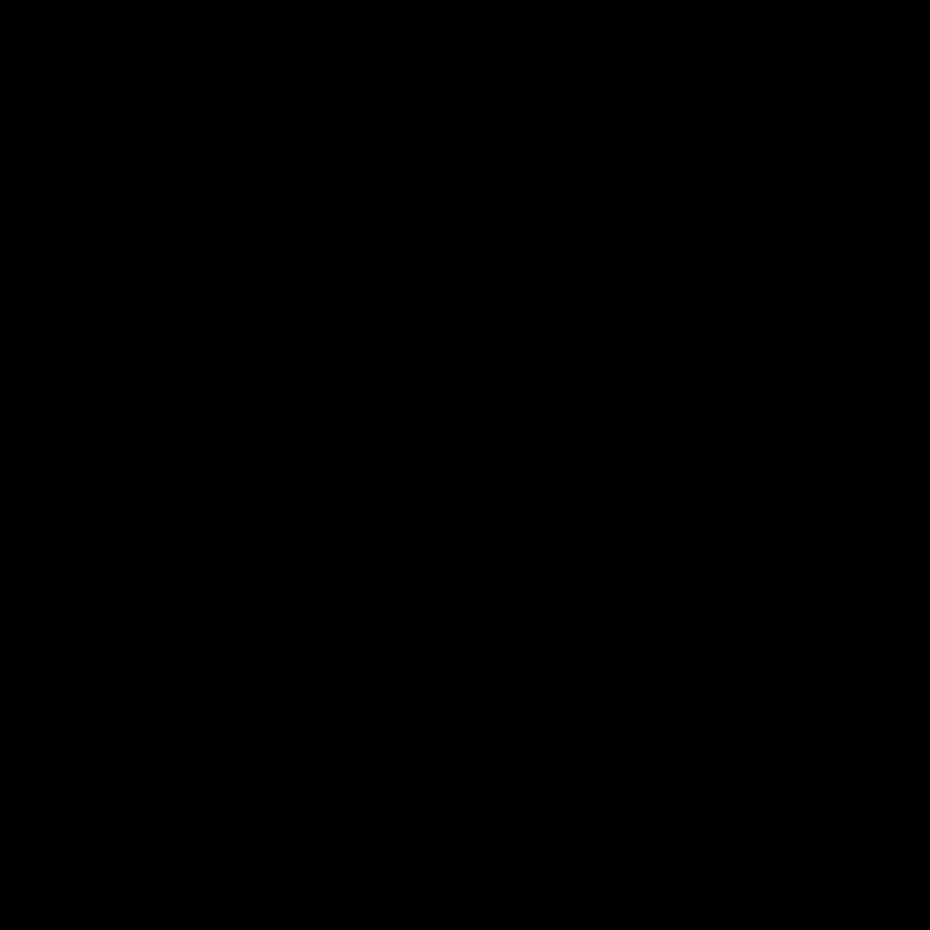}}\hfill
	\subfigure[]{\label{fig:effect_results_wb:b}\includegraphics[width=\picturewidth]{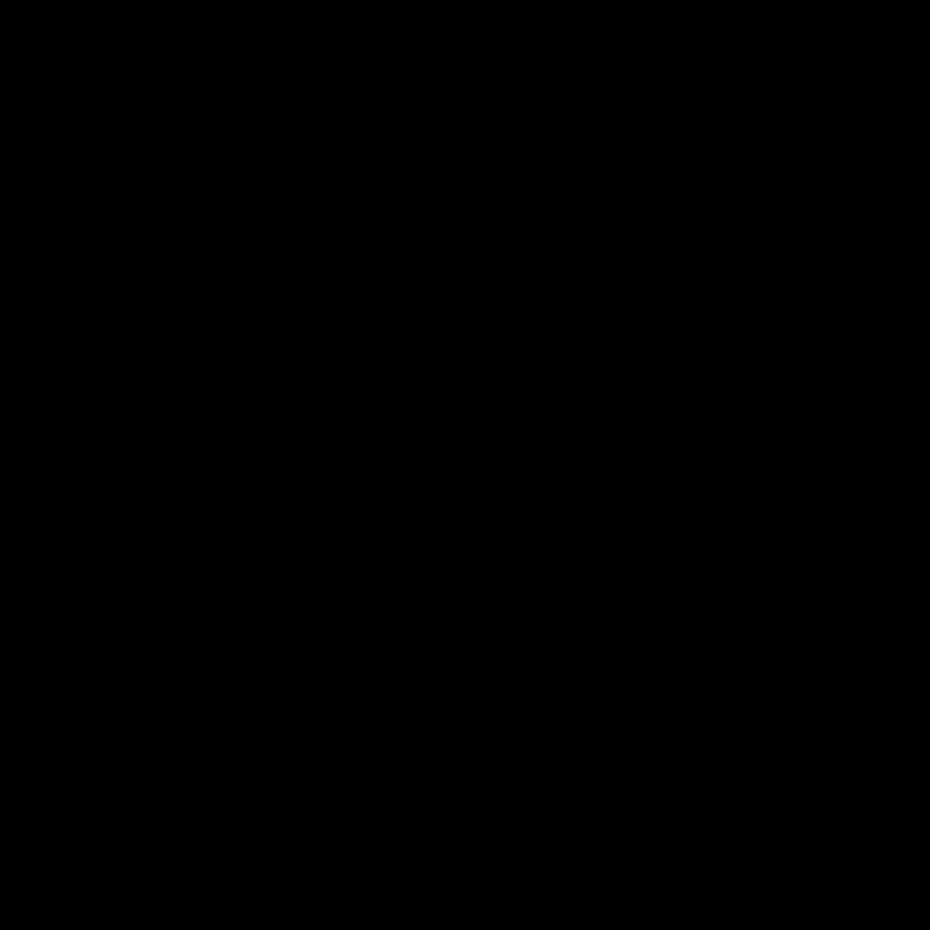}}\\[2ex]
	\subfigure[]{\label{fig:effect_results_wb:c}\includegraphics[width=\picturewidth]{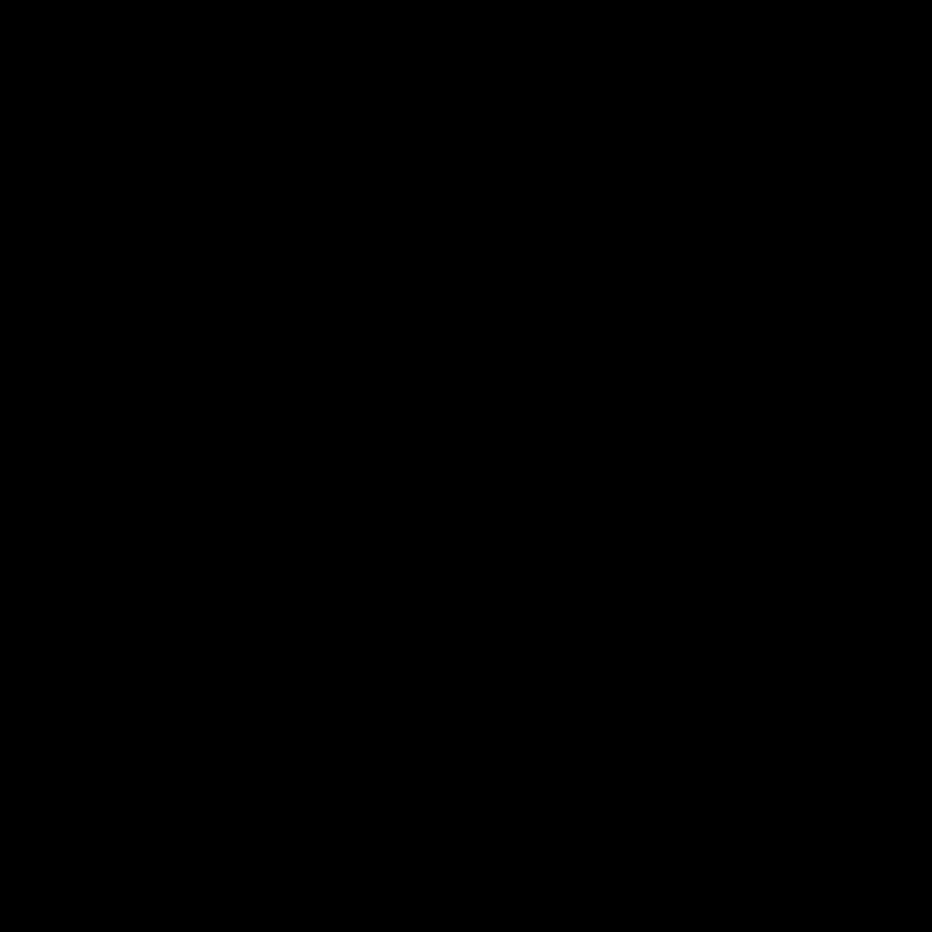}}\hfill%
	\subfigure[]{\label{fig:effect_results_wb:d}\includegraphics[width=\picturewidth]{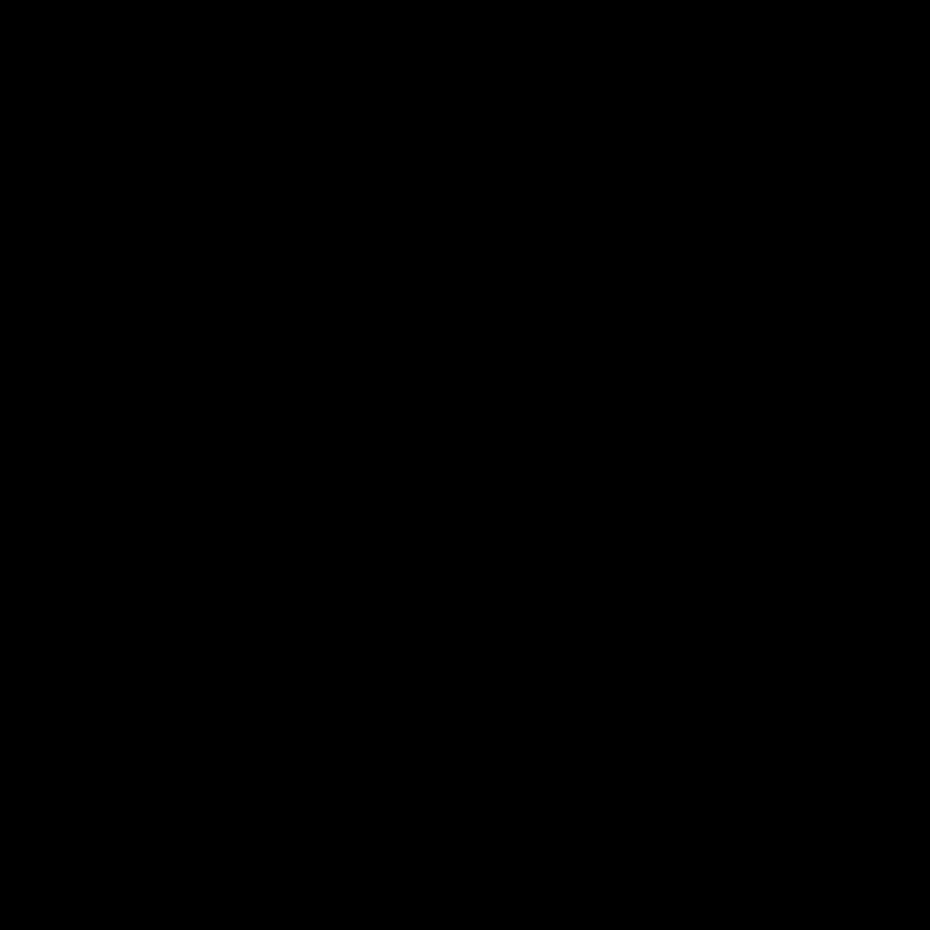}}%
	\caption{Large and complete version of the detail sections in \autoref{fig:effect_results:a}--\subref{fig:effect_results:d}: Use of intentional margin around edges. \imageproportion}
	\label{fig:effect_results_wb}
\end{figure*}

\begin{figure*}[!p]
	\centering
	\subfigure[]{\label{fig:effect_results_wide:a}\includegraphics[width=\picturewidth]{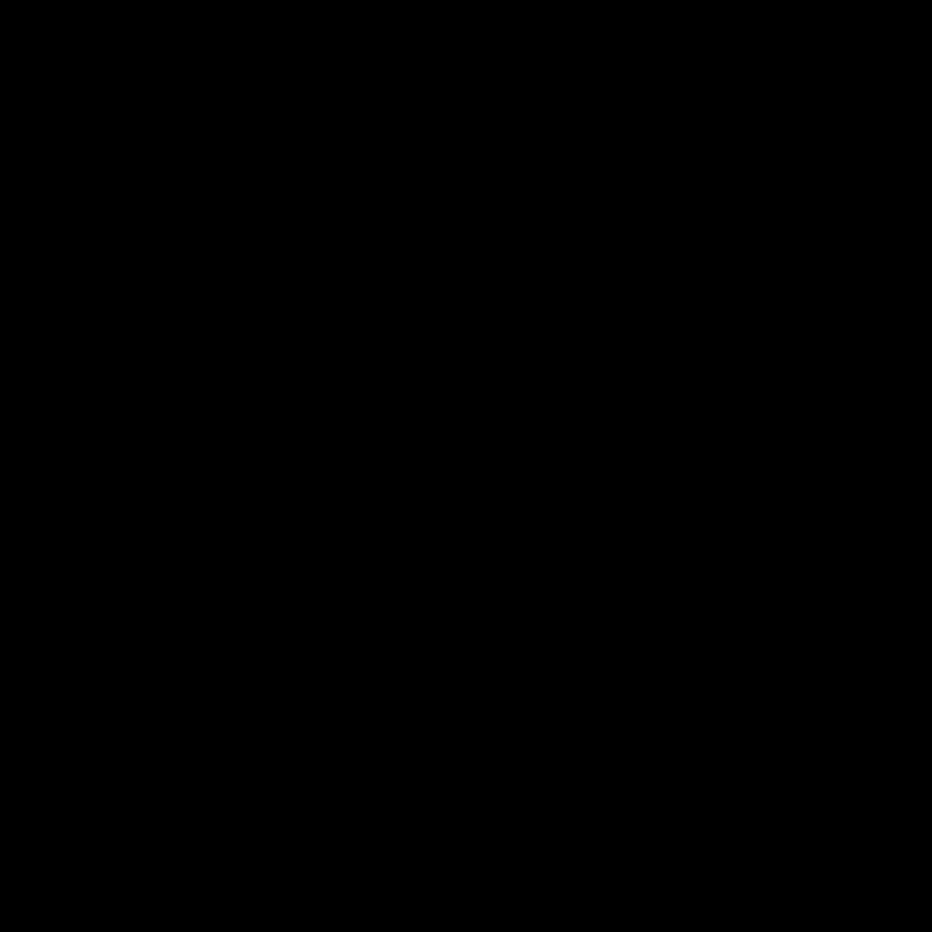}}\hfill
	\subfigure[]{\label{fig:effect_results_wide:b}\includegraphics[width=\picturewidth]{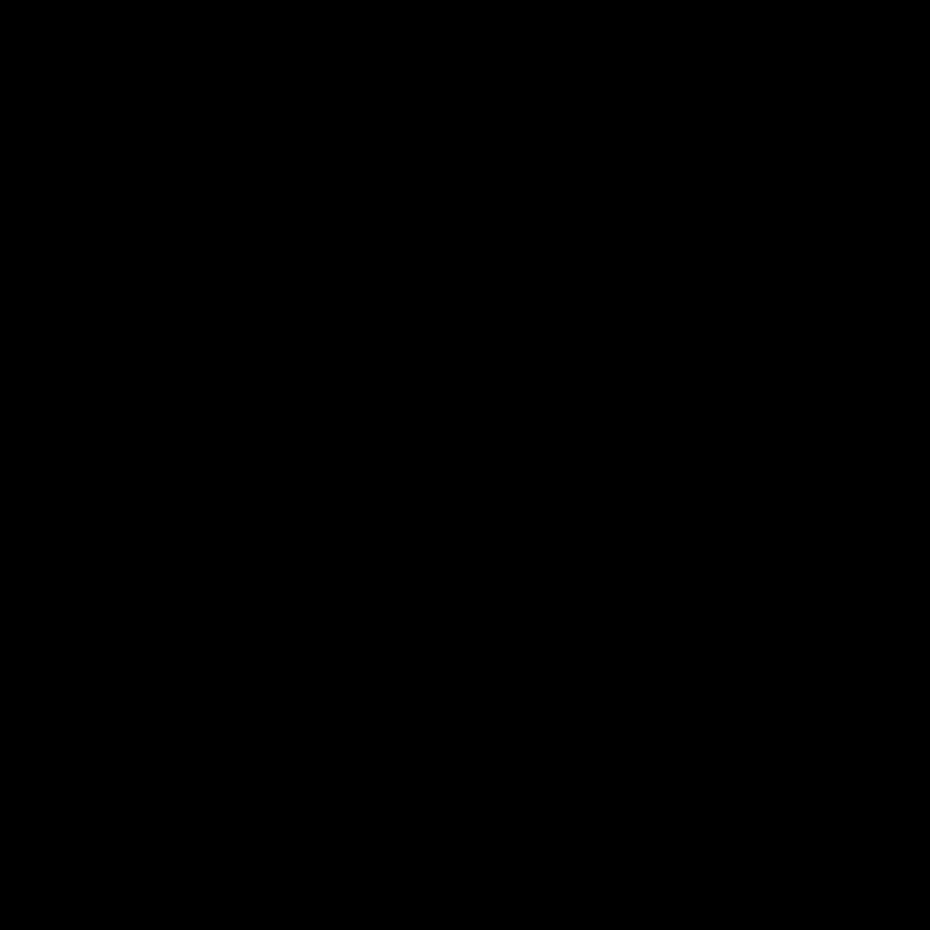}}\\[2ex]
	\subfigure[]{\label{fig:effect_results_wide:c}\includegraphics[width=\picturewidth]{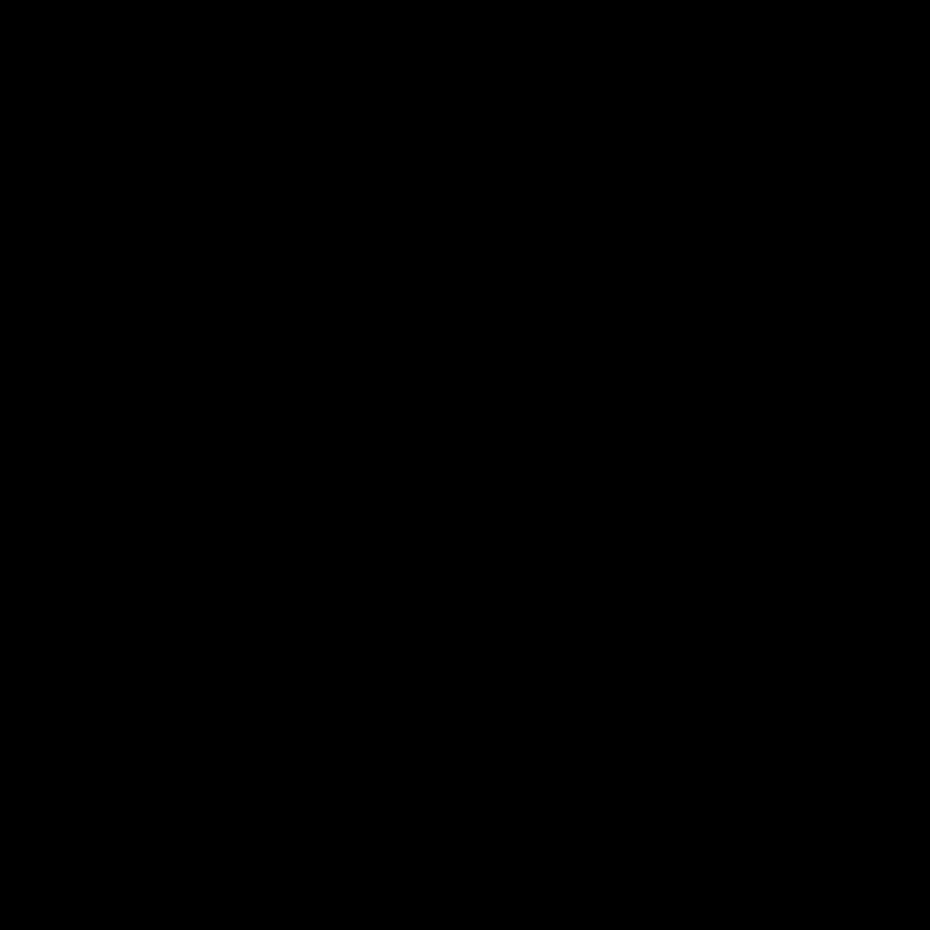}}\hfill%
	\subfigure[]{\label{fig:effect_results_wide:d}\includegraphics[width=\picturewidth]{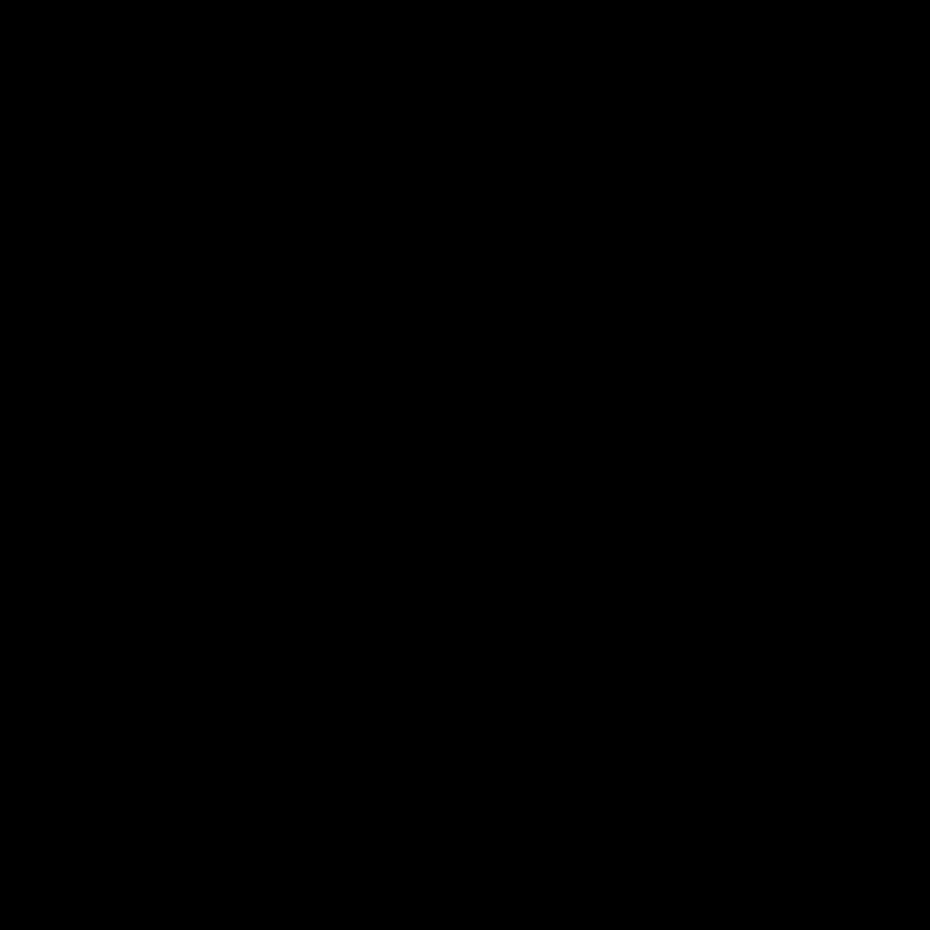}}\vspace{-1.0ex}
	\caption{Large and complete version of the detail sections in \autoref{fig:effect_results:e}--\subref{fig:effect_results:h}: DoG-based edge emphasis. \imageproportion}
	\label{fig:effect_results_wide}
\end{figure*}

\begin{figure*}[t]
	\centering
	\setlength{\subfigcapskip}{-3.5ex}%
	\subfigure[\hspace{\linewidth}]{\label{fig:other_effect_results_large:a}\includegraphics[width=\picturewidth]{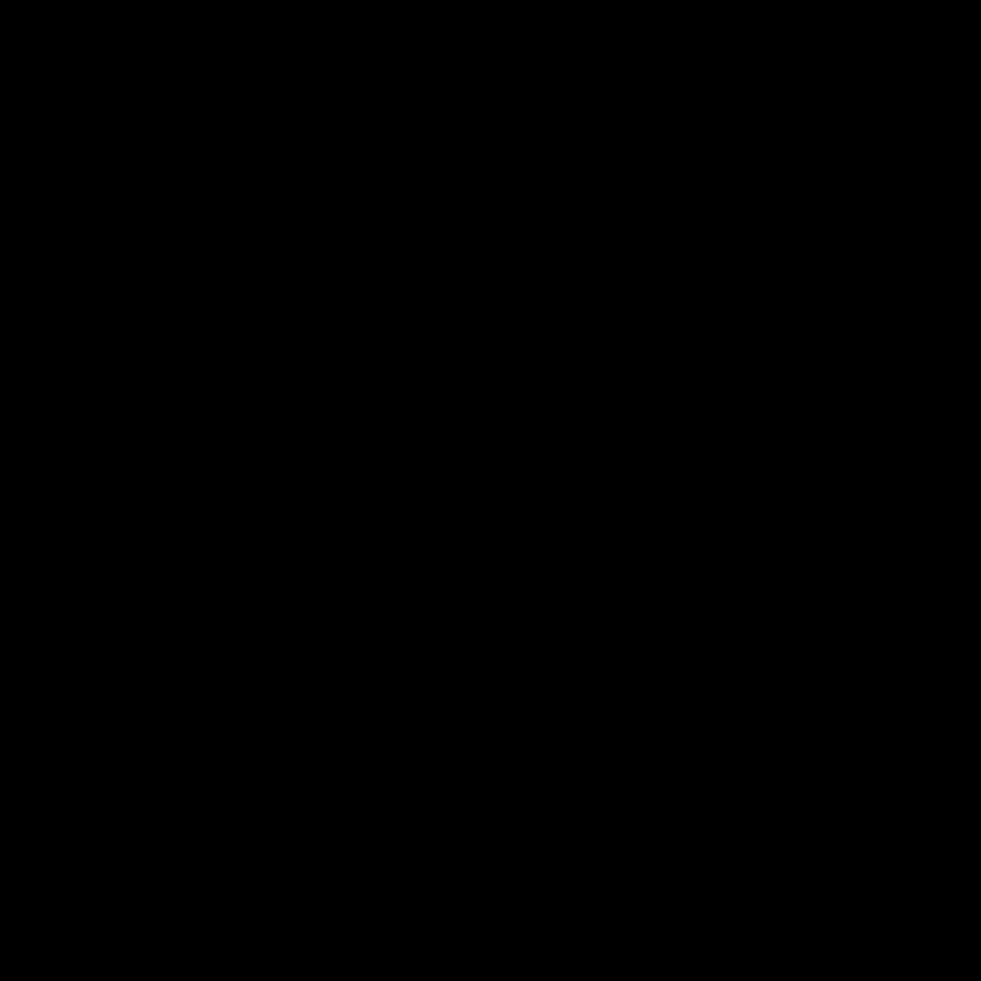}}\hfill
	\subfigure[\hspace{\linewidth}]{\label{fig:other_effect_results_large:b}\includegraphics[width=\picturewidth]{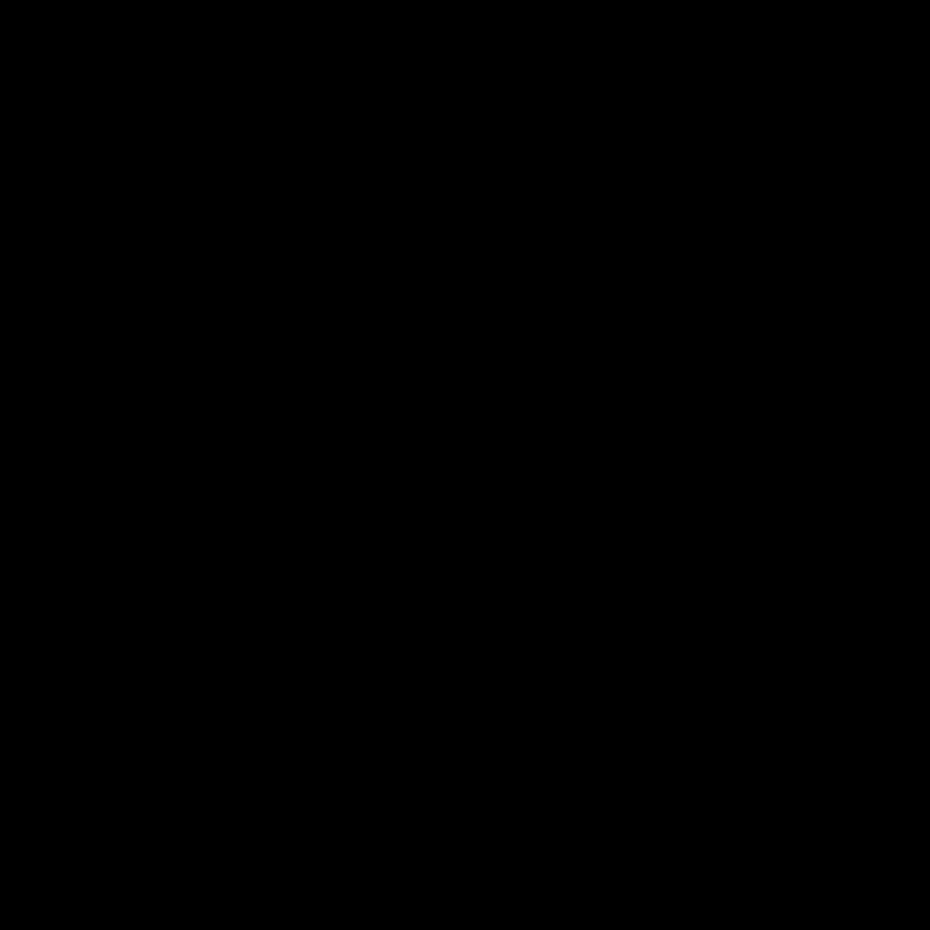}}\\[2ex]
	\subfigure[\hspace{\linewidth}]{~~~~~~\label{fig:other_effect_results_large:c}\includegraphics[width=\picturewidth]{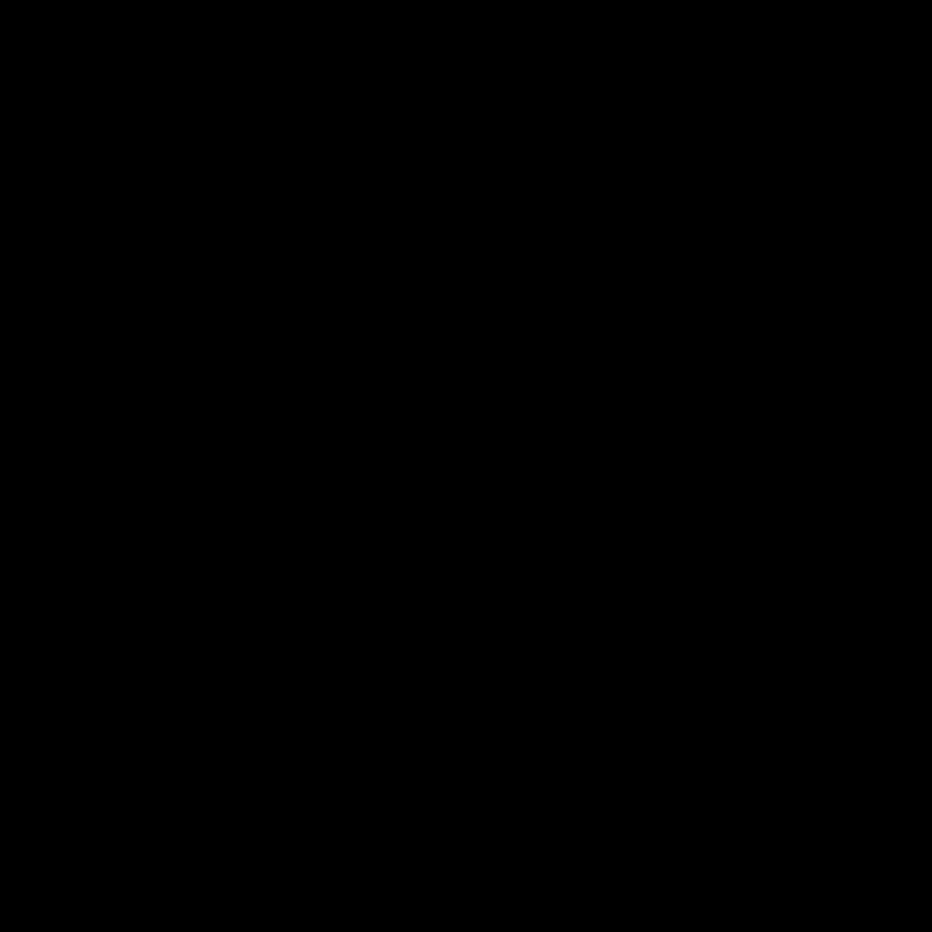}~~~~~~}
	\vspace{-1.0ex}
	\caption{Large versions of the examples in \autoref{fig:other_effect_results}: Examples for effects that make use of masks to adjust the distance function $\Delta$, to control the use of stipples from the input distributions. \imageproportion}
	\label{fig:other_effect_results_large}
\end{figure*}

In Figures~\ref{fig:results_constant_bw_WVS}--\ref{fig:filter_comparison_LoG} we show the complete versions of the images in Figures~\ref{fig:results_constant_bw}--\ref{fig:filter_comparison} we created for our comparison in \autoref{sec:results}. In addition, Figures~\ref{fig:effect_results_wb}--\ref{fig:effect_results_wide} show the complete versions of the examples in \autoref{fig:effect_results}. Finally, \autoref{fig:other_effect_results_large} shows larger versions of the images in \autoref{fig:other_effect_results} so that details are better visible. Note that we were not able to include them in their intended target size of A5 as this would have taken too much space. Instead, we show them at \nicefrac{2}{5} of A5 size.



\subsection{Additional results worthy of note}

\begin{figure*}[!p]
	\centering
	\subfigure[]{\label{fig:effect_results_wide2:a}\includegraphics[width=\picturewidth]{effect-results/new_2_4_3_4_wide/new_wide_full_plant_new_4_8_6_6_floats_14_wide1_fixed}}\hfill%
	\subfigure[]{\label{fig:effect_results_wide2:b}\includegraphics[width=\picturewidth]{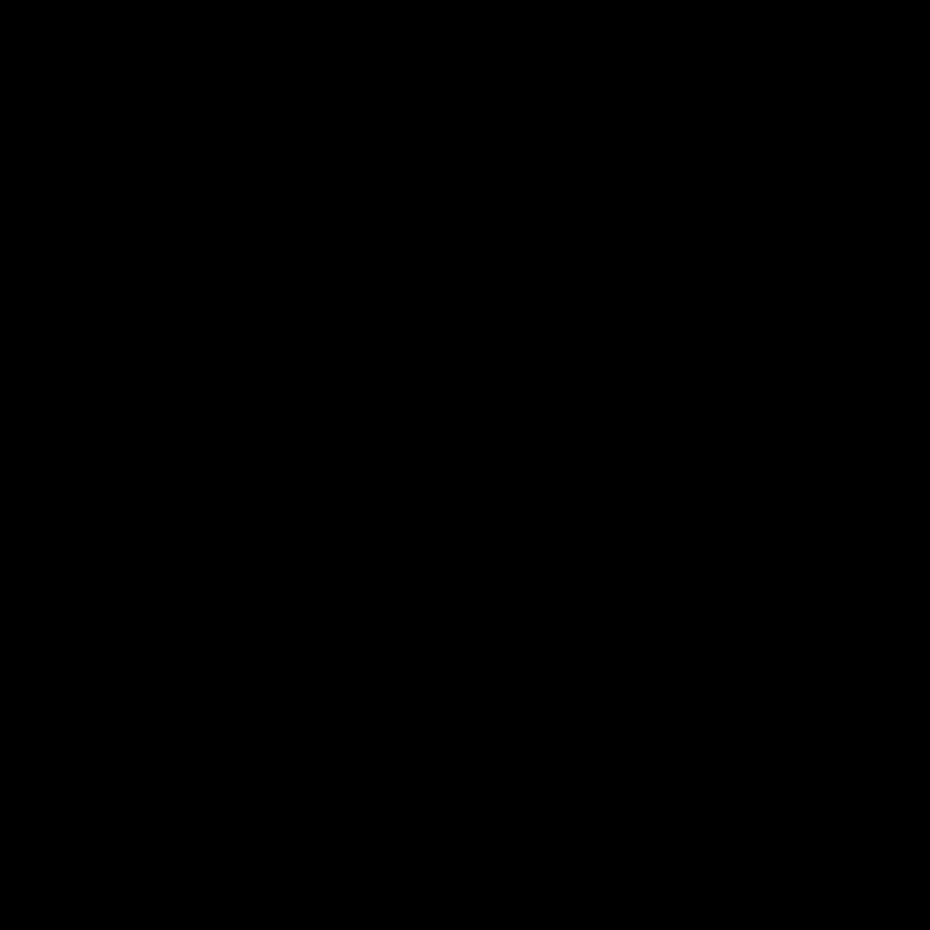}}\vspace{-1.0ex}
	\caption{Comparison of \autoref{fig:effect_results_wide:d} (shown in \subref{fig:effect_results_wide2:a}) with a different filter setup that places more emphasis on the edges (shown in \subref{fig:effect_results_wide2:b}). \imageproportion}
	\label{fig:effect_results_wide2}
\end{figure*}

\begin{figure*}[!p]
	\centering
	\subfigure[]{\label{fig:emphasis_comparison:a}\includegraphics[width=\picturewidth]{other_effects/new_lamp_full_headlight_new_mask2}}\hfill%
	\subfigure[]{\label{fig:emphasis_comparison:b}\includegraphics[width=\picturewidth]{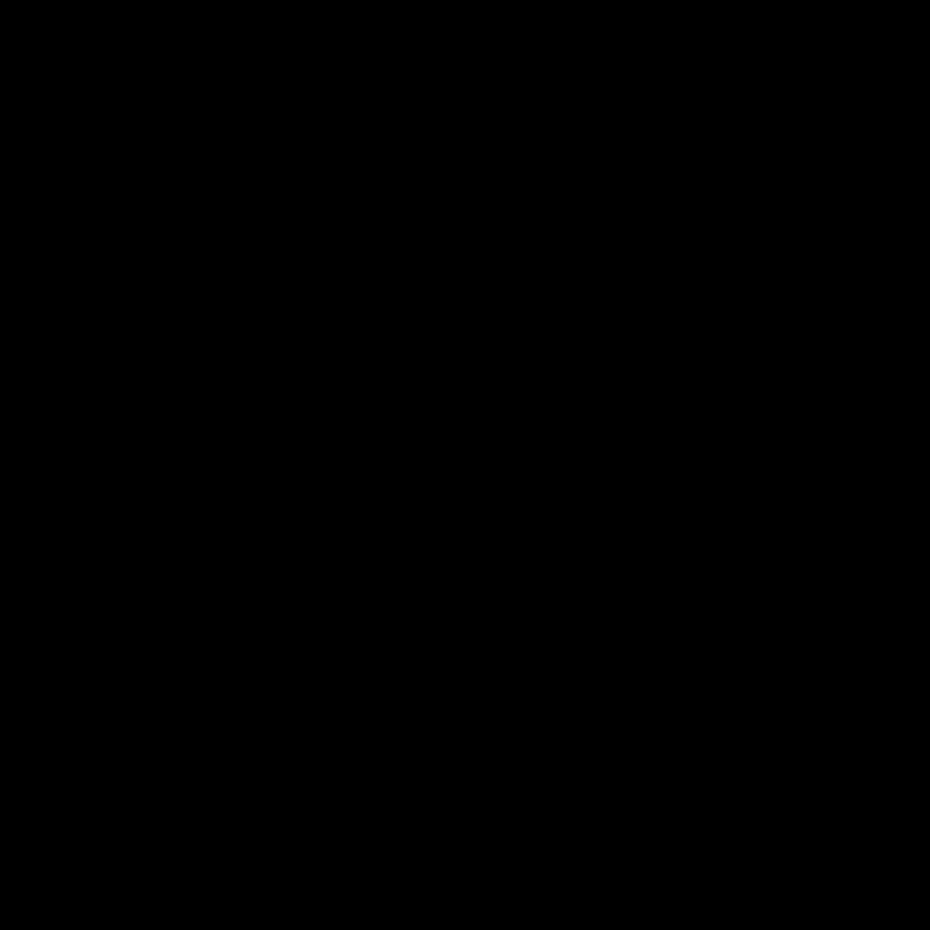}}\vspace{-1.0ex}
	\caption{Comparison of the control of the degree of emphasis with the same mask, based on changing the $b$ parameter of \autoref{eq:final-interp}. The image in \subref{fig:emphasis_comparison:a} is the same as in Figures~\ref{fig:other_effect_results:b} and~\ref{fig:other_effect_results_large:b}. \imageproportion}
	\label{fig:emphasis_comparison}
\end{figure*}

A few additional results can further illustrate our approach. \autoref{fig:effect_results_wide2} compares the approach from \autoref{fig:effect_results_wide} with a different filter setup that places more emphasis on the edges. We also show the possible control of emphasis in \autoref{fig:emphasis_comparison} which compares the previous result in \autoref{fig:other_effect_results:b} or \autoref{fig:other_effect_results_large:b} with a version that uses less emphasis by showing more area stipples. 


\subsection{Demo program and video figure}

We provide a demo implementation of our approach that we submit as additional material. We based it on \textit{StippleShop} (\href{https://github.com/dmperandres/StippleShop}{\texttt{github\discretionary{.}{}{.}com\discretionary{/}{}{/}dmperandres\discretionary{/}{}{/}StippleShop}}), which uses a block-based paradigm and already provides basic methods that are necessary to construct our solution. We added the necessary blocks to implement our new solution including the distance field, the interpolation, and the rendering. We illustrate the use of this demo for IPD in an additional video figure that serves as a tutorial. The tool can also be used to create images for other stippling approaches for a comparison.

	%

%

\end{document}